\documentclass[12pt,english]{article}
 \usepackage{ amsthm, amssymb}
\usepackage {epsfig}

\usepackage{babel}
\usepackage[usenames]{color}
  \voffset
-1.0cm \hoffset -3cm
\begin{document}
\newcommand{\beq}{\begin{equation}}
\newcommand{\eeq}{\end{equation}}
\newcommand{\beqn}{\begin{eqnarray}}
\newcommand{\eeqn}{\end{eqnarray}}
\newcommand{\imunit}{\mathrm i}
\newcommand{\e}{\mathrm e}
\newcommand{\mg}{\ensuremath{m_\frac{1}{2}}}
\newcommand{\tanbeta}{\tan\beta}
\newcommand{\rmGeV}{~\rm GeV}
\newcommand{\Omegachi}{\Omega_\chi}
\newcommand{\NCA}{{\em Nuovo Cimento}}
\newcommand{\NIM}{{\em Nucl. Instrum. Methods}}
\newcommand{\NIMA}{{\em Nucl. Instrum. Methods} {\bf A}}
\newcommand{\NPA}{{\em Nucl. Phys.} A}
\newcommand{\NPB}{{\em Nucl. Phys.} B}
\newcommand{\PLB}{{\em Phys. Lett.} B}
\newcommand{\PRL}{{\em Phys. Rev. Lett.}}
\newcommand{\PRB}{{\em Phys. Rev.} {\bf B}}
\newcommand{\PRD}{{\em Phys. Rev.} D}
\newcommand{\ZPC}{{\em Z. Phys.} C}
\newcommand{\Journal}[4]{#1{\bf #2}, #3 (#4)}
\catcode`\@=11
\def\lsim{\mathrel{\mathpalette\@versim<}}
\def\gsim{\mathrel{\mathpalette\@versim>}}
\def\@versim#1#2{\vcenter{\offinterlineskip
    \ialign{$\m@th#1\hfil##\hfil$\crcr#2\crcr\sim\crcr } }}
\def\btt#1{{\tt$\backslash$#1}}
\def\be{\begin{equation}}
\def\ee{\end{equation}}
\def\bea{\begin{eqnarray}}
\def\eea{\end{eqnarray}}
\def\ba{\begin{array}}
\def\ea{\end{array}}
\def\a{\alpha}
\def\aa{$A-A$ }
\def\ab{$A-B$ }
\def\b{\beta}
\def\bb{$B-B$ }
\def\c{\gamma}
\def\d{\delta}
\def\D{\Delta}
\def\G{\Gamma}
\def\gb{\overline{g}}
\def\gG{\overline{G}}
\def\L{${\cal L}$ }
\def\Ld{\Lambda}
\def\l{\lambda}
\def\Mb{M\"obius }
\def\k{kagom\'e }
\def\0{$\Gamma_0$}
\def\o{\omega}
\def\p{\phi}
\def\pd{\partial}
\def\r{\rho}
\def\r{{\bf R}^{p}}
\def\s{\sigma}
\def\t{\theta}
\def\tt{\tau}
\def\ub{\overline{u}}
\def\ubp{{\overline{u}}_{\pm}}
\def\ul{u^*}
\def\u#1{{\bf #1}}
\def\uu{u_{\pm}}
\def\uup{u_{\pm}}
\def\vt{\vartheta}
\def\wb{\overline{W}}
\def\x{\xi}
\def\xx{\times}
\def\Y{{\cal Y}}
\def\z{\zeta}
\def\zp{{\bf Z}^{p}}
\newcounter{saveeqn}
\newcommand{\alpheqn}{\setcounter{saveeqn}{\value{equation}} \stepcounter{saveeqn}\setcounter{equation}{0} \renewcommand{\theequation}
 {\mbox{\arabic{chapter}.\arabic{saveeqn}\alph{equation}}}}
\newcommand{\reseteqn}{\setcounter{equation}{\value{saveeqn}} \renewcommand{\theequation}{\arabic{chapter}.\arabic{equation}}}
\newcommand{\resetA}{\setcounter{equation}{\value{saveeqn}} \renewcommand{\theequation}{A\arabic{equation}}}
\newcommand{\resetB}{\setcounter{equation}{\value{saveeqn}} \renewcommand{\theequation}{B\arabic{equation}}}
\newcommand{\resetC}{\setcounter{equation}{\value{saveeqn}} \renewcommand{\theequation}{C\arabic{equation}}}

\renewcommand{\theenumi}{\bf\Roman{enumi}.}
\renewcommand{\labelenumi}{\theenumi}

\renewcommand{\theenumii}{\bf\Alph{enumii}.}
\renewcommand{\labelenumii}{\theenumii}

\renewcommand{\theenumiii}{\bf(\roman{enumiii})}
\renewcommand{\labelenumiii}{\theenumiii}

\renewcommand{\theenumiv}{\bf(\alph{enumiv})}
\renewcommand{\labelenumiv}{\theenumiv}
\definecolor{purple}{rgb}{1,0,1}
\definecolor{turquoise}{rgb}{0,1,1}
\definecolor{gold}{rgb}{1,1,0}
\definecolor{me}{cmyk}{1,0.4,0.4,0.4}

\begin{titlepage}

\begin{center}
{\Huge {\bf Problems and Solutions in a Graduate Course in Classical Electrodynamics  (1)}}\\
\end{center}
\vspace{6cm}
\renewcommand{\thefootnote}
{\fnsymbol{footnote}}
      \begin{center}
 \large{ \textsl{\bf\textsf{Raza M.
 Syed}}}\\
 \noindent {{Department of Physics, Northeastern University,
360 Huntington Ave., Boston, MA 02115-5000.}}
\end{center}
\begin{center}
\vspace{2cm}
\textsf{ABSTRACT}
\end{center}
\textsf{The following is the very first set of the series in 'Problems and Solutions in
a Graduate Course in Classical Electrodynamics'. In each of the sets of the problems we intend to follow a theme, which not only makes it unique but also deals with the investigation of a certain class of problems in a thorough and rigorous manner. Figures are not included.
}

 \end{titlepage}
\begin{center}
\huge{\bf{Two Point Charge System }}
\end{center}
\vspace{1cm}
\noindent\textsf{Two point charges $q$ and $\lambda q$ and masses $m$ and $\sigma m$, respectively in vacuum are located at the points A$(a,0)$ and B$(\mu a, 0)$ in the $xy$ plane as shown in the figure. Assume throughout that $a>0$ and $q>0$.}\\
\begin{center}
\textbf{PROBLEM 1}
\end{center}
\begin{enumerate}
  \item
  \begin{enumerate}
    \item \textsf{If the sum of the charges is fixed, what value of $\lambda$ will maximize the electric force between them? Is the force attractive or repulsive?}
    \item \textsf{If the midpoint of the line joining the charges is fixed, what value of $\mu$ will maximize the electric force between them? }
  \end{enumerate}
  \item
  \begin{enumerate}
    \item \textsf{Show that the equation for the lines of force in the $xy$ plane is given by}
    $$\label{1}\frac{\lambda(x-\mu a)}{\sqrt{(x-\mu a)^2+y^2}}+\frac{\lambda(x-a)}{\sqrt{(x-a)^2+y^2}}=C,$$
    \textsf{where $C$ is a constant.}
    \item
    \begin{enumerate}
    \item \textsf{Show that the equation for the line of force in part \ref{1}} \textsf{can be written as}
    $$\label{2}\cos\theta_1+\lambda\cos\theta_2=C.$$ \textsf{Interpret the angles $\theta_1$ and $\theta_2$ geometrically.}
    \item \textsf{Find the range of values of $C$ for which}
    \begin{enumerate}
    \item \textsf{the equation in part \ref{2}} \textsf{has a solution;}
    \item \textsf{the lines of force pass through charge $q$;}
    \item \textsf{the lines of force pass through charge $\lambda q$.}
   \end{enumerate}
   \end{enumerate}
   \item \textsf{What are the equations for the lines of force in the $xz$ and $yz$ planes?}
   \item
   \begin{enumerate}
   \item \textsf{Using the result of part \ref{1},}\textsf{show that the equation for the lines of force of an electric dipole is}
       $$\sqrt{(x^2+y^2)^3}=-\frac{2a}{C}y^2.$$
       \item \textsf{Hence, show that the equation describing the equipotential lines of an electric dipole is}
      $$\sqrt{(x^2+y^2)^3}=Kx,$$
      \textsf{where $K$ is a constant.}
   \end{enumerate}
  \end{enumerate}
  \item \textsf{Consider a plane, which is a perpendicular bisector of the line joining the charges.}
  \begin{enumerate}
  \item
  \begin{enumerate}
  \item \label{3} \textsf{For a fixed value of $\mu$ and $\lambda$, show that the electric field attains its maximum on a circle on the plane whose radius is}
      $$\frac{a}{2}\left|\frac{1-\mu}{1+\lambda}\right|\sqrt{3-(\lambda-2)^2}.$$
\item \textsf{Find the magnitude of the electric field everywhere on the circle.}
\item \textsf{State the necessary restrictions on the parameters $\mu$ and $\lambda$.}
  \end{enumerate}
  \item
  \begin{enumerate}
  \item\label{3'} \textsf{With $\lambda$ fixed, show that the electric field is maximized, if the parameter $\mu$ is chosen on a circle on the plane whose radius is}
      $$\frac{a}{2}\frac{\left|(1-\mu)(1-\lambda)\right|}{\sqrt{3-(\lambda+2)^2}}.$$
\item \textsf{Find the magnitude of the electric field in this case.}
\item \textsf{State the necessary restrictions on the parameter
$\lambda$.}
  \end{enumerate}
  \item
  \begin{enumerate}
  \item \label{3''}\textsf{With $\mu$ fixed, show that the electric field is maximized, if the parameter $\lambda$ is determined on a circle on the plane whose radius is}
      $$\frac{a}{2}\left|(1-\mu)\right|\sqrt{\frac{1-\lambda}{1+\lambda}}.$$
\item \textsf{What is the corresponding magnitude of the maximum electric field.}
\end{enumerate}
  \end{enumerate}
  \item
  \begin{enumerate}
  \item\textsf{Show that the electric flux passing through the circle described in parts \ref{3}, \ref{3'} and \ref{3''}} \textsf{is given by}
      $$2\pi q\sqrt{\frac{2}{3}} \left(\frac{1-\lambda}{\sqrt \lambda}\right)\left[\pm\lambda -\sqrt{6\lambda}\pm 1\right],$$
      $$2\pi q\sqrt{\frac{2}{3}} \left(\frac{1-\lambda}{\sqrt {-\lambda}}\right)\left[\pm\sqrt{3-(\lambda+2)^2}- 1\right],$$
      $$2{\sqrt 2}\pi q \left(1-\lambda\right)\left[\pm\sqrt{1+\lambda}- \sqrt{2}\right],$$
      \textsf{respectively. The $\pm$ sign corresponds to $\mu <1$ and $\mu >1$, respectively.}
      \item \textsf{Show that if,}
       \begin{eqnarray}
       \lambda=\displaystyle-\left[\frac{\displaystyle\frac{a}{\sqrt{a^2+b^2}}-1}{\displaystyle\frac{\mu a}{\sqrt{(\mu a)^2+b^2}}-1}\right],\nonumber
       \end{eqnarray}
      \textsf{then there is no net flux passing through the circle $y^2+z^2=b^2$ at $x=0$.}
  \end{enumerate}
  \item
  \begin{enumerate}\label{4'}
  \item\textsf{Show that the electric field is zero (\textit{neutral point}) at}
      \begin{eqnarray}
      \left(a\left[\frac{\mu\pm \sqrt{|\lambda|}}{1\pm\sqrt{|\lambda|}}\right],0,0\right).\nonumber
      \end{eqnarray}
      \item \textsf{Find the range of values of $\mu$ and $\lambda$ for which each of the location of the neutral points is valid.}
  \end{enumerate}
  \item
  \begin{enumerate}
  \item\label{4}\textsf{Show that the equipotential surface for which the electric potential is zero
  is a sphere on  $\overline{\textsf{I$_1$ I$_2$}}$ as diameter, where
 $$\frac{\textsf{AI$_1$}}{\textsf{BI$_1$}}=\frac{1}{\left|\lambda\right|}=\frac{\textsf{AI$_2$}}
 {\textsf{BI$_2$}}.$$
 That is, the points I$_1$ and I$_2$ divide $\overline{\textsf{AB}}$ internally and externally in the ratio
 $1:\left| \lambda\right|$}

\item\textsf{How would the answer to part \ref{4}} \textsf{change, when two charges of opposite signs and equal magnitudes are considered?}
      \item \textsf{State the necessary conditions on the parameters $\mu$ and $\lambda$ in order for the the result in part \ref{4}} \textsf{to be valid.}
  \end{enumerate}
  \item
  \begin{enumerate}
  \item\label{5}\textsf{Show that the asymptote (tangent at infinity) to any line of force must always
  pass through a fixed point G on $\overline{\textsf{AB}}$ such that
  $$\frac{\textsf{AG}}{\textsf{BG}}=|\lambda|.$$}
      \item\textsf{What conclusion can you draw from the result of part \ref{5}?}
      \item \textsf{Show that the point G is the ``center of gravity" of the charges.}
  \end{enumerate}
  \item \textsf{Assume $\lambda >0$ for this part of the problem.}
  \begin{enumerate}
  \item\label{6}\textsf{Show that for a line of force which starts from charge $q$ making an angle $\alpha$ with $\overline{\textsf{AB}}$ satisfies}
      $$\cos^2\left(\frac{1}{2}\theta_1\right)+|\lambda|\cos^2\left(\frac{1}{2}\theta_2\right)
      =\cos^2\left(\frac{1}{2}\alpha\right),$$
      $\forall$ values of $\theta_1$ and $\theta_2$ [\textsf{see part \ref{2}}].
      \item
      \begin{enumerate}
      \item\label{7}\textsf{Show that the asymptote to the line of force considered in part \ref{6}} \textsf{makes an angle}
          $$2\cos^{-1}\left[\frac{1}{\sqrt{1+|\lambda|}}\cos\left(\frac{1}{2}\alpha\right)\right]$$
          \textsf{with $\overline{\textsf{AB}}$.}
      \item \label{8}\textsf{For what values of $\alpha$ will the result in part \ref{7}} \textsf{hold true?}
      \item\textsf{Show that the equation of this asymptote is}
          $$y=\pm\sqrt{\left(\frac{1+|\lambda|}{\cos\alpha-|\lambda|}\right)^2-1}\left[x-\left(\frac{1+\mu|\lambda|}
          {1+|\lambda|}\right)a\right].$$
      \end{enumerate}
      \item \textsf{How would the result of parts \ref{6}, \ref{7}} \textsf{and} \textsf{\ref{8}} \textsf{change, if instead a line of force emanating from charge $\lambda q$ making an angle $\alpha$ with with $\overline{\textsf{AB}}$, is considered?}
  \end{enumerate}
  \item \textsf{Assume $\lambda <0$ for this part of the problem.}
  \begin{enumerate}
  \item\label{9}\textsf{Show that for a line of force which starts from charge $q$ making an angle $\alpha$ with $\overline{\textsf{AB}}$ satisfies}
      $$\cos^2\left(\frac{1}{2}\theta_1\right)-|\lambda|\cos^2\left(\frac{1}{2}\theta_2\right)
      =\cos^2\left(\frac{1}{2}\alpha\right),$$
      $\forall$ values of $\theta_1$ and $\theta_2$.
      \item
      \begin{enumerate}
      \item\label{10}\textsf{If the line of force considered in part \ref{9}} \textsf{is to end at charge $\lambda q$, then show that it must make an angle }
          $$2\cos^{-1}\left[\frac{1}{\sqrt{|\lambda|}}\sin\left(\frac{1}{2}\alpha\right)\right]$$
          \textsf{at B with $\overline{\textsf{AB}}$.}
      \item \textsf{Find the range of values of  $\alpha$ and $\lambda$ for which the result in part \ref{10}} \textsf{is valid.}
      \end{enumerate}
      \item
      \begin{enumerate}
      \item \label{11}\textsf{If the line of force considered in part \ref{9},} \textsf{is to go to infinity, then show that the asymptote to this line of force must make an angle }
          $$2\cos^{-1}\left[\frac{1}{\sqrt{{1-|\lambda|}}}\cos\left(\frac{1}{2}\alpha\right)\right]$$ \textsf{with $\overline{\textsf{AB}}$.}
          \item \textsf{Find the range of values of  $\alpha$ and $\lambda$ for which the result in part \ref{11}} \textsf{is valid.}
             \item\textsf{Show that the equation of this asymptote is}
          $$y=\pm\sqrt{\left(\frac{1-|\lambda|}{\cos\alpha+|\lambda|}\right)^2-1}\left[x-\left(\frac{1-\mu|\lambda|}
          {1-|\lambda|}\right)a\right].$$
          \end{enumerate}
          \item
      \begin{enumerate}
      \item \label{12}\textsf{If the line of force considered in part \ref{9},} \textsf{is an \textit{extreme} line of force from charge $q$ to charge $\lambda q$ (a line of force that separates the line of force going from charge $q$ to $\lambda q$ from those going from charge $q$ to infinity), then}
          $$\alpha=\cos^{-1}\left(1-2|\lambda|\right).$$
          \item \textsf{For what values of $\lambda$ will the result in part \ref{12}} \textsf{hold true?}
          \end{enumerate}
          \item
      \begin{enumerate}
      \item \label{13}\textsf{If the line of force considered in part \ref{9},} \textsf{is to meet the plane that bisects $\overline{\textsf{AB}}$ at right angles, in C, then show that the angle between $\overline{\textsf{AB}}$ and $\overline{\textsf{AC}}$ is }
          $$2\sin^{-1}\left[\frac{1}{\sqrt{{1+|\lambda|}}}\sin\left(\frac{1}{2}\alpha\right)\right].$$
          \item \textsf{For what values of  $\alpha$ will the  result in part \ref{13}} \textsf{hold true.}
             \item\textsf{Show that this line of force which crosses the plane is at a distance from $\overline{\textsf{AB}}$ not greater than }
          $$\left|\mu-1\right|\frac{\sqrt{|\lambda|}}{1-|\lambda|}a.$$
          \end{enumerate}
          \item
      \begin{enumerate}
      \item \label{14}\textsf{Show that for a line of force which terminates at charge $\lambda q$ making an angle $\alpha$ with $\overline{\textsf{AB}}$ satisfies }
          $$|\lambda|\sin^2\left(\frac{1}{2}\theta_2\right)-\sin^2\left(\frac{1}{2}\theta_1\right)
      =|\lambda|\sin^2\left(\frac{1}{2}\alpha\right),$$
      $\forall$ values of $\theta_1$ and $\theta_2$.
          \item \textsf{If the line of force considered in part \ref{14},} \textsf{is restricted to have been originated from charge $q$, then show that it must make an angle}
               $$2\sin^{-1}\left[\sqrt{|\lambda|}\cos\left(\frac{1}{2}\alpha\right)\right],$$
               \textsf{at A with $\overline{\textsf{AB}}$. For what values of $\alpha$ and $\lambda$ will this result hold true.}
             \item\textsf{If the line of force considered in \ref{14},} \textsf{is constrained to have been originated from infinity, then show that the asymptote to this line of force must make an angle }
                 $$2\sin^{-1}\left[\sqrt{\frac{|\lambda|}{|\lambda|-1}}\sin\left(\frac{1}{2}\alpha\right)\right],$$
                 with $\overline{\textsf{AB}}$.\textsf{ Find the range of values for $\alpha$ and $\lambda$ for which this result is valid.}
          \item \textsf{If the line of force considered in \ref{14},} \textsf{is an extreme line of force, then show that}
              $$\alpha=\cos^{-1}\left(\frac{2}{|\lambda|}-1 \right).$$
              \textsf{Find the range of values for $\lambda$ for which this result is valid.}
           \item \textsf{If the line of force considered in part \ref{14},} \textsf{has crossed the plane that bisects $\overline{\textsf{AB}}$ at right angles, in C, then show that the angle between $\overline{\textsf{AB}}$ and $\overline{\textsf{AC}}$ is}
               $$2\cos^{-1}\left[\sqrt{\frac{|\lambda|}{1+|\lambda|}}\cos\left(\frac{1}{2}\alpha\right)\right].$$
               \textsf{Find the range of values for $\alpha$ for which this result is valid. At what maximum distance from $\overline{\textsf{AB}}$  can this line of force cross the plane.}
          \end{enumerate}
          \end{enumerate}
         \item
  \begin{enumerate}
      \item\textsf{Assume $\lambda >0$ for this part of the problem.}
      \begin{enumerate}
      \item\textsf{Show that the (limiting) line of force through the neutral point on $\overline{\textsf{AB}}$ satisfies }
          $$\sin^2\left(\frac{1}{2}\theta_1\right)=|\lambda|\cos^2\left(\frac{1}{2}\theta_2\right),$$
      \textsf{$\forall$ values of $\theta_1$ and $\theta_2$.}
      \item \textsf{Show that this limiting line of force has the asymptote}
          $$y=\pm\frac{2\sqrt{|\lambda|}}{1-|\lambda|}\left[x-\left(\frac{1+\mu|\lambda|}
          {1+|\lambda|}\right)a\right].$$
      \end{enumerate}
      \item\textsf{Assume $\lambda <0$ for this part of the problem.}
      \begin{enumerate}
      \item\textsf{Show that the (limiting) line of force through the neutral point on $\overline{\textsf{AB}}$ satisfies }
            \begin{eqnarray}
          \sin^2\left(\frac{1}{2}\theta_1\right)&=&|\lambda|\sin^2\left(\frac{1}{2}\theta_2\right) ~~~\textsf{if}~~~-1<\lambda<0 ,\nonumber\\ \cos^2\left(\frac{1}{2}\theta_1\right)&=&|\lambda|\cos^2\left(\frac{1}{2}\theta_2\right) ~~~\textsf{if}~~~\lambda<-1,\nonumber
          \end{eqnarray}
      \textsf{$\forall$ values of $\theta_1$ and $\theta_2$.}
      \item
      \begin{enumerate}\label{15}
      \item \textsf{In the case when $-1<\lambda<0$, find the angle w.r.t. $\overline{\textsf{AB}}$ at A, the line of force through neutral point, leaves charge $q$.}
      \item \textsf{In the case when $\lambda < -1$, find the angle w.r.t. $\overline{\textsf{AB}}$ at B, the line of force through neutral point, ends at charge $\lambda q$.}
      \end{enumerate}
      \item\textsf{What conclusion can you draw from the results of part \ref{15}}
      \end{enumerate}
  \end{enumerate}
  \item \textsf{Assume $\lambda <0$ for this part of the problem.}
  \begin{enumerate}
  \item\textsf{In particular, consider the situation when $-1<\lambda<0$. Further let P be a point on the line of force through the neutral point, N. Show that, if bisectors of the $\angle$PAN and $\angle$PBN meet at point Q, the locus of Q is the circle on $\overline{\textsf{MN}}$ as diameter, where point M lies on $\overline{\textsf{AB}}$ and}
      $$\frac{\textsf{AM}}{\textsf{BM}}=\frac{1}{\sqrt{|\lambda|}}.$$
      \textsf{Compare this ratio with }$\frac{\textsf{AN}}{\textsf{BN}}.$
      \item \textsf{Show that the locus of the points at which the lines of force are parallel to  $\overline{\textsf{AB}}$ is a sphere of radius}
          $$\sqrt[3]{|\lambda|}\left|\frac{\mu-1}{1-\sqrt[3]{|\lambda|^2}}\right|a.$$
  \end{enumerate}
  \item
  \begin{enumerate}
  \item\textsf{Show that the ratio of the component of electric field perpendicular to $\overline{\textsf{AB}}$ to the component of electric field parallel to $\overline{\textsf{AB}}$ at a point on the limiting line of force close to neutral point is approximately}
      \begin{eqnarray}
        -\frac{2}{3}\left(\frac{1}{\pi-\theta_1-\theta_2}\right)&& ~~~ \textsf{if}~~~ \lambda>0,\nonumber\\
        -\frac{2}{3}\left(\frac{1}{\theta_1+\theta_2}\right)&& ~~~ \textsf{if}~~~ -1<\lambda<0,\nonumber\\
        -\frac{2}{3}\left(\frac{1}{\theta_1+\theta_2-2\pi}\right)&&~~~ \textsf{if}~~~ \lambda<-1.\nonumber
          \end{eqnarray}
      \item \textsf{At what angle does the line of force through N crosses $\overline{\textsf{AB}}$. }
  \end{enumerate}
  \item
  \begin{enumerate}
  \item\textsf{Show that the form of equipotential surfaces in the neighborhood of the neutral point are}
  \begin{itemize}
  \item \textsf{hyperboloid of one sheet}\\
  \textsf{or}
  \item\textsf{hyperboloid of two sheets}\\
  \textsf{or}
  \item \textsf{right circular cone}
  \end{itemize}
          \item \textsf{Show that the equipotential near the neutral point makes a constant angle
          $$\tan^{-1}(\sqrt2)$$ }
         \textsf{ with }$\overline{\textsf{AB}}$
         \item \textsf{Show that the lines of force near the neutral point are the curves}
         $$y^2\left[x-\left(\frac{\mu\pm\sqrt{|\lambda|}}{1\pm \sqrt{|\lambda|}}\right)a\right]=\textsf{Constant}$$
  \end{enumerate}
  \end{enumerate}
  \vspace{0.5cm}
\begin{center}
\textbf{PROBLEM 2}
\end{center}
\textsf{You may assume throughout this problem that $\mu>1$. }
\begin{enumerate}
\item
  \begin{enumerate}
  \item\textsf{If the charges in vacuum are now released from rest from their initial positions,}
  \begin{enumerate}
  \item  \textsf{show that at time $t$, their relative position $x(t)\equiv x_2(t)-x_1(t)$ is such that
        \begin{eqnarray}
        t=\sqrt{\frac{ma}{2q^2}\frac{\sigma(\mu-1)}{|\lambda|(\sigma+1)}}\times
        \left\{\displaystyle\matrix{\displaystyle\frac{a}{2}(\mu-1)\left[\pi-\displaystyle
        \sin^{-1}\sqrt{\frac{2x}{a(\mu-1)}}~\right]
        -\sqrt{x\left[a(\mu-1)-x\right]}&\textsf{if}&\lambda<0\cr
        \cr
        \displaystyle\frac{1}{2}a(\mu-1)\cosh^{-1}\left[\displaystyle\frac{2x}{a(\mu-1)}-1\right]+\sqrt{x\left[x-a(\mu-1)
        \right]}
        &\textsf{if}&\lambda>0}\right.\nonumber
        \end{eqnarray}
        where $x_2$ and $x_1$ are the positions of the charge $\lambda q$ and $q$ at time $t$, respectively.}
      \item \textsf{For the case of charges with unlike signs, find the collision time.}
\end{enumerate}

\item\textsf{Now assume that charges are immersed in a medium where the Coulomb force ($F_{1,2}$) is proportional to the velocity ($v_{1,2}$) rather than acceleration of the particle: $F_2=\beta v_2$ and $F_1=v_1$, where $\beta$ is a proportionality constant. The charges are now released from rest form their initial positions.}
  \begin{enumerate}
  \item \textsf{Show that the relative relative position $x$ and time $t$, in this case are related by
       \begin{eqnarray}
        t=\frac{1}{3q^2}\frac{\beta}{|\lambda|(1+\beta)}\times
        \left\{\displaystyle\matrix{\left[\displaystyle\frac{a^3}{8}(\mu-1)^3-x^3\right]
        &\textsf{if}&\lambda<0\cr
        \cr
        \left[x^3-\displaystyle\frac{a^3}{8}(\mu-1)^3\right]
        &\textsf{if}&\lambda>0}\right.\nonumber
        \end{eqnarray}
      }
\item \textsf{For the case of charges with unlike signs, find the collision time.}
\end{enumerate}
 \end{enumerate}
 \item \textsf{Assume that the charges are connected by means of a light, nonconducting inextensible string.}
  \begin{enumerate}
  \item  \textsf{The charge $q$ is fixed while the charge $\lambda q$ ($\lambda>0$) is free to rotate about $q$. Show that the minimum velocity with which the charge $\lambda q$ is projected so that it completes one revolution is given by
     $$\sqrt{ga(\mu-1)(\sigma+2+2\cos\alpha)-\frac{|\lambda|q^2}{ma(\mu-1)}},$$
     where   $\alpha$ is the angle the string makes with the downward vertical, just before the charge
     $\lambda q$ is projected.}
  \item  \textsf{Take $\lambda<0$. The entire charge and the string assembly is released and a uniform electric field, $-E_0\hat{\mathbf {x}}$ is turned on. Show that
      $$\mu=1+\frac{1}{a}\sqrt{\frac{q}{E_0}\frac{|\lambda|(1+\sigma)}{\sigma+|\lambda|}},$$
      in order for the string to remain taut at all times during the subsequent motion of the assembly.}
 \end{enumerate}
 \item \textsf{Assume that the charges $q$ and $\lambda q$ ($\lambda<0$) are held together at the ends of a massless rigid non-conducting rod. The whole arrangement is immersed in a region of a uniform electric field $-E_0\hat{\mathbf {x}}$, with the rod constrained to only rotate about an axis through its center and perpendicular to its length. If the rod is rotated through a small angle from its equilibrium position, show that it performs simple harmonic motion with time period
     $$2\pi\sqrt{\frac{ma}{2qE_0}\frac{(\mu-1)(\sigma+1)}{|\lambda+1|}}$$
     }
\end{enumerate}

\newpage
\begin{center}
\fbox{\Large{\bf{\textsc{SOLUTIONS }}}}
\end{center}
\vspace{0.5cm}
\begin{center}
\textbf{PROBLEM 1}
\end{center}
\vspace{0.5cm}
\begin{enumerate}
  \item
  \begin{enumerate}
\item \textsf{We are given that $\displaystyle q+\lambda q \equiv \textsf {\textsf{constant}}~(Q)$. This implies $\displaystyle q=\frac{Q}
{1+\lambda}$.
    The Coulomb force, $F_{1}$ between charges $q$ and $\lambda q$ is given by
\begin{eqnarray}
F_{1}&=&\frac{(q)(\lambda q)}{(a-\mu a)^2}= \left[\frac{Q}{a(1-\mu)}\right]^2\frac{\lambda}{(1+\lambda)^2}\nonumber\\
    \frac{\partial F_{1}}{\partial \lambda}&=&\left[\frac{Q}{a(1-\mu)}\right]^2\frac{(1+\lambda)^2-2\lambda(1+\lambda)}
    {(1+\lambda)^4}\nonumber\\
    &=&
   \left[\frac{Q}{a(1-\mu)}\right]^2\frac{1-\lambda}{(1+\lambda)^3}=0\nonumber\\
    &\Rightarrow& ~~ \fbox{$\displaystyle\lambda=1$}\nonumber\\
    \textsf{Further, note that}:~~ \frac{\partial^2 F_{1}}{\partial \lambda^2}&=&
\left[\frac{Q}{a(1-\mu)}\right]^2\frac{-(1+\lambda)^3-3(1\lambda)(1+\lambda)^2}{(1+\lambda)^6}\nonumber\\
&=&2\left[\frac{Q}{a(1-\mu)}\right]^2\frac{\lambda
-2}{(1+\lambda)^4}< 0 ~~\textsf{for}~~\lambda =1\nonumber\\
    \nonumber
    \end{eqnarray}
    Thus for $\displaystyle\lambda=1,~ q=\frac{Q}{2}$.
    This means the total fixed charge will be divided equally among the the two
    charges to give the maximum force of \emph{repulsion}:
   \begin{eqnarray}
   \fbox{$\displaystyle \left. F_{1}\right|_{\textsf{max}}=
    \left[\frac{q}{a(1-\mu)}\right]^2$.}
  \nonumber
    \end{eqnarray}}
     \vspace {0.3cm}
    \item \textsf{If the midpoint of the line joining the two charges is fixed, then $\displaystyle\frac{a+\mu a}{2}=
    \textsf{constant}~(d)$. Therefore, $\displaystyle a=\frac{2d}{1+\mu}$. The Coulomb force, $F_2$ in this case is\\
\begin{eqnarray}
F_{2}&=&\frac{(q)(\lambda q)}{(a-\mu a)^2}=\frac{\lambda
q^2}{4d^2}\left(\frac{1-\mu}{1+\mu}\right)^2\nonumber\\
 \frac{\partial F_{2}}{\partial \mu}&=&\left(\frac{\lambda
q^2}{4d^2}\right)\frac{(1-\mu)^22(1+\mu)-(1+\mu)^2(2\mu-2)}{(1-\mu)^4}\nonumber\\
&=& \frac{\lambda q^2}{d^2}\frac{1+\mu}{(1-\mu)^3}=0\nonumber\\
  &\Rightarrow& ~~\fbox{$\displaystyle\mu =-1 $}\nonumber\\
 \textsf{Further,}~~\frac{\partial^2 F_{2}}{\partial \lambda^2}&=&\left(\frac{\lambda
q^2}{4d^2}\right)\frac{(1-\mu)^3+3(1-\mu)^2(1+\mu)}{(1-\mu)^6}\nonumber\\
&=&\left(\frac{2\lambda
q^2}{d^2}\right)\frac{\mu+2}{(\mu-1)^4}\nonumber\\
\left.\frac{\partial^2 F_{2}}{\partial
\lambda^2}\right|_{\mu=-1}&=&\frac{\lambda q^2}{8d^2}\left\{\matrix{<0
~~\textsf{if}~~\lambda <0\cr
>0
~~\textsf{if}~~\lambda >0
}\right.
  \nonumber
\end{eqnarray}
Thus for $\mu=-1,~ d=0$. That is, the charges are placed
symmetrically about the origin, giving rise to a maximum
\emph{attractive} (\emph{repulsive}) force for $\lambda<0$ ($\lambda>0$):
\begin{eqnarray}
\fbox{$\displaystyle \left. F_{2}\right|_{\textsf{max}}=
    -|\lambda|\left(\frac{q}{2a}\right)^2$}.
\nonumber
    \end{eqnarray}}
  \end{enumerate}
  \vspace {0.6cm}
  \item
  \begin{enumerate}
    \item \textsf{The electric field, $\mathbf{E}$ for the system is given by
    $$\mathbf{E}=\frac{q(\mathbf{r}-\mathbf{r_1})}{|\mathbf{r}-\mathbf{r_1}|^3}+
    \frac{\lambda q(\mathbf{r}-\mathbf{r_2})}{|\mathbf{r}-\mathbf{r_2}|^3}.$$
The vectors $\mathbf{r_1}=a{\mathbf{\hat{x}}},~{\mathbf{r_2}}=\mu
a\mathbf{\hat{x}}$ and
$\mathbf{r}=x\mathbf{\hat{x}}+y\mathbf{\hat{y}}+z\mathbf{\hat{z}}$
are shown in Figure 1.
\begin{eqnarray}|\mathbf{r}-\mathbf{r_i}|&=&\left[(\mathbf{r}-\mathbf{r_i})
\cdot(\mathbf{r}-\mathbf{r_i})\right]^{\frac{1}{2}}=
\mathbf{r}^2-2\mathbf{r}\cdot\mathbf{r_i}+\mathbf{r_i}^2\nonumber\\
\textsf{Thus,}~~|\mathbf{r}-\mathbf{r_1}|^3&=&\left[(x-a)^2+y^2+z^2\right]^{\frac{3}{2}}\nonumber\\
\textsf{and}~~|\mathbf{r}-\mathbf{r_2}|^3&=&\left[(x-\mu
a)^2+y^2+z^2\right]^{\frac{3}{2}}.\nonumber
\end{eqnarray}
\begin{eqnarray}\label{s3}
\textsf{Writing,}~~\mathbf{E}&=&E_x{\mathbf{\hat{x}}}+E_y{\mathbf{\hat{y}}}+E_z{\mathbf{\hat{z}}}\nonumber\\
\textsf{where}~~E_x&=&\frac{q(x-a)}{\left[(x-a)^2+y^2+z^2\right]^{\frac{3}{2}}}+\frac{\lambda
q(x-\mu a)}{\left[(x-\mu
a)^2+y^2+z^2\right]^{\frac{3}{2}}};\nonumber\\
E_y&=&\frac{qy}{\left[(x-a)^2+y^2+z^2\right]^{\frac{3}{2}}}+\frac{\lambda
qy}{\left[(x-\mu
a)^2+y^2+z^2\right]^{\frac{3}{2}}};\nonumber\\
E_z&=&\frac{qz}{\left[(x-a)^2+y^2+z^2\right]^{\frac{3}{2}}}+\frac{\lambda
qz}{\left[(x-\mu a)^2+y^2+z^2\right]^{\frac{3}{2}}}.
\end{eqnarray}
The lines of force represent the direction of electric field in
space: $d\mathbf{l}=k\mathbf{E} ~~(k~\textsf{is a constant})$.
In rectangular coordinates this equation takes the form,
\begin{eqnarray}
dx{\mathbf{\hat{x}}}+dy{\mathbf{\hat{y}}}+dz{\mathbf{\hat{z}}}&=&
k\left(E_x{\mathbf{\hat{x}}}+E_y{\mathbf{\hat{y}}}+E_z{\mathbf{\hat{z}}}\right)\nonumber\\
\Rightarrow~~\frac{dx}{E_x}&=&\frac{dy}{E_y}~=~\frac{dz}{E_z}.\nonumber
\end{eqnarray}
Therefore, the differential equation for the lines of force in the
$xy$ plane ($z=0$) takes the form
\begin{eqnarray}
\frac{dy}{dx}&=&\frac{E_y}{E_x}\nonumber\\
\frac{dy}{dx}&=&\frac{y\left\{\left[(x-\mu
a)^2+y^2\right]^{\frac{3}{2}}+\lambda\left[(x-a)^2+y^2\right]^{\frac{3}{2}}\right\}}
{(x-a)\left[(x-\mu a)^2+y^2\right]^{\frac{3}{2}}+\lambda(x-\mu
a)\left[(x-a)^2+y^2\right]^{\frac{3}{2}}}\label{s4}\\
&=&\frac{(1+u^2)^{\frac{3}{2}}+
\lambda(1+v^2)^{\frac{3}{2}}}{v(1+u^2)^{\frac{3}{2}}+ \lambda
u(1+v^2)^{\frac{3}{2}}},\label{s4'}
\end{eqnarray}
where in the last equation, we have used the following sets of
transformations:
\begin{eqnarray}
u&=&\frac{x-\mu a}{y},~~~v=\frac{x-a}{y}\nonumber\\
 \textsf{Eliminating,}~x:~~~~~
 y&=&\frac{a(\mu-1)}{v-u}~~\Rightarrow~~dy=\frac{a(\mu-1)(du-dv)}{(v-u)^2}\nonumber\\
\textsf{Eliminating,}~y:~~~~~
 x&=&\frac{a(v\mu-u)}{v-u}~~\Rightarrow~~dx=\frac{a(\mu-1)(vdu-udv)}{(v-u)^2}\nonumber\\
\textsf{Thus Eq.(\ref{s4'}),}~~~~~\frac{du-dv}{vdu-udv}&=&\frac{(1+u^2)^{\frac{3}{2}}+
\lambda(1+v^2)^{\frac{3}{2}}}{v(1+u^2)^{\frac{3}{2}}+ \lambda
u(1+v^2)^{\frac{3}{2}}}\nonumber\\
\textsf{simplifying,}~~~~~\lambda\frac{du}{(1+u^2)^{\frac{3}{2}}}&=&-\frac{dv}{(1+v^2)^{\frac{3}{2}}}\nonumber\\
\textsf{integrating,}~~~~~\lambda\int\frac{du}{(1+u^2)^{\frac{3}{2}}}&=&-\int\frac{dv}{(1+v^2)^{\frac{3}{2}}}+
\textsf{constant}~(C)\nonumber\\
\lambda\frac{u}{(1+u^2)^{\frac{1}{2}}}&=&C-\frac{v}{(1+v^2)^{\frac{1}{2}}}~~~~[\textsf{see
appendix}].\nonumber
\end{eqnarray}
\begin{eqnarray}\label{s1}
\textsf{And
finally,}~~~~~\fbox{$\displaystyle \frac{(x-a)}{\left[(x-a)^2+y^2\right]^{\frac{1}{2}}}+
\frac{\lambda(x-\mu a)}{\left[(x-\mu a)^2+y^2\right]^{\frac{1}{2}}}=
C$}.
 \end{eqnarray}}
  \vspace {0.3cm}
\item
\begin{enumerate}
\item \textsf{Define $\theta_1$ and $\theta_2$ to be the angles of elevation (see Figure 2) as seen by the
    charges $q$ and $\lambda q$, respectively:
$$\fbox{$\displaystyle \cos\theta_1=\frac{(x-a)}{\left[(x-a)^2+y^2\right]^{\frac{1}{2}}}
$},~~~~\fbox{$\displaystyle\cos\theta_2=\frac{(x-\mu a)}{\left[(x-\mu
a)^2+y^2\right]^{\frac{1}{2}}} $}.$$
     Then Eq.(\ref{s1})} can be rewritten as
      \begin{eqnarray}\label{s2}
\cos\theta_1+\lambda\cos\theta_2=C
      \end{eqnarray}
      \vspace {0.3cm}
    \item
    \begin{enumerate}
    \item \textsf{Note that
\begin{eqnarray}
&-1\leq\cos\theta_1\leq 1&\nonumber\\
&-|\lambda|\leq\lambda\cos\theta_2\leq|\lambda|&\nonumber\\
\textsf{adding:}~~~~&-1-|\lambda|\leq\cos\theta_1+\lambda\cos\theta_2\leq|1+\lambda|&\nonumber\\
\textsf{Eq.(\ref{s2})}\Rightarrow~~~~~&\fbox{$\displaystyle -1-|\lambda|\leq
C\leq|1+\lambda|$}&\nonumber
\end{eqnarray}}
    \item \textsf{We write Eq.(\ref{s2}) as
$$\cos\theta_1+\lambda\frac{(x-\mu a)}{\left[(x-\mu
a)^2+y^2\right]^{\frac{1}{2}}}=C$$ and take the limit as
$(x,y)\rightarrow(a,0)$. Therefore,
\begin{eqnarray}
\lim_{(x,y)\rightarrow(a,0)}\cos\theta_1+\lambda\lim_{(x,y)\rightarrow(a,0)}\frac{(x-\mu
a)}{\left[(x-\mu
a)^2+y^2\right]^{\frac{1}{2}}}&=&\lim_{(x,y)\rightarrow(a,0)}C\nonumber\\
\lim_{(x,y)\rightarrow(a,0)}\cos\theta_1=C-\lambda ~sgn(1-\mu
)\nonumber
\end{eqnarray}
  where we have defined
$$sgn(1-\mu )\equiv \frac{1-\mu}{|1-\mu|}=\left\{\matrix{1;&&\mu<1 \cr
-1;&&\mu>1 }\right.$$ and since $-1\leq\cos\theta_1\leq 1$, we get
$$\fbox{$\displaystyle -1+\lambda~sgn(1-\mu)\leq C\leq 1+ \lambda ~sgn(1-\mu)$}.$$}
\vspace {0.3cm}
    \item \textsf{We now investigate the behavior of $\cos\theta_2$ as $(x,y)\rightarrow(\mu a,0)$ and find that
    \begin{eqnarray}
\lambda\lim_{(x,y)\rightarrow(\mu a,0)}\cos\theta_2+\frac{\mu
-1}{|\mu -1|}&=&C\nonumber\\
\lambda\lim_{(x,y)\rightarrow(\mu
a,0)}\cos\theta_2&=&\frac{C+sgn(1-\mu)}{\lambda}.\nonumber
\end{eqnarray}
Using the fact $-1\leq\cos\theta_2\leq 1$, we get
$$\fbox{$\displaystyle-1\leq\frac{C+sgn(1-\mu)}{\lambda}\leq 1 $}.$$  }
   \end{enumerate}
   \end{enumerate}
   \vspace {0.3cm}
   \item \textsf{Lines of force in the $yz$ plane can be obtained by simply setting $y=0$ in Eq.(\ref{s3}) to
   get,
   $$\frac{dz}{dx}=\frac{E_z}{E_x}=\frac{y\left\{\left[(x-\mu
a)^2+z^2\right]^{\frac{3}{2}}+\lambda\left[(x-a)^2+z^2\right]^{\frac{3}{2}}\right\}}
{(x-a)\left[(x-\mu a)^2+z^2\right]^{\frac{3}{2}}+\lambda(x-\mu
a)\left[(x-a)^2+z^2\right]^{\frac{3}{2}}}.$$ This equation is the
same as Eq.(\ref{s4}), with the replacement: $y\rightarrow z$.
Hence, the solution to this differential equation follows from
Eq.(\ref{s1})
$$\fbox{$\displaystyle \frac{(x-a)}{\left[(x-a)^2+z^2\right]^{\frac{1}{2}}}+
\frac{\lambda(x-\mu a)}{\left[(x-\mu a)^2+z^2\right]^{\frac{1}{2}}}=
C'$},$$
   where $C'$ is a new constant. \\
   \noindent Similarly, the equation in the $yz$ plane can be obtained by letting $x=0$ in Eq.(\ref{s3}). In this
   case, the differential equation take a very simple form:
   $$\frac{dy}{dz}=\frac{E_y}{E_z}=\frac{y}{z}.$$
 Therefore,
 \begin{eqnarray}
 \int\frac{dy}{y}&=&\int\frac{dz}{z}\nonumber\\
 \ln y&=&\ln z+\ln C''\nonumber
 \end{eqnarray}
 \begin{equation}
\fbox{$\displaystyle y=C''z$}~~(\textsf{straight lines}).\nonumber
\end{equation}}
\vspace {0.3cm}
   \item
   \begin{enumerate}
   \item \textsf{For an electric dipole, we set $\lambda =-1$ and $\mu=-1$ in Eq.(\ref{s1}):
$$\frac{(x+a)}{\left[x^2+a^2+2ax+y^2\right]^{\frac{1}{2}}}-
\frac{\lambda(x- a)}{\left[x^2+a^2-2ax+y^2\right]^{\frac{1}{2}}}=-
C.
$$
 Defining, $r^2=x^2+y^2$ expanding this equation in powers of $a/r$, we get
$$
\frac{x+a}{\displaystyle r\left[\displaystyle 1+\left(\frac{a}{r}\right)^2+\frac{2ax}{r}\right]^{\frac{1}{2}}}
-\frac{x-a}{\displaystyle r\left[\displaystyle 1+\left(\frac{a}{r}\right)^2-\frac{2ax}{r}\right]^{\frac{1}{2}}}=-C.
$$
  Neglecting terms like $\left(a/r\right)^2$ and higher ($a\rightarrow 0$) and using the binomial expansion
  $(1\pm\epsilon)^{1/2}\approx 1\pm 1/2\epsilon$ for $\epsilon\ll 1$, we get
\begin{eqnarray}
\frac{1}{r}\left\{\displaystyle \frac{x+a}{\displaystyle 1+\frac{ax}{r^2}}-\displaystyle \frac{x-a}{\displaystyle 1-\frac{ax}{r^2}}
\right\}&=&-C\nonumber\\
\frac{2a}{r}~\frac{\displaystyle 1-\left(\frac{x}{r}\right)^2}{\displaystyle 1-\left(\frac{a}{r}\right)^2\left(\frac{x}{r}\right)^2}&=&-C\nonumber\\
\frac{2a}{\displaystyle \sqrt{x^2+y^2}}~\left[1-\frac{x^2}{x^2+y^2}\right]&=&-C\nonumber
\end{eqnarray}
\begin{equation}\label{s5}
\fbox{$\displaystyle\left(x^2+y^2\right)^{\frac{3}{2}}=\left(\frac{-2a}{C}\right)y^2$}.
\end{equation}
Eq.(\ref{s5}) can be write in polar coordinates, if we make the
transformation: $x=r\cos\theta$ and $y=r\sin\theta$. Then the
equation for the lines of force take the more convenient form}
$$r=\left(\frac{-2a}{C}\right)\sin^2\theta.$$
\vspace {0.3cm}
       \item \textsf{Taking the natural logs of both sides of Eq.(\ref{s5}), we get
       $$\frac{3}{2}\ln(x^2+y^2)=\ln\left(\frac{-2a}{C}\right)+2\ln y.$$
       Differentiating this last equation w.r.t. $x$, we get
\begin{eqnarray}\label{s6}
\frac{3}{x^2+y^2}\left(x+y\frac{dy}{dx}\right)&=&\frac{2}{y}\frac{dy}{dx}\nonumber\\
\frac{dy}{dx}&=&\frac{3xy}{2x^2-y^2}.
       \end{eqnarray}
Eq.(\ref{s6}) represents the slope of the lines of force at the
point $(x,y)$. Therefore, the slope of the corresponding
equipotential line would be
\begin{eqnarray}
\frac{dy}{dx}&=&\frac{-1}{\displaystyle \frac{3xy}{\displaystyle 2x^2-y^2}}\nonumber\\
\textsf{simplifying,}~~~~~\frac{dy}{dx}&=&\frac{1}{3}\left(\frac{y}{x}\right)
-\frac{2}{3}\left(\frac{x}{y}\right).\nonumber
\end{eqnarray}
 To separate the variables we make the transformation:
 \begin{eqnarray}
w&=&\frac{y}{x}\nonumber\\
\Rightarrow~~~~~\frac{dw}{dx}&=&\displaystyle \frac{\displaystyle x\frac{dy}{dx}-y}{x^2}\nonumber\\
\frac{dy}{dx}&=&x\frac{dw}{dx}+w.\nonumber
\end{eqnarray}
Therefore, the differential equation for the equipotential lines in
the $wx$ plane are given by
\begin{eqnarray}
\frac{1}{3}w-\frac{2}{3}\frac{1}{w}&=&x\frac{dw}{dx}+w\nonumber\\
\textsf{integrating,}~~~~~3\int\frac{w}{w^2+1}&=&-2\int\frac{dx}{x}\nonumber\\
\frac{3}{2}\int\frac{dp}{p}&=&-2\int\frac{dx}{x}~~~~~~~~~[p\equiv w^2+1]\nonumber\\
\frac{3}{2}\ln(w^2+1)&=&-2\ln x+\ln K\nonumber\\
\left(\frac{y^2}{x^2}+1\right)^{\frac{3}{2}}&=&\frac{K}{x^2}\nonumber
\end{eqnarray}
\begin{equation}
\fbox{$\displaystyle\left(x^2+y^2\right)^{\frac{3}{2}}= Kx$}.\nonumber
\end{equation}
This last equation can be rewritten in polar coordinates as
 $$r^2=K \cos\theta.$$}
\end{enumerate}
 \end{enumerate}
 \vspace {0.3cm}
 \item
 \begin{enumerate}
\item
  \begin{enumerate}
  \item \textsf{Equation for the orthogonal bisecting plane is $\displaystyle x=\frac{(1+\mu)a}{2}$. With this value for $x$ in Eq.(\ref{s2}),
   we get
\begin{eqnarray}\label{s7}
E_x&=&\frac{4qa(1-\mu)(\lambda
-1)}{\left[a^2\left(1-\mu\right)^2+4\left(y^2+z^2\right)\right]^{\frac{3}{2}}}\nonumber\\
E_y&=&\frac{8qy(\lambda
+1)}{\left[a^2\left(1-\mu\right)^2+4\left(y^2+z^2\right)\right]^{\frac{3}{2}}}\nonumber\\
E_z&=&\frac{8qz(\lambda
+1)}{\left[a^2\left(1-\mu\right)^2+4\left(y^2+z^2\right)\right]^{\frac{3}{2}}}.
\end{eqnarray}
Now since,
$|{\mathbf{E}}|=\left(E_x^2+E_y^2+E_z^2\right)^{\frac{1}{2}}$.
Therefore,
\begin{eqnarray}
\frac{\partial |{\mathbf{E}}|}{\partial y}&=&\frac{E_x\frac{\partial
E_x}{\partial y}+E_y\frac{\partial E_y}{\partial
y}+E_z\frac{\partial E_z}{\partial
y}}{\left(E_x^2+E_y^2+E_z^2\right)^{\frac{1}{2}}}\nonumber\\
\frac{\partial |{\mathbf{E}}|}{\partial z}&=&\frac{E_x\frac{\partial
E_x}{\partial z}+E_y\frac{\partial E_y}{\partial
z}+E_z\frac{\partial E_z}{\partial
z}}{\left(E_x^2+E_y^2+E_z^2\right)^{\frac{1}{2}}}.\nonumber
\end{eqnarray}
For maximum value of the electric field, $$\frac{\partial
|{\mathbf{E}}|}{\partial y}=0,~~~~~~\frac{\partial
|{\mathbf{E}}|}{\partial z}=0,$$ and therefore
\begin{eqnarray}
E_x\frac{\partial E_x}{\partial y}+E_y\frac{\partial E_y}{\partial
y}+E_z\frac{\partial E_z}{\partial y}&=&0\label{s8}\\
 E_x\frac{\partial
E_x}{\partial z}+E_y\frac{\partial E_y}{\partial
z}+E_z\frac{\partial E_z}{\partial z}&=&0.\label{s8'}
\end{eqnarray}
The expressions for the derivatives of the component of electric fields are given by,
\begin{eqnarray}\label{s9}
\frac{\partial E_x}{\partial y}&=&\frac{48qa\left(1-\mu\right)\left(1-\lambda\right)y}{\left[a^2\left(1-\mu\right)^2+4\left(y^2+z^2\right)
\right]^{\frac{5}{2}}}\nonumber\\
\frac{\partial E_y}{\partial y}&=&\frac{8q\left(1+\lambda\right)\left[a^2\left(1-\mu\right)^2+4\left(z^2-2y^2\right)\right]}
{\left[a^2\left(1-\mu\right)^2+4\left(y^2+z^2\right)\right]^{\frac{5}{2}}}\nonumber\\
\frac{\partial E_z}{\partial y}&=&\frac{-96q(1+\lambda)yz}{\left[a^2\left(1-\mu\right)^2+4\left(y^2+z^2\right)\right]^{\frac{5}{2}}}.
\end{eqnarray}
The other derivatives can be simply obtained by the transformation, $y\rightarrow z$ in Eq.(\ref{s9}). Use of Eq.(\ref{s9}) in Eq.(\ref{s8}) gives
$$\frac{64qy\left\{-3a^2(\lambda-1)^2(1-\mu)^2+(\lambda+1)^2
\left[a^2(1-\mu)^2+4(z^2-2y^2)\right]-12(\lambda+1)^2z^2\right\}}
{\left[a^2(1-\mu)^2+4(y^2+z^2)
\right]^{4}}=0,$$
which can be simplified to
\begin{eqnarray}\label{s10'}
a^2(1-\mu)^2-8(y^2+z^2)=3a^2(1-\mu)^2\frac{(\lambda-1)^2}{(\lambda+1)^2}\nonumber\\
y^2+z^2=\left[\frac{a(1-\mu)}{2(1+\lambda)}\right]^2\left[3-(\lambda-2)^2\right]\equiv r_1^2.
\end{eqnarray}
Hence, the electric field is maximized on a circle, centered about the origin and lying on the $\displaystyle x=\frac{a(1+\mu)}{2}$ plane. The radius of the circle being
\begin{equation}\label{s10}
\fbox{$\displaystyle\frac{a}{2}\left|\frac{1-\mu}{1+\lambda}\right|\sqrt{3-(\lambda-2)^2}\displaystyle$}.
\nonumber
\end{equation}
Use of Eq.(\ref{s8'}) gives an identical result.
\vspace {0.3cm}}
\item \textsf{It is first useful to compute the quantity, $\displaystyle a^2(1-\mu)^2+4(y^2+z^2)$. Using the expression for $y^2+z^2$ from Eq.(\ref{s10'}), we get for $a^2(1-\mu)^2+4(y^2+z^2)= \displaystyle 6\lambda a^2\left(\displaystyle \frac{1-\mu}{1+\lambda}\right)^2$. Using this result in Eqs.(\ref{s7}), we get
    \begin{eqnarray}
    \left.E_{x}^2\right|_{\textsf{max}}&=&\frac{2q^2(\lambda-1)^2(\lambda+1)^6}{27a^4(1-\mu)^4\lambda^3}\nonumber\\
     \left.\left(E_{y}^2+E_z^2\right)\right|_{\textsf{max}}&=&
     \frac{2q^2(-\lambda^2+4\lambda-1)(\lambda+1)^6}{27a^4(1-\mu)^4\lambda^3}.\nonumber
    \end{eqnarray}
    Therefore,
    $$\fbox{$\displaystyle\left|{\mathbf{E}}\right|_{\textsf{max}}=\frac{2}{3\sqrt 3}\frac{q}{a^2(1-\mu)^2}\left|\frac{(\lambda+1)^3}{\lambda}\right|$}.$$}
    \vspace {0.3cm}
\item \textsf{Looking at the expressions for the radius and $\left|{\mathbf{E}}\right|$, we conclude
that
\begin{eqnarray}
\lambda\neq 0,~~\lambda\neq -1~~\& ~~3-(\lambda-2)^2>0\nonumber\\
\Rightarrow~~\fbox{$\displaystyle \lambda\in\left(2-\sqrt 3,2+\sqrt 3\right)\displaystyle$}.
\nonumber
\end{eqnarray}}
\end{enumerate}
\vspace {0.3cm}
  \item
  \begin{enumerate}
  \item \textsf{From the condition $\displaystyle \frac{\displaystyle \partial |{\mathbf{E}}|}{\partial \mu}=0$, one can derive an equation similar to Eqs.(\ref{s8}) and (\ref{s8'}):
      \begin{eqnarray}\label{s10''}
E_x\frac{\partial E_x}{\partial \mu}+E_y\frac{\partial E_y}{\partial
\mu}+E_z\frac{\partial E_z}{\partial \mu}&=&0.
\end{eqnarray}
Taking the derivatives of $E_x$, $E_y$ and $E_z$ w.r.t $\mu$ in Eqs.(\ref{s7}), we have
\begin{eqnarray}\label{s11}
\frac{\partial E_x}{\partial \mu}&=&\frac{8qa\left(\lambda-1\right)\left[a^2\left(1-\mu\right)^2-2\left(y^2+z^2\right)\right]}
{\left[a^2\left(1-\mu\right)^2+4\left(y^2+z^2\right)
\right]^{\frac{5}{2}}}\nonumber\\
\frac{\partial E_y}{\partial \mu}&=&\frac{24qa^2\left(1+\lambda\right)\left(1-\mu\right)y}
{\left[a^2\left(1-\mu\right)^2+4\left(y^2+z^2\right)\right]^{\frac{5}{2}}}\nonumber\\
\frac{\partial E_z}{\partial \mu}&=&\frac{24qa^2\left(1+\lambda\right)\left(1-\mu\right)z}
{\left[a^2\left(1-\mu\right)^2+4\left(y^2+z^2\right)\right]^{\frac{5}{2}}}.
\end{eqnarray}
Use of Eqs.(\ref{s7}) and (\ref{s11}) in Eq.(\ref{s10''}) gives,
$$\frac{32qa^2(1-\mu)\left\{(\lambda-1)^2
\left[a^2(1-\mu)^2-2(y^2+z^2)\right]+6(\lambda+1)^2(y^2+z^2)\right\}}
{\left[a^2(1-\mu)^2+4(y^2+z^2)
\right]^{4}}=0.$$
After simple algebraic manipulation, we arrive at
\begin{eqnarray}\label{s12'}
y^2+z^2&=&\left[\frac{a(1-\mu)(1-\lambda)}{2}\right]^2\frac{1}{3-(\lambda+2)^2}\equiv r_2^2.
\end{eqnarray}
where,
\begin{equation}\label{s12}
\fbox{$\displaystyle r_2=\frac{a}{2}\frac{\left|(1-\mu)(1-\lambda)\right|}{\sqrt{3-(\lambda+2)^2}}$}.
\end{equation}
From Eq.(\ref{s12'}), one can compute $\mu$ for a particular value of $y$ and $z$:
$$\mu=1\pm\frac{2}{a\left|1-\lambda\right|}\sqrt{\left(y^2+z^2\right)\left[3-\left(\lambda+2\right)^2\right]}.$$}
\vspace {0.3cm}
\item \textsf{As before we first compute $a^2(1-\mu)^2+4(y^2+z^2)$ using the expression for $y^2+z^2$ from Eq.(\ref{s12'}) to get $\displaystyle a^2(1-\mu)^2+4(y^2+z^2)= 6\lambda a^2\displaystyle \frac{\displaystyle (1-\mu)^2}{(\lambda+2)^2-3}$. Using this result in Eqs.(\ref{s7}), we get
    \begin{eqnarray}
    \left.E_{x}^2\right|_{\textsf{max}}&=&\frac{2q^2[(\lambda+2)^2-3]^3(\lambda-1)^2}{27a^4(1-\mu)^4}\nonumber\\
     \left.\left(E_{y}^2+E_z^2\right)\right|_{\textsf{max}}&=&
     \frac{-2q^2[3-(\lambda+2)]^2(\lambda-1)^2(\lambda+1)^2}{27a^4(1-\mu)^4\lambda^3}.\nonumber
    \end{eqnarray}
   $$ \textsf{Finally,}~~~~\fbox{$\displaystyle \left|{\mathbf{E}}\right|_{\textsf{max}}=\frac{2}{3\sqrt 3}\frac{q}{a^2(1-\mu)^2}\left|\frac{[3-(\lambda+2)^2](\lambda-1)}{\lambda}\right|$}.$$
   }
   \vspace {0.3cm}
\item \textsf{From the results of the previous two parts, we infer that
\begin{eqnarray}
\lambda\neq 0,~~\lambda\neq 1~~\& ~~3-(\lambda+2)^2>0\nonumber\\
\Rightarrow~~\fbox{$\displaystyle \lambda\in\left(-2-\sqrt 3,-2+\sqrt 3\right)$}.
\nonumber
\end{eqnarray}}
  \end{enumerate}
  \vspace {0.3cm}
  \item
  \begin{enumerate}
  \item \textsf{In this case the extremum condition,
   $\displaystyle \frac{\displaystyle \partial |{\mathbf{E}}|}{\partial \lambda}=0$, leads to
      \begin{eqnarray}\label{s13}
E_x\frac{\partial E_x}{\partial \lambda}+E_y\frac{\partial E_y}{\partial
\lambda}+E_z\frac{\partial E_z}{\partial \lambda}&=&0.
\end{eqnarray}
Taking the derivatives of $E_x$, $E_y$ and $E_z$ w.r.t. $\lambda$ in Eqs.(\ref{s7}), we have
\begin{eqnarray}\label{s14}
\frac{\partial E_x}{\partial \lambda}&=&\frac{4qa\left(1-\mu\right)}
{\left[a^2\left(1-\mu\right)^2+4\left(y^2+z^2\right)
\right]^{\frac{3}{2}}}\nonumber\\
\frac{\partial E_y}{\partial \lambda}&=&\frac{8qy}
{\left[a^2\left(1-\mu\right)^2+4\left(y^2+z^2\right)\right]^{\frac{3}{2}}}\nonumber\\
\frac{\partial E_z}{\partial \lambda}&=&\frac{8qz}
{\left[a^2\left(1-\mu\right)^2+4\left(y^2+z^2\right)\right]^{\frac{3}{2}}}.
\end{eqnarray}
Use of Eqs.(\ref{s7}) and (\ref{s14}) in Eq.(\ref{s13}) gives,
$$\frac{16q^2\left[a^2(1-\mu)^2(\lambda-1)
+4(\lambda+1)(y^2+z^2)\right]}
{\left[a^2(1-\mu)^2+4(y^2+z^2)
\right]^{3}}=0.$$
Simplifying this expression gives,
\begin{eqnarray}\label{s15'}
y^2+z^2&\equiv&r_3^2
\end{eqnarray}
\begin{equation}\label{s15}
\fbox{$\displaystyle r_3=\frac{a}{2}\left|(1-\mu)\right|\sqrt{\frac{1-\lambda}{1+\lambda}}$}.
\end{equation}
From Eqs.(\ref{s15'}) and (\ref{s15}), one can compute $\lambda$ for a particular value of $y$ and $z$:
$$\lambda=\frac{a^2(1-\mu)^2-4(y^2+z^2)}{a^2(1-\mu)^2+4(y^2+z^2)}$$.}
\vspace {0.3cm}
\item \textsf{Using Eqs.(\ref{s15'}) and (\ref{s15}) in Eq.(\ref{s7}), we get
\begin{eqnarray}
    \left.E_{x}^2\right|_{\textsf{max}}&=&\frac{2q^2(\lambda-1)^2(\lambda+1)^3}{a^4(1-\mu)^4}\nonumber\\
     \left.\left(E_{y}^2+E_z^2\right)\right|_{\textsf{max}}&=&
     \frac{2q^2(\lambda+1)^5}{a^4(1-\mu)^4}.\nonumber
    \end{eqnarray}
\begin{eqnarray}
\textsf{Now since,}~~~~~\left|{\mathbf{E}}\right|_{\textsf{max}}&=&
\sqrt{\left.E_{x}^2\right|_{\textsf{max}}+\left.E_{y}^2\right|_{\textsf{max}}
+\left.E_{z}^2\right|_{\textsf{max}}}\nonumber
\end{eqnarray}
\begin{equation}
\fbox{$\displaystyle\left|{\mathbf{E}}\right|_{\textsf{max}}= \frac{2q}{a^2(1-\mu)^2}\sqrt{(\lambda^2+1)(\lambda+1)^3} $}\nonumber
\end{equation}}
\end{enumerate}
\end{enumerate}
 \vspace {0.6cm}
\item
  \begin{enumerate}
  \item\textsf{The electric flux, $N$ passing through the circles (see Figure 3)
  \begin{eqnarray}\label{s16}
  y^2+z^2=R^2\equiv\left\{\matrix{r_1^2&=&
  \displaystyle \left[\displaystyle \frac{a(1-\mu)}{2(1+\lambda)}\right]^2\left[3-(\lambda-2)^2\right]\cr
  r_2^2&=&\displaystyle \left[\displaystyle \frac{a(1-\mu)(1-\lambda)}{2}\right]^2\frac{1}{3-(\lambda+2)^2}\cr
  r_3^2&=&\displaystyle \left[\displaystyle \frac{a(1-\mu)}{2}\right]^2\frac{1-\lambda}{1+\lambda}}\right.
  \end{eqnarray}
  at $\displaystyle x=\frac{\displaystyle a\left(1+\mu\right)}{2}$, can be readily computed from
  \begin{eqnarray}
   N&=&\int\int{\bf{D}}\cdot d{\bf{s}},\nonumber\\
  \textsf{where}~~~~~{\bf{D}}={\bf{E}}~~&\textsf{and}&~~d{\bf{s}}={\bf{\hat{x}}}~dydz\nonumber\\
  \textsf{Therefore}~~~~~N &=&\int\int \left.E_x \right|_{ \displaystyle x=\frac{ a\left( 1+\mu\right)}{\displaystyle 2}} \,dydz.\nonumber
  \end{eqnarray}
  Using Eq.(\ref{s7}), we obtain
  \begin{eqnarray}\label{s17}
  N &=&2[4qa(1-\mu)(\lambda-1)]\int_{y=0}^{y=R}dy\int_{z=-\sqrt{R^2-y^2}}^{z=+\sqrt{R^2-y^2}}
   \frac{dz}{\left[a^2\left(1-\mu\right)^2+4\left(y^2+z^2\right)\right]^{\frac{3}{2}}}\nonumber\\
  &=&16qa(1-\mu)(\lambda-1)\int_{y=0}^{y=R}\frac{dy}{a^2\left(1-\mu\right)^2+4y^2}
   \left\{\frac{z}{\left[a^2\left(1-\mu\right)^2+4\left(y^2+z^2\right)\right]^{\frac{1}{2}}}\right\}_{z=-\sqrt{R^2-y^2}}
   ^{z=+\sqrt{R^2-y^2}}\nonumber\\
   &=&\frac{8qa(1-\mu)(\lambda-1)}{\left[a^2\left(1-\mu\right)^2+4R^2\right]^{\frac{1}{2}}}
    \int_{y=0}^{y=R}\frac{\sqrt{R^2-y^2}}{\left[\frac{a\left(1-\mu\right)}{2}\right]^2+y^2}\, dy
   \nonumber\\
    &=&\frac{8qa(1-\mu)(\lambda-1)}{\left[a^2\left(1-\mu\right)^2+4R^2\right]^{\frac{1}{2}}}
    \left\{-\sin^{-1}\left(\frac{y}{R}\right)-\frac{\sqrt{R^2+\frac{a^2\left(1-\mu\right)^2}{4}}}{\frac{a\left(1-\mu
    \right)}{2}}\sin^{-1}\left[\frac{\frac{a\left(1-\mu\right)}{2}}{R}
    \sqrt{\frac{R^2-y^2}{\frac{a^2\left(1-\mu\right)^2}{4}+y^2}}\right]\right\}_{y=0}^{y=R}\nonumber\\
    &=&4\pi q\left(\lambda-1\right)\left\{
    1-\frac{a\left(1-\mu\right)}{\left[a^2\left(1-\mu\right)^2+4R^2\right]^{\frac{1}{2}}}
    \right\},
    \end{eqnarray}
    where in the second and fourth steps above, we have made use of Eqs.(\#) and (\#), respectively, from the appendix. One can now fully compute $N$ in Eq.(\ref{s17}) for $R=r_1,~r_2$ and $r_3$ given by Eq.(\ref{s16}). Thus
    \begin{eqnarray}
    \frac{a\left(1-\mu\right)}{\left[a^2\left(1-\mu\right)^2+4r_1^2\right]^{\frac{1}{2}}}
    &=& sgn{(1-\mu)}\frac{\left|1+\lambda\right|}{\sqrt{6\lambda}}\nonumber\\
     \frac{a\left(1-\mu\right)}{\left[a^2\left(1-\mu\right)^2+4r_2^2\right]^{\frac{1}{2}}}
    &=& sgn{(1-\mu)}\sqrt\frac{3-\left(\lambda+2\right)^2}{{-6\lambda}}\nonumber\\
    \frac{a\left(1-\mu\right)}{\left[a^2\left(1-\mu\right)^2+4r_3^2\right]^{\frac{1}{2}}}
    &=& sgn{(1-\mu)}\sqrt\frac{1+\lambda}{{2}}.\nonumber
    \end{eqnarray}
    Finally,
    $$\fbox{$\displaystyle N_1=2\sqrt{\frac{2}{3}}\pi q\left(\frac{1-\lambda}{\sqrt{\lambda}}\right)
\left[sgn(1-\mu)~\left(1+\lambda\right)-\sqrt{6\lambda}\right] $}$$
$$\fbox{$\displaystyle N_2=2\sqrt{\frac{2}{3}}\pi q\left(\frac{1-\lambda}{\sqrt{-\lambda}}\right)
\left[sgn(1-\mu)~\sqrt{3-\left(\lambda+2\right)^2}-1\right] $}$$
  $$\fbox{$\displaystyle N_3=2\sqrt{2}\pi q\left(1-\lambda\right)
\left[sgn(1-\mu)~\sqrt{1+\lambda}-\sqrt{2}\right] $}.$$}
\vspace{0.3cm}
  \item
      \textsf{For this part of the problem, we first calculate the flux passing through the circle $y^2+z^2=b^2$ at $x=0$ (see Figure 4). Using Eq.(\ref{s3}), we have
      \begin{eqnarray}
      N &=&\int\int \left.E_x\right|_{\displaystyle x=0}\, dydz\nonumber\\
      &=&-2qa\int_{y=0}^{y=b}\, dy\int_{z=-\sqrt{b^2-y^2}}^{z=+\sqrt{b^2-y^2}}dz\,
   \left\{\frac{1}{\left[a^2+y^2+z^2\right]^{\frac{3}{2}}}+\frac{\lambda\mu}{\left[\left(\mu a\right)^2+y^2+z^2\right]^{\frac{3}{2}}}\right\}\nonumber\\
   &=&-2qa\int_{y=0}^{y=b}\, dy
   \left\{\frac{z}{\left[a^2+y^2\right]\left[a^2+y^2+z^2\right]^{\frac{1}{2}}}
   +\frac{\lambda\mu z}{\left[\left(\mu a\right)^2+y^2\right]\left[\left(\mu a\right)^2+y^2+z^2\right]^{\frac{1}{2}}}\right\}_{z=-\sqrt{b^2-y^2}}^{z=+\sqrt{b^2-y^2}}\nonumber\\
   &=&-4qa\left\{\frac{1}{\sqrt{a^2+b^2}}\int_{y=0}^{y=b}\, dy \,
   \frac{\sqrt{b^2-y^2}}{a^2+y^2}
   +\frac{\lambda\mu }{\sqrt{\left(\mu a\right)^2+b^2}}\int_{y=0}^{y=b}\, dy \,\frac{\sqrt{b^2-y^2}}{\left(\mu a\right)^2+y^2}\right\}\nonumber\\
   &=&-4qa\left\{\frac{1}{\sqrt{a^2+b^2}}\left[-\sin^{-1}\left(\frac{y}{b}\right)-\frac{\sqrt{a^2+b^2}}{a}
   \sin^{-1}\left(\frac{a}{b}\sqrt{\frac{b^2-y^2}{a^2+y^2}}\right)\right]_{y=0}^{y=b}\right.\nonumber\\
   &&\left.+\frac{\lambda\mu}{\sqrt{\left(\mu a\right)^2+b^2}}\left[-\sin^{-1}\left(\frac{y}{b}\right)-\frac{\sqrt{\left(\mu a\right)^2+b^2}}{\mu a}
   \sin^{-1}\left(\frac{\mu a}{b}\sqrt{\frac{b^2-y^2}{\left(\mu a\right)^2+y^2}}\right)\right]_{y=0}^{y=b}
   \right\}\nonumber
   \end{eqnarray}
    where again, in the second and fourth steps above, we have made use of Eqs.(\#) and (\#), respectively, from the appendix. Finally, evaluating the last expression between the limits, yields
    $$N=2\pi q\left\{\frac{a}{\sqrt{a^2+b^2}}-1+\lambda\left[\frac{\mu a}{\sqrt{\left(\mu a\right)^2+b^2}}-1\right]\right\}$$
  Now if, $N=0$, then
  $$\fbox{$\displaystyle \lambda=-\left[\displaystyle\frac{\displaystyle\frac{a}{\sqrt{a^2+b^2}}-1}{\displaystyle\frac{\mu a}{\sqrt{(\mu a)^2+b^2}}-1}\right]$}.$$}
  \end{enumerate}
  \vspace{0.6cm}
  \item
  \begin{enumerate}
  \item\textsf{The neutral point or the equilibrium position can be located by setting $|{\mathbf{E}}|=0$. This implies $E_x=0, ~E_y=0,~E_z=0$. Use of Eqs.(\ref{s3}) gives
      \begin{eqnarray}
      E_y=0 &\Rightarrow & qy\left[\frac{1}{\left[(x-a)^2+y^2+z^2\right]^{\frac{3}{2}}}+\frac{\lambda
}{\left[(x-\mu
a)^2+y^2+z^2\right]^{\frac{3}{2}}}\right]=0~\Rightarrow~y=0\nonumber\\
   E_z=0 &\Rightarrow & qz\left[\frac{1}{\left[(x-a)^2+y^2+z^2\right]^{\frac{3}{2}}}+\frac{\lambda
}{\left[(x-\mu
a)^2+y^2+z^2\right]^{\frac{3}{2}}}\right]=0~\Rightarrow~ z=0.\nonumber
      \end{eqnarray}
   Then,
   \begin{eqnarray}
   \left.E_x\right|_{{x=0,
   y=0}}~\Rightarrow~\left.\frac{q(x-a)}{\left[(x-a)^2+y^2+z^2\right]^{\frac{3}{2}}}+\frac{\lambda
q(x-\mu a)}{\left[(x-\mu
a)^2+y^2+z^2\right]^{\frac{3}{2}}}\right|_{{x=0,
   y=0}}=0.\nonumber
   \end{eqnarray}
   This last equation gives,
   \begin{eqnarray}\label{s18}
   \frac{x-a}{\left|x-a\right|^3}+\frac{\lambda\left(x-a\right)}{\left|x-\mu a\right|^3}=0.
   \end{eqnarray}
   \[
\begin{array} {ccc}
\textsf{For}~(x-a>0,x-\mu a>0)&~~~~~~~~~~~~&\textsf{For}~(x-a<0,x-\mu a>0)\nonumber\\
   \textsf{or}&~~~~~~~~~~~~&\textsf{or}\nonumber\\
   (x-a<0,x-\mu a<0)&~~~~~~~~~~~~&(x-a>0,x-\mu a<0)\nonumber\\
   \nonumber\\
   \textsf{Eq.}(\ref{s18})\Rightarrow\displaystyle\frac{1}{\displaystyle\left(x-a\right)^2}
   +\displaystyle\frac{\displaystyle\lambda}{\left(x-\mu a\right)^2}=0&~~~~~~~~~~~~&\textsf{Eq.}(\ref{s18})\Rightarrow\displaystyle\frac{1}{\displaystyle
   \left(x-a\right)^2}-\frac{\displaystyle\lambda}{\displaystyle\left(x-\mu a\right)^2}=0\nonumber\\
   \nonumber\\
   \Rightarrow\left(\displaystyle\frac{x-\mu a}{x-a}\right)^2=-\lambda&~~~~~~~~~~~~&\Rightarrow\left(\displaystyle\frac{x-\mu a}{x-a}\right)^2=\lambda\nonumber\\
   \Rightarrow\lambda<0,~\textsf{set}~\lambda=-|\lambda|&~~~~~~~~~~~~&\Rightarrow\lambda>0,
   ~\textsf{set}~\lambda=|\lambda|
   \end{array}
\]
\newline
In either case, one gets ($\lambda<0$ or $\lambda>0$)
$$\displaystyle\left(\frac{x-\mu a}{x-a}\right)^2=\left|\lambda\right|.$$
Solving for $x$ gives,
\begin{eqnarray}\label{s18'}
\fbox{$\displaystyle x=a\left[\frac{\mu\pm \sqrt{|\lambda|}}{1\pm\sqrt{|\lambda|}}\right]$}
\end{eqnarray}}
\vspace{0.3cm}
      \item \textsf{\begin{center}\underline{CASE 1:~~~~$\mu > 1$}\end{center}
      \begin{itemize}
      \item  $\lambda<0$  \\
      $$\Rightarrow x< a ~~\textsf{or}~~ x>\mu a$$
      \[
\begin{array} {ccc}
      \textsf{For}~ x=a\left[\displaystyle\frac{\displaystyle\mu-\sqrt{|\lambda|}}{1-\sqrt{|\lambda|}}\right]&~~~~~~~~~~~~&
      \textsf{For}~ x=a\left[\displaystyle\frac{\displaystyle\mu+\sqrt{|\lambda|}}{1+\sqrt{|\lambda|}}\right]\nonumber\\
      x< a\Rightarrow \mu>1 ~\textsf{and}~ \lambda<-1&~~~~~~~~~~~~& x< a\Rightarrow \textsf{unacceptable solutions} \nonumber\\
       x>\mu a\Rightarrow \mu>1 ~\textsf{and}~ -1<\lambda<0&~~~~~~~~~~~~& x>\mu a\Rightarrow \textsf{unacceptable solutions} \nonumber\\
      \end{array}
\]
\\
 \item  $\lambda>0$  \\
      $$\Rightarrow ~~ a<x<\mu a$$
      \[
\begin{array} {ccc}
      \textsf{For}~ x=a\left[\displaystyle\frac{\displaystyle\mu-\sqrt{|\lambda|}}{1-\sqrt{|\lambda|}}\right]&~~~~~~~~~~~~&
      \textsf{For}~ x=a\left[\displaystyle\frac{\displaystyle\mu+\sqrt{|\lambda|}}{1+\sqrt{|\lambda|}}\right]\nonumber\\
      a<x<\mu a\Rightarrow \textsf{unacceptable solutions}&~~~~~~~~~~~~& a<x<\mu a\Rightarrow \mu >1\nonumber\\
      \end{array}
\]
\end{itemize}
\vspace{0.6cm}
\begin{center}\underline{CASE 2:~~~~$0<\mu< 1$}\end{center}
      \begin{itemize}
      \item  $\lambda<0$  \\
      $$\Rightarrow x> a ~~\textsf{or}~~ x<\mu a$$
      \[
\begin{array} {ccc}
      \textsf{For}~ x=a\left[\displaystyle\frac{\mu-\sqrt{|\lambda|}}{1-\sqrt{|\lambda|}}\right]&~~~~~~~~~~~~&
      \textsf{For}~ x=a\left[\displaystyle\frac{\mu+\sqrt{|\lambda|}}{1+\sqrt{|\lambda|}}\right]\nonumber\\
      x> a\Rightarrow 0<\mu<1 ~\textsf{and}~ \lambda<-1&~~~~~~~~~~~~& x> a\Rightarrow \textsf{unacceptable solutions} \nonumber\\
       x<\mu a\Rightarrow 0<\mu<1 ~\textsf{and}~ -1<\lambda<0&~~~~~~~~~~~~& x<\mu a\Rightarrow \textsf{unacceptable solutions} \nonumber\\
      \end{array}
\]
\\
 \item  $\lambda>0$  \\
      $$\Rightarrow ~~ \mu a<x<a$$
      \[
\begin{array} {ccc}
      \textsf{For}~ x=a\left[\displaystyle\frac{\mu-\sqrt{|\lambda|}}{1-\sqrt{|\lambda|}}\right]&~~~~~~~~~~~~&
      \textsf{For}~ x=a\left[\displaystyle\frac{\mu+\sqrt{|\lambda|}}{1+\sqrt{|\lambda|}}\right]\nonumber\\
      \mu a<x< a\Rightarrow \textsf{unacceptable solutions}&~~~~~~~~~~~~& \mu a<x<a\Rightarrow 0<\mu <1\nonumber\\
      \end{array}
\]
\end{itemize}
\vspace{0.6cm}
\begin{center}\underline{CASE 3:~~~~$\mu< 0$}\end{center}
      \begin{itemize}
      \item  $\lambda<0$  \\
      $$\Rightarrow x<-|\mu| a ~~\textsf{or}~~ x> a$$
      \[
\begin{array} {ccc}
      \textsf{For}~ x=a\left[\displaystyle\frac{-|\mu|-\sqrt{|\lambda|}}{1-\sqrt{|\lambda|}}\right]&~~~~~~~~~~~~&
      \textsf{For}~ x=a\left[\displaystyle\frac{-|\mu|+\sqrt{|\lambda|}}{1+\sqrt{|\lambda|}}\right]\nonumber\\
      x<-|\mu|a\Rightarrow \mu<1 ~\textsf{and}~ -1<\lambda<0&~~~~~~~~~~~~& x<-|\mu| a\Rightarrow \textsf{unacceptable solutions} \nonumber\\
       x> a\Rightarrow \mu<1 ~\textsf{and}~ \lambda<-1&~~~~~~~~~~~~& x> a\Rightarrow \textsf{unacceptable solutions} \nonumber\\
      \end{array}
\]
\\
 \item  $\lambda>0$  \\
      $$\Rightarrow ~~ -|\mu| a<x<a$$
      \[
\begin{array} {ccc}
      \textsf{For}~ x=a\left[\displaystyle\frac{-|\mu|-\sqrt{|\lambda|}}{1-\sqrt{|\lambda|}}\right]&~~~~~~~~~~~~&
      \textsf{For}~ x=a\left[\displaystyle\frac{-|\mu|+\sqrt{|\lambda|}}{1+\sqrt{|\lambda|}}\right]\nonumber\\
      -|\mu| a<x< a\Rightarrow \textsf{unacceptable solutions}&~~~~~~~~~~~~& -|\mu| a<x<a\Rightarrow \mu <0\nonumber\\
      \end{array}
\]
\end{itemize}
\vspace{0.6cm}
\begin{center}\underline{CASE 4:~~~~$\forall\mu$}\end{center}
      \begin{itemize}
      \item  $\lambda=-1$  \\
      No solutions
\\
 \item  $\lambda=+1$  \\
      $x=\displaystyle\frac{a}{2}(1+\mu)$, which follows directly from $x=a\left[\displaystyle\frac{\mu+\sqrt{|\lambda|}}{1+\sqrt{|\lambda|}}\right]$
\end{itemize}
Therefore,
$$
\fbox{$\displaystyle x=a\left[\frac{\mu-\sqrt{|\lambda|}}{1-\sqrt{|\lambda|}}\right]\textsf{is
valid for} \mu\in (-\infty,1)\cup (1,\infty)~\textsf{and}~
\lambda\in (-\infty,-1)\cup (-1,0)$}$$
$$
\fbox{$\displaystyle x=a\left[\frac{\mu+\sqrt{|\lambda|}}{1+\sqrt{|\lambda|}}\right]\textsf{is
valid for} \mu\in (-\infty,1)\cup (1,\infty)~\textsf{and}~
\lambda\in (0,\infty)$}$$}
\end{enumerate}
\vspace{0.8cm}
\item
  \begin{enumerate}
  \item\label{4}\textsf{The equation for the electric potential, $V=0$,
  the two charge system takes the form:
\begin{eqnarray}
V=\frac{q}{\left[(x-a)^2+y^2+z^2\right]}+\frac{\lambda
q}{\left[(x-\mu a)^2+y^2+z^2\right]}&=&0 \nonumber\\
\Rightarrow\left[\frac{(x-\mu
a)^2+y^2+z^2}{(x-a)^2+y^2+z^2}\right]^{\frac{1}{2}}&=&-\lambda \label{s19}\\
\nonumber
\end{eqnarray}
 This last equation will hold true if $\lambda<0$. Squaring and simplifying we get,
\begin{eqnarray}
x^2(\lambda^2-1)+2a(\mu-\lambda^2)x+y^2(\lambda^2-1)+z^2(\lambda^2-1)&=&a^2(\mu^2-\lambda^2)\nonumber\\
x^2+2a\left(\frac{\mu-\lambda^2}{\lambda^2-1}\right)x+
\left[a\left(\frac{\mu-\lambda^2}{\lambda^2-1}\right)\right]^2+
y^2+z^2&=&\frac{a^2(\mu^2-\lambda^2)}{\lambda^2-1}+\left[a\left(\frac{\mu-\lambda^2}{\lambda^2-1}\right)\right]^2
\nonumber\\
\left[x+a\left(\frac{\mu-\lambda^2}{\lambda^2-1}\right)\right]^2+y^2+z^2&=&
\left[\frac{a\lambda\left(\mu-1\right)}{\lambda^2-1}\right]^2\equiv
r^2\nonumber
\end{eqnarray}
 Thus the equipotential surface is a sphere with center, O' and  radius, $r$, where
 $$\textsf{O'}:
 \left(a\left[\frac{\lambda^2-\mu}{\lambda^2-1}\right],0,0\right)~~~~\textsf{and}
 ~~~~r=\left|\frac{a\lambda\left(\mu-1\right)}{\lambda^2-1}\right|$$
If we denote the origin by O, then (see Figure 5)
\[
\begin{array} {ccccccc}
     \textsf{OI$_1$}&=&\left|\textsf{OO'}-r\right| &~~~~~~~~~~~&
     \textsf{OI$_2$}&=&\left|\textsf{OO'}+r\right|\nonumber\\
     &=&\left|\displaystyle a\left(\frac{\lambda^2-\mu}{\lambda^2-1}\right)-\frac{a\lambda\left(\mu-1\right)}
     {\lambda^2-1}\right|&~~~~~~~~~~~&
     &=&\left|\displaystyle a\left(\frac{\lambda^2-\mu}{\lambda^2-1}\right)+
     \frac{a\lambda\left(\mu-1\right)}
     {\lambda^2-1}\right| \nonumber\\
&=&a\displaystyle\left|\frac{\lambda-\mu}{\lambda-1}\right|&~~~~~~~~~~~&
&=&a\displaystyle\left|\frac{\lambda+\mu}{\lambda+1}\right|
      \end{array}
\]
One can now use \textsf{OI$_1$} and \textsf{OI$_1$}, calculated
above, to find \textsf{AI$_1$}, \textsf{BI$_1$}, \textsf{AI$_2$} and
\textsf{BI$_2$} as follows (see Figure 6):
\[
\begin{array} {ccccccc}
     \textsf{AI$_1$}&=&\left|\textsf{OI$_1$}-\textsf{OA}\right| &~~~~~~~~~~~&
     \textsf{BI$_1$}&=&\left|\textsf{OB}-\textsf{OI$_1$}\right|\nonumber\\
     &=&\displaystyle\left|a\left(\frac{\lambda-\mu}{\lambda-1}\right)-a\right|&~~~~~~~~~~~&
     &=&\displaystyle\left|\mu a-a\left(\frac{\lambda-\mu}{\lambda-1}\right)\right|
      \nonumber\\
&=&a\displaystyle\left|\frac{1-\mu}{\lambda-1}\right|&~~~~~~~~~~~&
&=&a\displaystyle\left|\frac{\lambda\left(\mu-1\right)}{\lambda-1}\right|
      \end{array}
\]
Therefore, it follows from this last result
$$\fbox{$\displaystyle\frac{\textsf{AI$_1$}}{\textsf{BI$_1$}}=\frac{1}{\left|\lambda\right|}$}$$
Further,
\\
\[
\begin{array} {ccccccc}
     \textsf{AI$_2$}&=&\left|\textsf{OI$_2$}-\textsf{OA}\right| &~~~~~~~~~~~&
     \textsf{BI$_2$}&=&\left|\textsf{OB}-\textsf{OI$_1$}\right|\nonumber\\
     &=&\displaystyle\left|a\left(\frac{\lambda+\mu}{\lambda+1}\right)-a\right|&~~~~~~~~~~~&
     &=&\displaystyle\left|a\left(\frac{\lambda+\mu}{\lambda+1}-\mu a\right)\right|
      \nonumber\\
&=&a\displaystyle\left|\frac{\mu-1}{\lambda+1}\right|&~~~~~~~~~~~&
&=&a\displaystyle\left|\frac{\lambda\left(1-\mu\right)}{\lambda+1}\right|
      \end{array}
\]
Hence,
$$\fbox{$\displaystyle\frac{\textsf{AI$_2$}}{\textsf{BI$_2$}}=\frac{1}{\left|\lambda\right|}$}$$}
\item\textsf{Setting $\lambda=-1$ in Eq.(\ref{s19}), we get
\begin{eqnarray}
2a(\mu-1)x&=&a^2(\mu^2-1)\nonumber\\
x&=&\frac{a}{2}(\mu-1)
\end{eqnarray}
Thus for equal and opposite charges,\emph{ the equipotential is a plane,
bisecting the line of charges orthogonally}. }
      \item \textsf{Since $\lambda<0$ and further since the radius of the sphere is
      $\displaystyle\left|\frac{a\lambda\left(\mu-1\right)}{\lambda^2-1}\right|$, $\lambda$ cannot be equal to $-1$.
      Therefore $$\fbox{$\displaystyle\lambda\in(-\infty,-1)\cup(-1,0)$}$$}
  \end{enumerate}
  \vspace{0.6cm}
  \item
  \begin{enumerate}
  \item\textsf{
       \begin{center}\underline{CASE 1:~~~~$\lambda > 0~~\left(\lambda=\left|\lambda\right|\right)$}\end{center}
Since both charges have same sign (positive), all the lines of force
emanating from either charges must go off to infinity. \noindent Let
P$_1$ be a point on the lines of force originating from A (see Figure
7). Assume that the tangent at P$_1$ passes through
$\overline{\textsf{AB}}$ in C$_1$ \footnote{\noindent Considering point
P$_1$ on a line of force, originating from B instead of A would still
lead to a point C$_1$ in between A and B.}. By Law of sines one has
\begin{eqnarray}
\frac{\sin\beta_1}{\textsf{AC$_1$}}&=&\frac{\sin\delta_1}{\textsf{AP$_1$}}\nonumber\\
\frac{\sin\gamma_1}{\textsf{BC$_1$}}&=&\frac{\sin(\pi-\delta_1)}{\textsf{BP$_1$}}\nonumber
\end{eqnarray}
Dividing first equation by the second, we get
\begin{eqnarray}\label{s20}
\frac{\sin\beta_1}{\sin\gamma_1}=\left(\frac{\textsf{BP$_1$}}{\textsf{AP$_1$}}\right)\left(\frac{\textsf{AC$_1$}}{\textsf{BC$_1$}}\right).
\end{eqnarray}
Also, at P$_1$, the resultant electric field is in the tangential
direction, hence the normal component of the resultant electric
field must vanish at P$_1$ (see Figure 8):
\begin{eqnarray}\label{s21}
\frac{\left|\lambda\right|q}{\textsf{BP$_1$}^2}\sin\gamma_1&=&\frac{q}{\textsf{AP$_1$}^2}\sin\beta_1\nonumber\\
\Rightarrow
\frac{\sin\beta_1}{\sin\gamma_1}&=&\left|\lambda\right|\left(\frac{\textsf{AP$_1$}}{\textsf{BP$_1$}}\right)^2
\end{eqnarray}
Eqs.(\ref{s20}) and (\ref{s21}) implies
\begin{eqnarray}\label{s22}
\frac{\textsf{AC$_1$}}{\textsf{BC$_1$}}&=&\left|\lambda\right|\left(\frac{\textsf{AP$_1$}}{\textsf{BP$_1$}}\right)^3
\end{eqnarray}
\begin{eqnarray}\textsf{As}~\textsf{P$_1$}\rightarrow \infty~ \textsf{along the line of
force,}~ \frac{\textsf{AP$_1$}}{\textsf{BP$_1$}}\rightarrow 1 ~\textsf{and
C$_1$}\rightarrow~\textsf{some fixed point G$_1$ (say)
in}~\overline{\textsf{AB}}.\nonumber
\end{eqnarray}
With this limit, Eq.(\ref{s22}) becomes
\begin{eqnarray}\label{s27}
\fbox{$\displaystyle\frac{\textsf{AG$_1$}}{\textsf{BG$_1$}}=\left|\lambda\right|$}.
\end{eqnarray}
 For this
case, lines of force are sketched in Figure 9.
\vspace{0.3cm}
\begin{center}\underline{CASE 2:~~~~$\lambda < 0~~\left(\lambda=-\left|\lambda\right|\right)$}\\
\vspace{0.3cm}
Here two situations occur (see Figures 10a and 10b)\end{center}
\begin{itemize}
\item $0<\left|\lambda\right|<1$\\
\\
In this case consider a point P$_2$ on the line of force starting from A and going off to infinity (see Figure 11a). Then from $\triangle$P$_2$C$_2$A and $\triangle$P$_2$C$_2$B, we have
\begin{eqnarray}\label{s23}
\frac{\sin\beta_2}{\textsf{AC$_2$}}&=&\frac{\sin(\pi-\delta_2)}{\textsf{AP$_2$}} \nonumber\\
\frac{\sin(\beta_2+\gamma_2)}{\textsf{BC$_2$}}&=&\frac{\sin(\pi-\delta_2)}{\textsf{BP$_2$}}\nonumber\\
\Rightarrow\frac{\sin\beta_2}{\sin(\beta_2+\gamma_2)}&=&\left(\frac{\textsf{BP$_2$}}{\textsf{AP$_2$}}\right)\left(\frac{\textsf{AC$_2$}}
      {\textsf{BC$_2$}}\right)
\end{eqnarray}
Further, we know that the normal component of the electric field at P$_2$ must vanish (see Figure 12 a):
\begin{eqnarray}\label{s24}
\frac{q}{\textsf{AP$_2^2$}}\sin{\beta_2}-\frac{|\lambda|q}{\textsf{BP$_2^2$}}\sin{(\beta_2+\gamma_2)}=0
\end{eqnarray}
Then Eqs.(\ref{s23}) and (\ref{s24}) gives
\begin{eqnarray}\label{s25}
\frac{\textsf{AC$_2$}}{\textsf{BC$_2$}}=|\lambda|\left(\frac{\textsf{AP$_2$}}{\textsf{BP$_2$}}\right)^3
\end{eqnarray}
As before, if we let \textsf{P$_2$} $\rightarrow$ $\infty$ along the line of
force, then
 $\frac{\textsf{AP$_2$}}{\textsf{BP$_2$}}$ $\rightarrow$ 1
 and
C$_2$ $\rightarrow$ some fixed point G$_2$ (say)
in $\overline{\textsf{AB}}$,
leading Eq.(\ref{s25}) to
\begin{eqnarray}\label{s27'}
\fbox{$\displaystyle\frac{\textsf{AG$_2$}}{\textsf{BG$_2$}}=\left|\lambda\right|$}.
\end{eqnarray}
The lines of force are sketched in Figure 10a.\\
\item $\left|\lambda\right|>1$\\
\\
In this case we need to consider a point P$_3$ on the line of force coming from infinity and terminating at B (see Figure 11b).
From $\triangle$P$_3$C$_3$A and $\triangle$P$_3$C$_3$B, we have
\begin{eqnarray}
\frac{\sin(\beta_3+\gamma_3)}{\textsf{AC$_3$}}&=&\frac{\sin(\pi-\delta_3)}{\textsf{AP$_3$}} \nonumber\\
\frac{\sin\beta_3}{\textsf{BC$_3$}}&=&\frac{\sin(\pi-\delta_3)}{\textsf{BP$_3$}}\nonumber\\
\Rightarrow\frac{\sin(\beta_3+\gamma_3)}{\sin\beta_3}&=&\left(\frac{\textsf{BP$_3$}}{\textsf{AP$_3$}}\right)\left(\frac{\textsf{AC$_3$}}
      {\textsf{BC$_3$}}\right)
\end{eqnarray}
Again the normal component of the electric field at P$_3$ must vanish (see Figure 12 b):
\begin{eqnarray}
\frac{q}{\textsf{AP$_3^2$}}\sin(\beta_3+\gamma_3)-\frac{|\lambda|q}{\textsf{BP$_3^2$}}\sin\beta_3=0
\end{eqnarray}
These last two equations again lead to the familiar result
\begin{eqnarray}\label{s26}
\frac{\textsf{AC$_3$}}{\textsf{BC$_3$}}=|\lambda|\left(\frac{\textsf{AP$_3$}}{\textsf{BP$_3$}}\right)^3
\end{eqnarray}
Let \textsf{P$_3$} approach infinity along the line of
force, then
 $\frac{\textsf{AP$_3$}}{\textsf{BP$_3$}}$ would approach 1
 and
C$_2$ tends to some fixed point G$_3$ (say)
in $\overline{\textsf{AB}}$,
leading Eq.(\ref{s26}) to
\begin{eqnarray}\label{s27''}
\fbox{$\displaystyle\frac{\textsf{AG$_3$}}{\textsf{BG$_3$}}=\left|\lambda\right|$}.
\end{eqnarray}
The lines of force are sketched in Figure 10b.\\
\end{itemize}}
 \item\textsf{From Eqs.(\ref{s27}), (\ref{s27'}) and (\ref{s27''}), the ratio $\frac{\textsf{AG}}{\textsf{BG}}\left(=\left|\lambda\right|\right)$ is independent of any angle, it follows that \emph{tangents to all the lines of force at infinity (asypmtotes) must pass through the fixed point G}.}
     \vspace{0.3cm}
   \item \textsf{Referring to Figure 13, one can define a point G (in analogy with the center of mass of a system) as follows:
       \begin{eqnarray}\label{s28}
       \textsf{OG} &=& \frac{q\left(\textsf{OA}\right)\pm |\lambda|q\left(\textsf{OB}\right)}{q\pm\lambda q}\nonumber\\
       &=& \frac{\textsf{OA}\pm|\lambda|\left(\textsf{OB}\right)}{1\pm|\lambda|}
   \end{eqnarray}
   where $\pm$ refers to $\lambda=\pm|\lambda|$ for $\lambda>0$ and $\lambda<0$, respectively.
   Further, from Eq.(\ref{s28}), one can calculate AG and BG
   \begin{eqnarray}\label{s29}
   \textsf{AG}&=&\left|\textsf{OG}-\textsf{OA}\right|\nonumber\\
   &=&\left|\frac{\textsf{OA}\pm|\lambda|(\textsf{OB})-\textsf{OA}\mp |\lambda|(\textsf{OA})}{1\pm|\lambda|}               \right|\nonumber\\
   &=& \frac{|\lambda|}{1\pm|\lambda|}\textsf{AB} ~~~~~~~~~~(\textsf{since}~\textsf{AB}=\left|\textsf{OB}-\textsf{OA}\right|)
   \end{eqnarray}
   \begin{eqnarray}\label{s29'}
   \textsf{BG}&=&\left|\textsf{OB}-\textsf{OG}\right|\nonumber\\
   &=&\left|\frac{\textsf{OB}\pm|\lambda|(\textsf{OB})-\textsf{OA}\mp |\lambda|(\textsf{OB})}{1\pm|\lambda|}               \right|\nonumber\\
   &=& \frac{1}{1\pm|\lambda|}\textsf{AB}
   \end{eqnarray}
   Thus from Eqs.(\ref{s29}) and (\ref{s29'}) , it follows immediately
   $$\fbox{$\displaystyle\frac{\textsf{AG}}{\textsf{BG}}=|\lambda|$}$$
   For future purposes we record OG, AG and BG purely in terms of the given parameters, $\lambda$ and $\mu$. Since, OA $=a$ and OB $=\mu a$, this implies AB=$|\mu-1|a$ and thus, it follows from Eqs. (\ref{s28}), (\ref{s29}) and (\ref{s29'})
   \begin{eqnarray}
    \textsf{OG}&=&\left(\frac{1\pm\mu|\lambda|}{1\pm|\lambda|}\right)a\label{s30'}\\
    \textsf{AG}&=&\left|\frac{\lambda(\mu-1)}{1\pm|\lambda|}\right| a\label{s30''}\\
    \textsf{BG}&=&\left|\frac{\mu-1}{1\pm|\lambda|}\right| a\label{s30'''}
   \end{eqnarray}}
  \end{enumerate}
  \vspace{0.6cm}
  \item
  \begin{enumerate}
  \item\textsf{First recall from part  \ref{2}}\textsf{, that the equation for the lines of force can be written as
   \begin{eqnarray}\label{s30}
   \cos\theta_1+|\lambda|\cos\theta_2=C,
   \end{eqnarray}
   where we have set $\lambda=|\lambda|$, since $\lambda>0$.
   Let Q be a point on a particular line of force emanating from A at an angle $\alpha$ to $\overline{\textsf{AB}}$ as shown in Figure 14. When Q $\rightarrow$ A on the particular line of force considered, then $\theta_1\rightarrow \alpha$ and $\theta_2\rightarrow \pi$. Eq.(\ref{s30}) gives $C=\cos\alpha-|\lambda|$. This Value of $C$ in Eq.(\ref{s30}) gives
   $$\cos\theta_1+|\lambda|\cos\theta_2=\cos\alpha-|\lambda|.$$
   On using the half angle formula, the last equation takes the form
   \begin{eqnarray}
2\cos^2\left(\frac{1}{2}\theta_1\right)-1+|\lambda|\left[2\cos^2\left(\frac{1}{2}\theta_2\right)-1\right]
   &=&2\cos^2\left(\frac{1}{2}\alpha\right)-1-|\lambda|\label{s31}\nonumber
   \end{eqnarray}
    \begin{equation}\label{s32}
          \Rightarrow\fbox{$\displaystyle\cos^2\left(\frac{1}{2}\theta_1\right)+|\lambda|\cos^2\left(\frac{1}{2}\theta_2\right)
   =\cos^2\left(\frac{1}{2}\alpha\right)$}
   \end{equation}
    }
    \vspace{0.3cm}
    \item
      \begin{enumerate}
      \item\label{7}\textsf{When Q has receded to infinity on the particular line of force, then $\theta_1=\theta_2\equiv\theta$ (see Figure 16). This is because AQ$\parallel$BQ at infinity. Therefore, Eq.(\ref{s32}) gives
          \begin{eqnarray}
          \cos^2\left(\frac{1}{2}\theta\right)+|\lambda|\cos^2\left(\frac{1}{2}\theta\right)
   &=&\cos^2\left(\frac{1}{2}\alpha\right)\nonumber
   \end{eqnarray}}
          \begin{equation}\label{s33}
          \Rightarrow\fbox{$\displaystyle\theta=2\cos^{-1}\left[\frac{1}{\sqrt{1+|\lambda|}}
          \cos\left(\frac{1}{2}\alpha\right)\right]$}
          \end{equation}
          \vspace{0.3cm}
      \item \label{8}\textsf{To determine the range of values of $\alpha$, it is convenient to rewrite Eq.(\ref{s33}) as
          \begin{eqnarray}\label{s34}
          \left(1+|\lambda|\right)\left(\frac{1+\cos\theta}{2}\right)&=&\left(\frac{1+\cos\alpha}{2}
          \right)\nonumber\\
          \Rightarrow\cos\theta &=&\frac{\cos\alpha-|\lambda|}{1+|\lambda|}
          \end{eqnarray}
    Thus in order for $\theta$ to be real
          \begin{eqnarray}
          -1<\frac{\cos\alpha-|\lambda|}{1+|\lambda|}<1\nonumber\\
          \Rightarrow-1<\cos\alpha<1+2|\lambda|\nonumber
          \end{eqnarray}
          The second inequality is impossible and therefore $\cos\alpha >-1$ which gives \fbox{$\displaystyle\alpha <180^0$}.}
          \vspace{0.3cm}
      \item\textsf{Using the expression for $\cos\theta$ in Eq.(\ref{s34}) and the right triangle shown in Figure 15, one can write down
          $$\tan\theta=\sqrt{\left(\frac{1+|\lambda|}{\cos\alpha-|\lambda|}\right)^2-1}.$$
          The slope of the asymptote is $\tan\theta$. By symmetry there are two asymptotes (see Figure 16) and therefore the slopes of the lines are given by $m=\pm\tan\theta$. We have also proved that all asymptotes must pass through the center of gravity, G of charges whose coordinates are given by Eq.(\ref{s30'}):
          $$\left(\frac{1+\mu|\lambda|}{1+|\lambda|}\right)a.$$
          Therefore, the general equation of a line $y=mx+b$ immediately yields }
          $$\fbox{$\displaystyle y=\pm\sqrt{\left(\frac{1+|\lambda|}{\cos\alpha-|\lambda|}\right)^2-1}\left[x-\left(\frac{1+\mu|\lambda|}
          {1+|\lambda|}\right)a\right]$}.$$
      \end{enumerate}
      \item \textsf{In this case, assume a point  Q on a  line of force originating from B at an angle $\alpha$ to $\overline{\textsf{AB}}$ (see Figure 17).
$$\textsf{When Q}\rightarrow \textsf{B}:~~~~~ \theta_1\rightarrow\alpha ~~\textsf{and}~~ \theta_2\rightarrow 0$$
 For these values of $\theta_1$ and $\theta_2$, Eq.(\ref{s30}) gives for $C=1+|\lambda|\cos\alpha$. Therefore, Eq.(\ref{s30}) gives
   \begin{eqnarray}\cos\theta_1+|\lambda|\cos\theta_2&=&1+|\lambda|\cos\alpha\nonumber\\
   1-2\sin^2\left(\frac{1}{2}\theta_1\right)+|\lambda|\left[1-2\sin^2\left(\frac{1}{2}\theta_2\right)\right]
   &=&1+|\lambda|\left[1-2\sin^2\left(\frac{1}{2}\alpha\right)\right]\nonumber
   \end{eqnarray}
\begin{equation}
          \Rightarrow \fbox{$\displaystyle\sin^2\left(\frac{1}{2}\theta_1\right)+|\lambda|\sin^2\left(\frac{1}{2}\theta_2\right)
   =|\lambda|\sin^2\left(\frac{1}{2}\alpha\right)\label{s35}$}
      \end{equation}
   Now, when Q$\rightarrow \infty$, $\theta_1=\theta_2\equiv\theta$. With these values of $\theta_1$ and $\theta_2$
   in Eq.(\ref{s35}), we get
   $$\sin^2\left(\frac{1}{2}\theta\right)+|\lambda|\sin^2\left(\frac{1}{2}\theta\right)
   =|\lambda|\sin^2\left(\frac{1}{2}\alpha\right)$$
   $$\Rightarrow\fbox{$\displaystyle\theta=2\sin^{-1}\left[\frac{|\lambda|}{\sqrt{1+|\lambda|}}
   \sin\left(\frac{1}{2}\alpha\right)\right]$}$$
To find the range of values of $\alpha$, we rewrite this last equation as
$$\left(1+|\lambda|\right)\left(\frac{1-\cos\theta}{2}\right)=|\lambda|\left(\frac{1-\cos\alpha}{2}\right)$$
$$\Rightarrow\cos\theta =\frac{1+|\lambda|\cos\alpha}{1+|\lambda|}$$
Therefore, $\theta$ to be real
$$-1<\frac{1+|\lambda|\cos\alpha}{1+|\lambda|}<1$$
$$\Rightarrow-1-\frac{2}{|\lambda|}<\cos\alpha<1.$$
The first inequality is not valid and therefore, $\cos\alpha<1$, that is, \fbox{$\displaystyle\alpha>0^0$} }
  \end{enumerate}
  \vspace{0.6cm}
  \item
  \begin{enumerate}
  \item\label{9}\textsf{Since $\lambda<0$, we put $\lambda=-|\lambda|$ in the equation given in part \ref{2}}. \textsf{\begin{equation}\label{s36'}
      \cos\theta_1-|\lambda|\cos\theta_2=C.
      \end{equation}
  Next, consider a point P on a line of force which originates from A at an angle $\alpha$ with respect to $\overline{\textsf{AB}}$. Thus, as P approaches A along the particular line of force we are considering (see Figure 18), $\theta_1\rightarrow\alpha$ and $\theta_2\rightarrow\pi$. This gives $C=\cos\alpha+|\lambda|$. Therefore Eq.(\ref{s36'}) takes the form,
\begin{eqnarray}
  \cos\theta_1-|\lambda|\cos\theta_2&=&\cos\alpha+|\lambda|\nonumber\\
  \Rightarrow 2\cos^2\left(\frac{1}{2}\theta_1\right)-1-|\lambda|\left[2\cos^2\left(\frac{1}{2}\theta_2\right)-1\right]
   &=&2\cos^2\left(\frac{1}{2}\alpha\right)-1+|\lambda|\nonumber
   \end{eqnarray}
   \begin{equation}\label{s36}
   \Rightarrow\fbox{$\displaystyle\cos^2\left(\frac{1}{2}\theta_1\right)-|\lambda|\cos^2\left(\frac{1}{2}\theta_2\right)
   =\cos^2\left(\frac{1}{2}\alpha\right)$}
   \end{equation}}
\vspace{0.3cm}
  \item
      \begin{enumerate}
      \item\label{10}\textsf{Assume that the line of force considered in part \ref{9}} \textsf{terminates at B at an angle $\beta$ w.r.t. $\overline{\textsf{AB}}$ (see Figure 19). Then as P $\rightarrow$ B, $\theta_2=\beta, ~\theta_1=0$ and Eq.(\ref{s36}) gives
       \begin{eqnarray}\label{s37}
          1-|\lambda|\cos^2\left(\frac{1}{2}\beta\right)&=&1-\sin^2\left(\frac{1}{2}\alpha\right)\nonumber
          \end{eqnarray}
          \begin{equation}\label{s38}
          \Rightarrow \fbox{$\displaystyle \beta= 2\cos^{-1}\left[\frac{1}{\sqrt{|\lambda|}}\sin\left(\frac{1}{2}\alpha\right)\right]$}\nonumber
          \end{equation}
          }
          \vspace{0.3cm}
      \item \textsf{Rewrite Eq.(\ref{s38}) as
      \begin{eqnarray}
      |\lambda|\left(\frac{1+\cos\beta}{2}\right)&=&\frac{1-\cos\beta}{2}\nonumber\\
      \Rightarrow \cos\beta &=&\frac{1-\cos\alpha-|\lambda|}{|\lambda|}\nonumber
      \end{eqnarray}
      Thus, in order for $\beta$ to exist
      \begin{eqnarray}
      -1<\frac{1-\cos\alpha-|\lambda|}{|\lambda|}<1\nonumber\\
      \Rightarrow 1-2|\lambda|<\cos\alpha<1\nonumber
      \end{eqnarray}
      \begin{eqnarray}
      \textsf{Finally,}~~\fbox{$\displaystyle 0<\alpha< \cos^{-1}\left(1-2|\lambda|\right)$}\nonumber
      \end{eqnarray}
      Further, from this last equation,
      \begin{eqnarray}
      -1<1-2|\lambda|<1\nonumber\\
      \Rightarrow\fbox{$\displaystyle -1<\lambda<0 $}\nonumber
      \end{eqnarray}}
      \end{enumerate}
      \vspace{0.3cm}
      \item
      \begin{enumerate}
      \item \label{11}\textsf{ When the point P on the  line of force considered in part \ref{9},} \textsf{goes to infinity (see Figure 20), AP$\parallel$BP and $\theta_1=\theta_1\equiv\theta $. Then, Eq.(\ref{s36}) implies
          \begin{eqnarray}
          \cos^2\left(\frac{1}{2}\theta\right)-|\lambda|\cos^2\left(\frac{1}{2}\theta\right)
   =\cos^2\left(\frac{1}{2}\alpha\right)\nonumber\\
   \fbox{$\displaystyle\theta=2\cos^{-1}\left[\frac{1}{\sqrt{1-|\lambda|}}
   \cos\left(\frac{1}{2}\alpha\right)\right]$}\nonumber
          \end{eqnarray}}
          \vspace{0.3cm}
          \item \textsf{Rewriting the last equation, we get
          \begin{eqnarray}\label{s39}
          \left(1-|\lambda|\right)\left(\frac{1+\cos\theta}{2}\right)&=&\frac{1+\cos\alpha}{2}\nonumber\\
          \cos\theta&=&\frac{|\lambda|+\cos\alpha}{1-|\lambda|}
          \end{eqnarray}
          Therefore,
          $$-1<\frac{|\lambda|+\cos\alpha}{1-|\lambda|}<1$$
          \\
              \[
\begin{array} {ccc}
\displaystyle\frac{|\lambda|+\cos\alpha}{1-|\lambda|}>-1 &\bigcap&
      \displaystyle\frac{|\lambda|+\cos\alpha}{1-|\lambda|}<1\nonumber\\
      \displaystyle\frac{\cos\alpha+1}{1-|\lambda|}>0&~~~~~\bigcap~~~~~&\displaystyle\frac{2|\lambda|-1+\cos\alpha}{1-|\lambda|}<0 \nonumber\\
       \Rightarrow \alpha<180^0~~\textsf{and}~~-1<\lambda<0&~~~~~\bigcap~~~~~&\Rightarrow \left\{\matrix{ \alpha>\cos^{-1}\left(1-2|\lambda|\right)~~\textsf{and}~~-1<\lambda<0\cr
       \textsf{or}\cr
      \alpha<\cos^{-1}\left(1-2|\lambda|\right)~~\textsf{and}~~\lambda<-1}\right.  \nonumber\\
      \end{array}
\]
\begin{eqnarray}
\textsf{Finally,}~~~~~~~~\fbox{$\displaystyle \cos^{-1}\left(1-2|\lambda|\right)<\alpha<180^0 $}~~\textsf{and}~~\fbox{$\displaystyle-1<\lambda<0$}\nonumber
\end{eqnarray}}
\vspace{0.3cm}
             \item\textsf{Recall that all asymptotes must through the center of gravity, G of charges which has the $x$-coordinates (see Eq.(\ref{s30'})):
                 $$\left(\frac{1-\mu|\lambda|}{1-|\lambda|}\right)a$$
             The equation of the asymptote is $y=mx+b$, where $m=\pm\tan\theta$. Then from Eq.(\ref{s39}), we have
             $$\tan\theta=\sqrt{\sec^2\theta-1}=\sqrt{\frac{1}{\cos^2\theta}-1}=
             \sqrt{\left(\frac{1-|\lambda|}{\cos\alpha+|\lambda|}\right)^2-1}
             $$}
          $$\Rightarrow \fbox{$\displaystyle y=\pm\sqrt{\left(\frac{1-|\lambda|}{\cos\alpha+|\lambda|}\right)^2-1}\left[x-\left(\frac{1-\mu|\lambda|}
          {1-|\lambda|}\right)a\right]$}.$$
          \end{enumerate}
          \vspace{0.3cm}
          \item
      \begin{enumerate}
      \item \label{12}\textsf{If the line of force  originating from A, considered above, is an \textit{extreme} line of force (see Figure 21), then as P$\rightarrow$B, $\theta_1=0$ and $\theta_2=0$. This is because the lines of force leave tangentially $\overline{\textsf{AB}}$. Eq.(\ref{s36}) implies
          \begin{eqnarray}
          \cos^2\left(\frac{1}{2}\alpha\right)&=&1-|\lambda|\nonumber\\
          \Rightarrow \frac{1+\cos\alpha}{2}&=&1-|\lambda|\nonumber
          \end{eqnarray}
          \begin{equation}\label{s39'}
          \Rightarrow\fbox{$\displaystyle\alpha=\cos^{-1}\left(1-2|\lambda|\right)$}
          \end{equation}}
       \vspace{0.3cm}
\item \textsf{For real values of $\alpha$}
$$
-1<1-2|\lambda|<1
\Rightarrow \fbox{$\displaystyle -1<\lambda<0$}$$
\end{enumerate}
\vspace{0.3cm}
          \item
      \begin{enumerate}
      \item \label{13}\textsf{Let the angle between $\overline{\textsf{AB}}$ and $\overline{\textsf{AC}}$ be denoted by $\theta_0$ (see Figure 22). Then as P$\rightarrow$ C, $\theta_1=\theta_0$ and $\theta_2=\pi-\theta_0$.  Eq.(\ref{s36}) implies
          \begin{eqnarray}
          \cos^2\left(\frac{1}{2}\theta_0\right)-|\lambda|\cos^2\left(\frac{1}{2}\left[\pi-\theta_0\right]\right)
   &=&\cos^2\left(\frac{1}{2}\alpha\right)\nonumber\\
   \Rightarrow 1-\sin^2\left(\frac{1}{2}\theta_0\right)-|\lambda|\sin^2\left(\frac{1}{2}\theta_0\right)
      &=&    1-\sin^2\left(\frac{1}{2}\alpha\right)\nonumber
          \end{eqnarray}
          \begin{equation}\label{s40}
          \Rightarrow \fbox{$\displaystyle\theta_0=2\sin^{-1}\left[\frac{1}{\sqrt{{1+|\lambda|}}}
          \sin\left(\frac{1}{2}\alpha\right)\right]$}
          \end{equation}}
          \vspace{0.3cm}
          \item \textsf{One can rewrite Eq.(\ref{s40}) as
          \begin{eqnarray}\label{s41}
          \left(\frac{1-\cos\theta_0}{2}\right)\left(1+|\lambda|\right)&=&\frac{1-\cos\alpha}{2}\nonumber\\
          \Rightarrow \cos\theta_0=\frac{|\lambda|+\cos\alpha}{1+|\lambda|}
          \end{eqnarray}
          In order for $\theta_0$ to be valid
          \begin{eqnarray}
          -1<\frac{|\lambda|+\cos\alpha}{1+|\lambda|}<1\nonumber\\
          \Rightarrow-1-2|\lambda|<\cos\alpha<1\nonumber
          \end{eqnarray}
          First inequality is not valid, therefore $\cos\alpha<1 \Rightarrow$ \fbox{$\displaystyle \alpha>0 $}}
           \vspace{0.3cm}
             \item\textsf{From Figure 22, $\frac{\textsf{OC}}{\textsf{OA}}=\tan\theta_0$. Thus
             \begin{eqnarray}
             \left(\textsf{OC}\right)_{\textsf{max}}&=&\textsf{OA}\tan\left(\theta_{0}\right)_{\textsf{max}}\label{s42'}\\ \textsf{ Eq.(\ref{s41})}\Rightarrow
             \cos\left(\theta_{0}\right)_{\textsf{max}}&=&\frac{|\lambda|+\cos\left(\alpha\right)
             _{\textsf{max}}}{1+|\lambda|}\label{s42''}\\
             \textsf{Eq.(\ref{s39'})}\Rightarrow
      \cos\left(\alpha\right)
             _{\textsf{max}}&=&1-2|\lambda|\label{s42'''}
             \end{eqnarray}
           Eqs.(\ref{s42''}) and (\ref{s42'''}) gives
             \begin{equation}\label{s43}
             \cos\left(\theta_{0}\right)_{\textsf{max}}=\frac{1-|\lambda|}{1+|\lambda|}
             \end{equation}
             and that
             \begin{eqnarray}
             \tan\left(\theta_{0}\right)_{\textsf{max}}&=&\sqrt{\sec^2\left(\theta_{0}\right)_{\textsf{max}}-1}\nonumber\\
             &=&\sqrt{\frac{1}{\cos^2\left(\theta_{0}\right)_{\textsf{max}}}-1}\nonumber\\
              \textsf{Eqs.(\ref{s43})}\Rightarrow\tan\left(\theta_{0}\right)_{\textsf{max}}&=&\frac{2\sqrt{|\lambda|}}
              {1-|\lambda|}\label{s44}
             \end{eqnarray}
             Using OA$=\displaystyle\left|\frac{a}{2}(1+\mu)-a\right|=\displaystyle\left|\frac{1}{2}(1-\mu)\right|a$, and Eq.(\ref{s44}) in Eq.(\ref{s42'}), we get
             $$\fbox{$\displaystyle\left(\textsf{OC}\right)_{\textsf{max}}=
             \left|\mu-1\right|\frac{\sqrt{|\lambda|}}{1-|\lambda|}a$}$$}
          \end{enumerate}
          \vspace{0.3cm}
          \item
      \begin{enumerate}
      \item \label{14}\textsf{Let there be a point R on a line of force that ends at B at an angle $\alpha$ w.r.t. $\overline{\textsf{AB}}$ (see figure 23). When R$\rightarrow$B: $\theta_1=0$ and $\theta_2=\alpha$. With this set of information, Eq.(\ref{s36'}) gives $C=1-|\lambda|\cos\alpha$ and for this specific value of $C$, Eq.(\ref{s36'}) becomes
          \begin{eqnarray}
          \cos\theta_1-|\lambda|\cos\theta_2&=&1-|\lambda|\cos\alpha\nonumber\\
          \Rightarrow 1-\sin^2\left(\frac{1}{2}\theta_1\right)-|\lambda|
          \left[1-\sin^2\left(\frac{1}{2}\theta_2\right)\right]
          &=& 1-|\lambda|\left[1-\sin^2\left(\frac{1}{2}\alpha\right)\right]\nonumber
          \end{eqnarray}
          \begin{equation}\label{s45}
          \Rightarrow\fbox{$\displaystyle|\lambda|\sin^2\left(\frac{1}{2}\theta_2\right)-\sin^2\left(\frac{1}{2}\theta_1\right)=
           |\lambda|\sin^2\left(\frac{1}{2}\alpha\right)$}
          \end{equation}}
           \vspace{0.3cm}
          \item \textsf{If the line of force considered in part \ref{14},} \textsf{is restricted to have been originated from A, then this implies that as R$\rightarrow$A, $\theta_1=\beta$ (say) and $\theta_2=\pi$ (see Figure 24). Eq.(\ref{s45}) then takes the form
              \begin{eqnarray}
              |\lambda|-\sin^2\left(\frac{1}{2}\beta\right)=|\lambda|
              \sin^2\left(\frac{1}{2}\alpha\right)\nonumber
              \end{eqnarray}
              \begin{equation}\label{s46}
              \Rightarrow\fbox{$\displaystyle\beta=2\sin^{-1}\left[\sqrt{|\lambda|}
              \cos\left(\frac{1}{2}\alpha\right)\right]$}
              \end{equation}
              As in previous cases, let us rewrite Eq.(\ref{s46}) as
              \begin{eqnarray}
              \sin^2\left(\frac{1}{2}\beta\right)&=&|\lambda|\cos^2\left(\frac{1}{2}\alpha\right)\nonumber\\
              \Rightarrow\frac{1-\cos\beta}{2}&=&\left(\frac{1+\cos\alpha}{2}\right)\nonumber\\
              \Rightarrow \cos\beta&=&1-|\lambda|-|\lambda|\cos\alpha\nonumber
              \end{eqnarray}
              Therefore in order for $\beta$ to be real,
              \begin{eqnarray}
              -1<1-|\lambda|-|\lambda|\cos\alpha<1\nonumber\\
              \Rightarrow \frac{2}{|\lambda|}-1>\cos\alpha>-1\nonumber
              \end{eqnarray}
              $$\fbox{$\displaystyle \Rightarrow\cos^{-1}\left(\frac{2}{|\lambda|}-1\right)<\alpha<180^0$}$$
              Further, in order for $\alpha$ to exist
              $$\frac{2}{|\lambda|}-1<1\Rightarrow\fbox{$\displaystyle\lambda<-1$}$$}
                \vspace{0.3cm}
             \item\textsf{In order for the line of force, that ends at B, to have been originated from infinity, $\theta_1=\theta_2\equiv\theta$ (see Figure 25). This is because as R$\rightarrow\infty$, AR becomes parallel to BR. Therefore Eq.(\ref{s45}) yields
                 \begin{eqnarray}
                 \left[|\lambda|-1\right]\sin^2\left(\frac{1}{2}\theta\right)=
           |\lambda|\sin^2\left(\frac{1}{2}\alpha\right)\nonumber
                 \end{eqnarray}
                 $$\Rightarrow\theta=2\sin^{-1}\left[\sqrt{\frac{|\lambda|}
                 {|\lambda|-1}}\sin\left(\frac{1}{2}\alpha\right)\right]$$
                  To find the restriction on the angle $\alpha$, it would be convenient if we write the last equation as
                  \begin{eqnarray}
                  \left(|\lambda|-1\right)\left(\frac{1-\cos\theta}{2}\right)&=&|\lambda|
                  \left(\frac{1-\cos\alpha}{2}\right)\nonumber\\
                  \Rightarrow\cos\theta&=&\frac{1-|\lambda|\cos\alpha}{1-|\lambda|}\nonumber
           \end{eqnarray}
                $$
               \textsf{Therefore,}~~~~-1<\frac{1-|\lambda|\cos\alpha}{1-|\lambda|}<1\nonumber
             $$
             \\
               \[
\begin{array} {ccc}
\nonumber\\
\displaystyle\frac{1-|\lambda|\cos\alpha}{1-|\lambda|}>-1 &~~~~~~\bigcap~~~~~~&
      \displaystyle\frac{1-|\lambda|\cos\alpha}{1-|\lambda|}<1\nonumber\\
      \displaystyle\frac{2-|\lambda|-|\lambda|\cos\alpha}{1-|\lambda|}>0&~~~~~~\bigcap~~~~~~&
      \displaystyle\frac{|\lambda|\left(1-\cos\alpha|\right)}{1-|\lambda|}<0 \nonumber\\
      \nonumber\\
         \Rightarrow \left\{\matrix{ \displaystyle\cos\alpha<\frac{2}{|\lambda|}-1~~\textsf{and}~~-1<\lambda<0\cr
       \textsf{or}\cr
      \displaystyle\cos\alpha>\frac{2}{|\lambda|}-1~~\textsf{and}~~\lambda<-1}\right.&~~~~~~\bigcap~~~~~~& \Rightarrow \cos\alpha<1~~\textsf{and}~~\lambda<-1 \nonumber\\
      \end{array}
\]
\begin{eqnarray}
\textsf{Therefore,}~~~~~~~\frac{2}{|\lambda|}-1<\cos\alpha<1 ~~\textsf{and}~~\lambda<-1\nonumber\\
\Rightarrow\fbox{$\displaystyle 0<\alpha<\cos^{-1}\left(\frac{2}{|\lambda|}-1\right) $}~~\textsf{and}~~\fbox{$\displaystyle\lambda<-1$}\nonumber
\end{eqnarray}}
                  \vspace{0.3cm}
          \item \textsf{If the line of force considered in \ref{14},} \textsf{is an extreme line of force, then as R$\rightarrow$A, $\theta_1=\theta_2\rightarrow\pi$ (see Figure 26). Eq.(\ref{s45})
              \begin{eqnarray}
           |\lambda|\sin^2\left(\frac{1}{2}\pi\right)-\sin^2\left(\frac{1}{2}\pi\right)&=&
           |\lambda|\sin^2\left(\frac{1}{2}\alpha\right)\nonumber\\
           \Rightarrow |\lambda|-1&=&|\lambda|\left(\frac{1-\cos\alpha}{2}\right)\nonumber
              \end{eqnarray}
              \begin{equation}\label{s48}
              \fbox{$\displaystyle\alpha=\cos^{-1}\left(\frac{2}{|\lambda|}-1 \right)$}.
              \end{equation}
              In order for $\alpha$ to exist
              $$-1<\frac{2}{|\lambda|}-1<1\Rightarrow \fbox{$\displaystyle\lambda<-1$}$$
              }
           \item \textsf{Let the angle between  $\overline{\textsf{AB}}$ and $\overline{\textsf{AC}}$ be denoted by $\theta_0$ (see Figure 27). As R$\rightarrow$C, $\theta_1=\theta_0$ and $\theta_2=\pi-\theta_0$. Eq.(\ref{s45}) implies
               \begin{eqnarray}
               |\lambda|\sin^2\left(\frac{1}{2}\theta_0\right)-\cos^2\left(\frac{1}{2}\theta_0\right)&=&
           |\lambda|\sin^2\left(\frac{1}{2}\alpha\right)\nonumber\\
           \Rightarrow|\lambda|\left[1-\cos^2\left(\frac{1}{2}\theta_0\right)\right]
           -\cos^2\left(\frac{1}{2}\theta_0\right)&=&
           |\lambda|\left[1-\cos^2\left(\frac{1}{2}\alpha\right)\right]\nonumber
               \end{eqnarray}
               \begin{eqnarray}
               \Rightarrow\fbox{$\displaystyle\theta_0=
               2\cos^{-1}\left[\sqrt{\frac{|\lambda|}{1+|\lambda|}}\cos\left(\frac{1}{2}\alpha\right)\right]$}
               \end{eqnarray}
               Rewriting this last equation, we get
               \begin{eqnarray}
               \left(1+|\lambda|\right)\left(\frac{1+\cos\theta_0}{2}\right)&=&|\lambda|\left(
               \frac{1+\cos\alpha}{2}\right)\nonumber\\
               \cos\theta_0=\frac{|\lambda|\cos\alpha-1}{|\lambda|+1}\label{s49}
               \end{eqnarray}
               In the above equation, $\theta_0$ would be real, if
               \begin{eqnarray}
               -1<\frac{|\lambda\cos\alpha|-1}{|\lambda|+1}<1\nonumber\\
               \Rightarrow -1<\cos\alpha<1+\frac{2}{|\lambda|}\nonumber
               \end{eqnarray}
               The second inequality is not possible. The only solution is $\cos\alpha>-1\Rightarrow$ \fbox{$\displaystyle \alpha<180^0$}.
               \\
               From Figure 27, $\frac{\textsf{OC}}{\textsf{OA}}=\tan\theta_0$. Thus
             \begin{eqnarray}
             \left(\textsf{OC}\right)_{\textsf{max}}&=&\textsf{OA}\tan\left(\theta_{0}\right)_{\textsf{max}}\label{s50'}\\ \textsf{ Eq.(\ref{s49})}\Rightarrow
             \cos\left(\theta_{0}\right)_{\textsf{max}}&=&\frac{|\lambda|\cos\left(\alpha\right)
             _{\textsf{max}}-1}{|\lambda|+1}\label{s50''}\\
             \textsf{Eq.(\ref{s48})}\Rightarrow
      \cos\left(\alpha\right)
             _{\textsf{max}}&=&\frac{2}{|\lambda|}-1\label{s50'''}
             \end{eqnarray}
           Eqs.(\ref{s50''}) and (\ref{s50'''}) gives
             \begin{equation}\label{s51}
             \cos\left(\theta_{0}\right)_{\textsf{max}}=\frac{1-|\lambda|}{1+|\lambda|}
             \end{equation}
             and that
             \begin{eqnarray}
             \tan\left(\theta_{0}\right)_{\textsf{max}}&=&\sqrt{\sec^2\left(\theta_{0}\right)_{\textsf{max}}-1}\nonumber\\
             &=&\sqrt{\frac{1}{\cos^2\left(\theta_{0}\right)_{\textsf{max}}}-1}\nonumber\\
              \textsf{Eqs.(\ref{s51})}\Rightarrow\tan\left(\theta_{0}\right)_{\textsf{max}}&=&\frac{2\sqrt{|\lambda|}}
              {1-|\lambda|}\label{s52}
             \end{eqnarray}
             Using OA$=|\frac{a}{2}(1+\mu)-a|=|\frac{1}{2}(1-\mu)|a$, and Eq.(\ref{s52}) in Eq.(\ref{s50'}), we get
             $$\fbox{$\displaystyle\left(\textsf{OC}\right)_{\textsf{max}}=
             \left|\mu-1\right|\frac{\sqrt{|\lambda|}}{1-|\lambda|}a$}$$ }
          \end{enumerate}
 \end{enumerate}
 \vspace{0.6cm}
         \item
  \begin{enumerate}
      \item\textsf{Since, $\lambda >0$ for this part of the problem, we set $\lambda=|\lambda|$}
      \begin{enumerate}
      \item\textsf{Starting from the equation for the lines of force, $\cos\theta_1+|\lambda|\cos\theta_2=C$, one can determine the constant $C$ for the limiting line of force through the neutral point N by setting $\theta_1=0$ and $\theta_2=\pi$ (see Figure 28). This is because for like charges the neutral point N is on the line joining the charges ($\overline{\textsf{AB}}$) and is in between the two charges (see part \ref{4'}).} \textsf{Thus, $C=\cos0+|\lambda|\cos\pi\Rightarrow C=1-|\lambda|$. Equation for the line of force in this case, then takes the form
          \begin{eqnarray}
          \cos\theta_1+|\lambda|\cos\theta_2&=&1-|\lambda|\label{s53}\\
          \Rightarrow1-2\sin^2\left(\frac{1}{2}\theta_1\right)+|\lambda|
          \left[2\cos^2\left(\frac{1}{2}\theta_2\right)-1\right]&=&1-|\lambda|\nonumber
          \end{eqnarray}
          \begin{equation}\label{s54}
          \Rightarrow\fbox{$\displaystyle\sin^2\left(\frac{1}{2}\theta_1\right)=
          |\lambda|\cos^2\left(\frac{1}{2}\theta_2\right)$}
          \end{equation}}
           \vspace{0.3cm}
      \item \textsf{When a point on this  limiting line of force through N has receded to infinity,
      $\theta_1=\theta_2\equiv\theta_0$. We now use this information in Eq.{\ref{s53}} to determine the slope of the asymptote through N:
      \begin{eqnarray}
      \cos\theta_0 &=&\frac{1-|\lambda|}{1+|\lambda|}\label{s55}\\
      \textsf{Further,}~~~~~~\tan\theta_0&=&\sqrt{\frac{1}{\cos^2\theta_0}-1}\nonumber\\
      \textsf{Eq.(\ref{s55})}\Rightarrow~~~~~~\tan\theta_0&=&\frac{2\sqrt{|\lambda|}}{1-|\lambda|}\nonumber
        \end{eqnarray}
        And by symmetry it has two asymptotes with slopes,
         $$m=\pm\tan\theta_0.$$
        As before the asymptote to this limiting line of force must pass through center of gravity of charges  G, whose coordinates are
      (see Eq.(\ref{s30'})):
                 $$\left(\left[\frac{1-\mu|\lambda|}{1-|\lambda|}\right]a,0\right)$$
        Then, the $y-$intercept, $b$ in the equation for an asymptote ($y=mx+b$) is given by
        $$b=-m\left(\frac{1+\mu|\lambda|}{1+|\lambda|}\right)a$$ and finally
        $$\fbox{$\displaystyle y=\pm\frac{2\sqrt{|\lambda|}}{1-|\lambda|}\left[x-\left(\frac{1+\mu|\lambda|}
          {1+|\lambda|}\right)a\right]$}.$$}
          \end{enumerate}
          \vspace{0.3cm}
      \item\textsf{Since $\lambda <0$ for this part of the problem, we set $\lambda=-|\lambda|$}
      \begin{enumerate}
      \item\textsf{Without loss of generality, we assume $\mu>1$ for this part of the question. This means that point B (charge $-|\lambda|q$) will always to the right of point A (charge $q$) Further, recall from part \ref{4'}}, \textsf{that in the case of unlike charges, the neutral point N  will always be either to the left of A or to the right of B and that N has $x$-coordinates:
          $$\frac{\mu-\sqrt{|\lambda|}}{1-\sqrt{|\lambda|}a}$$
           We now determine the permissible values of $\lambda$ for each of the two locations.
           \[
\begin{array} {ccc}
     \frac{\mu-\sqrt{|\lambda|}}{1-\sqrt{|\lambda|}}a>\mu a &~~~~~~~~~~~&
     \frac{\mu-\sqrt{|\lambda|}}{1-\sqrt{|\lambda|}}a<\mu a\nonumber\\
    \frac{\left(\mu-1\right)\sqrt{|\lambda|}}{1-\sqrt{|\lambda|}} >0 &~~~~~~~~~~~&
     \frac{\left(\mu-1\right)}{1-\sqrt{|\lambda|}} <0
      \nonumber\\
\Rightarrow-1<\lambda<0  &~~~~~~~~~~~&
 \Rightarrow\lambda<-1
      \end{array}
\]
The situation is depicted in  Figure 29.\\
\begin{itemize}
\item $\lambda<-1$ (see Figure 30a)\\
\\
When S$_1$$\rightarrow$N$_1$, $\theta_1\rightarrow\pi$ and  $\theta_2\rightarrow\pi$ . Then $\cos\theta_1-|\lambda|\cos\theta_2=C$ leads to $C=\cos\pi-|\lambda|\cos\pi=-1+|\lambda|$. Therefore the equation for the line of force which passes through the neutral point N$_1$ take form
\begin{eqnarray}
\cos\theta_1-|\lambda|\cos\theta_2&=&-1+|\lambda|\nonumber\\
 \Rightarrow 2\cos^2\left(\frac{1}{2}\theta_1\right)-1-|\lambda|\left[2\cos^2\left(\frac{1}{2}\theta_2\right)-1\right]
&=&-1+|\lambda|\nonumber
\end{eqnarray}
\begin{equation}\label{s56}
\fbox{$\displaystyle\cos^2\left(\frac{1}{2}\theta_1\right)=|\lambda|\cos^2\left(\frac{1}{2}\theta_2\right)$}
\end{equation}\\
\item $-1<\lambda<0$ (see Figure 30b)\\
\\
When S$_2$$\rightarrow$N$_2$, $\theta_1\rightarrow0$ and  $\theta_2\rightarrow0$ . Then $\cos\theta_1-|\lambda|\cos\theta_2=C$ leads to $C=\cos0-|\lambda|\cos0=1-|\lambda|$. Therefore the equation for the line of force which passes through the neutral point N$_2$ take form
\begin{eqnarray}
\cos\theta_1-|\lambda|\cos\theta_2&=&1-|\lambda|\nonumber\\
 \Rightarrow 1-2\sin^2\left(\frac{1}{2}\theta_1\right)-|\lambda|\left[1-2\sin^2\left(\frac{1}{2}\theta_2\right)\right]
&=&1-|\lambda|\nonumber
\end{eqnarray}
\begin{equation}\label{s57}
\fbox{$\displaystyle\sin^2\left(\frac{1}{2}\theta_1\right)=|\lambda|\sin^2\left(\frac{1}{2}\theta_2\right)$}
\end{equation}
\end{itemize}}
\vspace{0.3cm}
      \item
      \begin{enumerate}
      \item \textsf{When S$_2$$\rightarrow$A, $\theta_1\rightarrow\theta_{02}$ (say) and  $\theta_2\rightarrow\pi$ (see Figure 30b). Then Eq.(\ref{s57}) implies
\begin{eqnarray} \sin^2\left(\frac{1}{2}\theta_{01}\right)&=&|\lambda|\left[\sin^2\left(\frac{1}{2}\pi\right)\right]
\nonumber\\
\Rightarrow\frac{1-\cos\theta_{01}}{2}&=&|\lambda|\nonumber
\end{eqnarray}
\begin{equation}\label{s58}
\fbox{$\displaystyle\theta_{01}=\cos^{-1}\left(1-2|\lambda|\right)$}
\end{equation}}
\vspace{0.3cm}
      \item \textsf{When S$_1$$\rightarrow$B, $\theta_1\rightarrow0$  and  $\theta_2\rightarrow\theta_{02}$ (say) (see Figure 30a). Then Eq.(\ref{s56}) implies
\begin{eqnarray} \cos^2\left(0\right)&=&|\lambda|\left[\cos^2\left(\frac{1}{2}\theta_{02}\right)\right]
\nonumber\\
\Rightarrow1&=&|\lambda|\left[\frac{1+\cos\theta_{02}}{2}\right]\nonumber
\end{eqnarray}
\begin{equation}\label{s59}
\fbox{$\displaystyle\theta_{02}=\cos^{-1}\left(\frac{2}{|\lambda|}-1\right)$}
\end{equation}}
      \end{enumerate}
      \item\textsf{Comparing Eq.(\ref{s58}) with Eq.(\ref{s39'}) and Eq.(\ref{s59}) with Eq.(\ref{s48}), we conclude that \emph{the line of force which passes through the neutral point (N$_1$) N$_2$ separates the lines going from A to B from those (coming to B) going from A  (from infinity) to  infinity}. Complete situation is  shown in Figures 31a and 31b. }
      \end{enumerate}
  \end{enumerate}
  \vspace{0.6cm}
  \item
  \begin{enumerate}
  \item\textsf{The statement of the problem is depicted in Figure 32. Here we are assuming that $\mu>1$, without loss of generality. Now on using Sine Rule in $\triangle$ABQ (see Figure 33), we get
      \begin{eqnarray}
      \frac{\sin\left(\pi-\frac{1}{2}\theta_2\right)}{\textsf{AQ}}&=&
      \frac{\sin\left(\frac{1}{2}\theta_1\right)}{\textsf{BQ}}\nonumber\\
      \frac{\sin^2\left(\frac{1}{2}\theta_1\right)}{\sin^2\left(\frac{1}{2}\theta_2\right)}&=&
      \frac{\textsf{BQ}^2}{\textsf{AQ}^2}\nonumber\\
      \textsf{Eq.(\ref{s57})}\Rightarrow |\lambda|&=& \frac{\textsf{BQ}^2}{\textsf{AQ}^2}\nonumber\
       \end{eqnarray}
       Thus,
       \begin{eqnarray}
      \frac{\left(x-\mu a\right)^2+y^2}{\left(x- a\right)^2+y^2}&=&|\lambda|\nonumber\\
     \textsf{expanding,}~~~~ x^2-2a\left(\frac{\mu-|\lambda|}{1-|\lambda|}\right)x+y^2&=&
     a^2\left(\frac{|\lambda|-\mu^2}{1-|\lambda|}\right)\nonumber\\
     \textsf{completing the square,}~~~~\left[x-\left(\frac{\mu-|\lambda|}{1-|\lambda|}\right)a\right]^2+y^2&=&
     a^2\left(\frac{|\lambda|-\mu^2}{1-|\lambda|}\right)+
     \left[\left(\frac{\mu-|\lambda|}{1-|\lambda|}\right)a\right]^2\nonumber\\
     \textsf{and finally,}~~~~\left[x-\left(\frac{\mu-|\lambda|}{1-|\lambda|}\right)a\right]^2+y^2&=&
     |\lambda|\left[\left(\frac{\mu-1}{1-|\lambda|}\right)a\right]^2\nonumber
      \end{eqnarray}
      This circle has a (see Figure 34)
      $$\textsf{Center}:~~ \left(\mathcal C,0\right)~~~~~\textsf{and}~~~~~\textsf{Radius},~R=~~
      \sqrt{|\lambda|}\left[\frac{\mu-1}{1-|\lambda|}\right]a,$$
      $$\textsf{where}~~\mathcal C=\left[\frac{\mu-|\lambda|}{1-|\lambda|}\right]a$$
      One must now, from the above information, show that the x-coordinate of N is
      $$ \frac{\mu-\sqrt{|\lambda|}}{1-\sqrt{|\lambda|}}a~~~~~~~\textsf{(see Eq.(\ref{s18'}))} $$}
      \textsf{To that end},
      \begin{eqnarray}
      x\textsf{-coordinate of N}&=&\mathcal C+R\nonumber\\
      &=&\left[\frac{\mu-|\lambda|}{1-|\lambda|}\right]a+\sqrt{|\lambda|}\left[\frac{\mu-1}{1-|\lambda|}\right]a
      \nonumber\\
      &=&\frac{a\left(\mu-\sqrt{|\lambda|}\right)\left(1+\sqrt{|\lambda|}\right)}{1-|\lambda|}\nonumber\\
      &=&\frac{\mu-{|\lambda|}}{1-\sqrt{|\lambda|}}a\nonumber
      \end{eqnarray}
      Also,
      \begin{eqnarray}
      x\textsf{-coordinate of M}&=&\mathcal C-R\nonumber\\
      &=&\left[\frac{\mu-|\lambda|}{1-|\lambda|}\right]a-\sqrt{|\lambda|}\left[\frac{\mu-1}{1-|\lambda|}\right]a
      \nonumber\\
      &=&\frac{a}{1-{|\lambda|}}\left[\mu-|\lambda|-\mu\sqrt{|\lambda|}+\sqrt{|\lambda|}\right]\nonumber
      \end{eqnarray}
      Further,
      \begin{eqnarray}
     \textsf{AM}&=&\left| x\textsf{-coordinate of M}-x\textsf{-coordinate of A}\right|\nonumber\\
      &=&\left|\frac{a}{1-\sqrt{|\lambda|}}\left[\mu-|\lambda|-\mu\sqrt{|\lambda|}+\sqrt{|\lambda|}\right]-a\right|
      \nonumber\\
      &=&\left|\frac{a\left(\mu-1\right)\left(1-\sqrt{|\lambda|}\right)}{1-|\lambda|}\right|\nonumber\\
      &=&\frac{\left|\mu-1\right|}{1+\sqrt{|\lambda|}}a\nonumber
      \end{eqnarray}
      Similarly,
      \begin{eqnarray}
     \textsf{BM}&=&\left| x\textsf{-coordinate of M}-x\textsf{-coordinate of B}\right|\nonumber\\
      &=&\left|\frac{a}{1-{|\lambda|}}\left[\mu-|\lambda|-\mu\sqrt{|\lambda|}+\sqrt{|\lambda|}\right]-\mu a\right|
      \nonumber\\
      &=&\left|\frac{a\left(\mu-1\right)\left(\lambda-\sqrt{|\lambda|}\right)}{1-|\lambda|}\right|\nonumber\\
      &=&\sqrt{|\lambda|}\frac{\left|\mu-1\right|}{1+\sqrt{|\lambda|}}a\nonumber
      \end{eqnarray}
      Therefore,
      \begin{eqnarray}
      \frac{\textsf{AM}}{\textsf{BM}}&=&\frac{a\left|\mu-1\right|}{1+\sqrt{|\lambda|}}\cdot
      \frac{1+\sqrt{|\lambda|}}{a\sqrt{|\lambda|}\left|\mu-1\right|}\nonumber\\
      &=&\fbox{$\displaystyle\frac{1}{\sqrt{|\lambda|}}$}\nonumber
      \end{eqnarray}
      \begin{eqnarray}
     \textsf{AN}&=&\left| x\textsf{-coordinate of N}-x\textsf{-coordinate of A}\right|\nonumber\\
      &=&\left|\frac{\mu-\sqrt{|\lambda|}}{1-\sqrt{|\lambda|}}a-a\right|
      \nonumber\\
      &=&\left|\frac{\mu-1}{1-\sqrt{|\lambda|}}\right|a\nonumber
      \end{eqnarray}
       \begin{eqnarray}
     \textsf{BN}&=&\left| x\textsf{-coordinate of N}-x\textsf{-coordinate of B}\right|\nonumber\\
      &=&\left|\frac{\mu-\sqrt{|\lambda|}}{1-\sqrt{|\lambda|}}a-\mu a\right|
      \nonumber\\
      &=&\sqrt{|\lambda|}\left|\frac{\mu-1}{1-\sqrt{|\lambda|}}\right|a\nonumber
      \end{eqnarray}
      \begin{eqnarray}
      \frac{\textsf{AN}}{\textsf{BN}}&=&\frac{a\left|\mu-1\right|}{\left|1-\sqrt{|\lambda|}\right|}\cdot
      \frac{\left|1-\sqrt{|\lambda|}\right|}{a\sqrt{|\lambda|}\left|\mu-1\right|}\nonumber\\
      &=&\fbox{$\displaystyle\frac{1}{\sqrt{|\lambda|}}$}\label{s60}
      \end{eqnarray}
      \vspace{0.3cm}
      \item \textsf{In order to look for the locus of points at which the lines of force are parallel to $\overline{\textsf{AB}}$, the component of electric field perpendicular to $\overline{\textsf{AB}}$ must vanish. Thus setting $E_y$ and $E_z$ to zero in Eq.(\ref{s3}) and replacing $\lambda$ by $-|\lambda|$, we get
          \begin{eqnarray}
E_y&=&qy\left\{\frac{1}{\left[(x-a)^2+y^2+z^2\right]^{\frac{3}{2}}}-\frac{|\lambda|
}{\left[(x-\mu
a)^2+y^2+z^2\right]^{\frac{3}{2}}}\right\}=0\nonumber\\
E_z&=&qz\left\{\frac{1}{\left[(x-a)^2+y^2+z^2\right]^{\frac{3}{2}}}-\frac{|\lambda|
}{\left[(x-\mu a)^2+y^2+z^2\right]^{\frac{3}{2}}}\right\}=0\nonumber
\end{eqnarray}
Either of the equations gives
\begin{eqnarray}
\frac{\left(x-\mu a\right)^2+y^2+z^2}{\left(x- a\right)^2+y^2+z^2}&=&|\lambda|^{\frac{2}{3}}\nonumber\\
x^2\left(1-|\lambda|^{\frac{2}{3}}\right)
+2a\left(|\lambda|^{\frac{2}{3}}-\mu\right)x+y^2\left(1-|\lambda|^{\frac{2}{3}}\right)
+z^2\left(1-|\lambda|^{\frac{2}{3}}\right)
&=&a^2\left(|\lambda|^{\frac{2}{3}}-\mu^2\right)\nonumber
\end{eqnarray}
Completing the square,
\begin{eqnarray}
x^2-2a\left[\frac{\mu-|\lambda|^{\frac{2}{3}}}{1-|\lambda|^{\frac{2}{3}}}\right]x+
\left[a\left(\frac{\mu-|\lambda|^{\frac{2}{3}}}{1-|\lambda|^{\frac{2}{3}}}\right)\right]^2+y^2+z^2
&=&\frac{a^2\left(|\lambda|^{\frac{2}{3}}-\mu^2\right)}{1-|\lambda|^{\frac{2}{3}}}+
\left[a\left(\frac{\mu-|\lambda|^{\frac{2}{3}}}{1-|\lambda|^{\frac{2}{3}}}\right)\right]^2\nonumber\\
\left[x-a\left(\frac{\mu-|\lambda|^{\frac{2}{3}}}{1-|\lambda|^{\frac{2}{3}}}\right)\right]^2+y^2+z^2
&=&\left[\frac{|\lambda|^{\frac{1}{3}}\left(\mu-1\right)a}{1-|\lambda|^{\frac{2}{3}}}\right]^2\nonumber
\end{eqnarray}
This is the equation of a sphere with radius,
$$\fbox{$\displaystyle\sqrt[3]{|\lambda|}\left|\frac{\mu-1}{1-\sqrt[3]{|\lambda|^2}}\right|a$}$$
and center
$$ \left(a\left[\frac{\mu-|\lambda|^{\frac{2}{3}}}{1-|\lambda|^{\frac{2}{3}}}\right],0,0\right)$$
}
\end{enumerate}
\vspace{0.6cm}
\item
  \begin{enumerate}
  \item
  \textsf{Refer to Figures 35a, 35b and 35c.
  \[
\begin{array} {ccccccc}
     \textsf{AS}&=&\textsf{AR}\cos\theta_1 &~~~~~~~~~~~&
     \textsf{BS}&=&\textsf{BR}\cos\left(\pi-\theta_2\right)\nonumber\\
    \textsf{AR}&=&\frac{\textsf{AS}}{\displaystyle\cos\theta_1} &~~~~~~~~~~~&
     \textsf{BR}&=&-\frac{\textsf{BS}}{\displaystyle\cos\theta_2}
      \nonumber\\
 && &\textsf{For R close to N: S}\rightarrow \textsf{N}& &&\nonumber
 \end{array}
\]
\begin{eqnarray}\label{s61}
 \textsf{AR}\approx\frac{\textsf{AN}}{\cos\theta_1},~~~~~~~~~~~~~
     \textsf{BR}\approx-\frac{\textsf{BN}}{\cos\theta_2}
  \end{eqnarray}
  For the component of electric field parallel to $\overline{\textsf{AB}}$ at a point R on the limiting line of force is
  \begin{eqnarray}
  E_{\displaystyle\parallel{\textsf{AB}}}&=&\frac{q}{\textsf{AR}^2}\cos\theta_1-
  \frac{\left(\pm|\lambda|q\right)}{\textsf{BR}^2}\cos\left(\pi-\theta_2\right)
 \nonumber\\
  &=&q\left[\frac{\cos\theta_1}{\textsf{AR}^2}\pm |\lambda|\frac{\cos\theta_2}{\textsf{BR}^2}\right]
  \label{s62}
  \end{eqnarray}
Similarly, the component of electric field perpendicular to $\overline{\textsf{AB}}$ at a point R on the limiting line of force is
\begin{eqnarray}
  E_{\displaystyle\perp{\textsf{AB}}}&=&\frac{q}{\textsf{AR}^2}\sin\theta_1+
  \frac{\left(\pm|\lambda|q\right)}{\textsf{BR}^2}\sin\left(\pi-\theta_2\right)
 \nonumber\\
  &=&q\left[\frac{\sin\theta_1}{\textsf{AR}^2}\pm |\lambda|\frac{\sin\theta_2}{\textsf{BR}^2}\right]
  \label{s63}
  \end{eqnarray}
Eqs.(\ref{s62}) and (\ref{s63}) implies
\begin{eqnarray}
\frac{E_{\displaystyle\perp{\textsf{AB}}}}{E_{\displaystyle\parallel{\textsf{AB}}}}&=&
\frac{\displaystyle\frac{\sin\theta_1}{\textsf{AR}^2}\pm |\lambda|\frac{\sin\theta_2}{\textsf{BR}^2}}
{\frac{\displaystyle\cos\theta_1}{\textsf{AR}^2}\pm |\lambda|\frac{\displaystyle\cos\theta_2}{\textsf{BR}^2}}\nonumber\\
\nonumber\\
\textsf{Eq.(\ref{s61})}\Rightarrow~~~~~~~~~~~~~~~~~~
&\approx&
\frac{\displaystyle\frac{\sin\theta_1\cos^2\theta_1}{\textsf{AN}^2}\pm |\lambda|\frac{\sin\theta_2\cos^2\theta_2}{\textsf{BN}^2}}
{\frac{\displaystyle\cos^3\theta_1}{\textsf{AN}^2}\pm |\lambda|\frac{\displaystyle\cos^3\theta_2}{\textsf{BN}^2}}\nonumber\\
\nonumber\\
\textsf{Eq.(\ref{s60})}\Rightarrow~~~~~~~~~~~~~~~~~~&=&
\frac{\displaystyle\frac{\sin\theta_1\cos^2\theta_1}{\textsf{AN}^2}\pm |\lambda|\frac{\sin\theta_2\cos^2\theta_2}{\left(\sqrt{|\lambda|}\textsf{AN}\right)^2}}
{\frac{\displaystyle\cos^3\theta_1}{\textsf{AN}^2}\pm |\lambda|\frac{\displaystyle\cos^3\theta_2}{\displaystyle\left(\sqrt{|\lambda|}\textsf{AN}\right)^2}}\nonumber
\end{eqnarray}
Finally,
\begin{eqnarray}
\frac{E_{\displaystyle\perp{\textsf{AB}}}}{E_{\displaystyle\parallel{\textsf{AB}}}}
\approx\frac{\sin\theta_1\cos^2\theta_1\pm\sin\theta_2\cos^2\theta_2}{\cos^3\theta_1\pm\cos^3\theta_2}
~~~~~\textsf{for}\left\{\matrix{\lambda>0\cr
\lambda<0}\right.\label{s64}
\end{eqnarray}
We need to evaluate Eq.(\ref{s64}) when R os close to N, i.e., S$\rightarrow$N, so that for
\[
\begin{array} {ccccc}
     &\lambda>0:\left\{\matrix{\theta_1\rightarrow 0\cr
     \theta_2\rightarrow \pi}\right.,&~~~~ -1<\lambda<0:\left\{\matrix{\theta_1\rightarrow 0\cr
     \theta_2\rightarrow 0}\right.,&~~~~ \lambda<-1:\left\{\matrix{\theta_1\rightarrow \pi\cr
     \theta_2\rightarrow \pi}\right.&
 \end{array}
\]
  Taylor series expansion of a function, $f(\theta)$ about the point $\theta=\theta_0$:
  $$f(\theta)=f(\theta_0)+(\theta-\theta_0)\left.f'(\theta)\right|_{\theta=\theta_0}
  +\frac{(\theta-\theta_0)^2}{2!}\left.f''(\theta)\right|_{\theta=\theta_0}+...$$
\[
\begin{array}{cccc}
  &f(\theta)=\cos\theta,&f'(\theta)=-\sin\theta,&f''(\theta)=-\cos\theta\nonumber\\
  &\left.f(\theta)\right|_{\theta=0}=1,&\left.f'(\theta)\right|_{\theta=0}=0,&  \left.f'(\theta)\right|_{\theta=0}=-1\nonumber\\
  &\left.f(\theta)\right|_{\theta=\pi}=-1,&\left.f'(\theta)\right|_{\theta=\pi}=0,&  \left.f'(\theta)\right|_{\theta=\pi}=1\nonumber\\
\end{array}
\]
\[
\begin{array}{cccc}
  &f(\theta)=\sin\theta,&f'(\theta)=\cos\theta,&f''(\theta)=-\sin\theta\nonumber\\
  &\left.f(\theta)\right|_{\theta=0}=0,&\left.f'(\theta)\right|_{\theta=0}=1,&  \left.f'(\theta)\right|_{\theta=0}=-1\nonumber\\
  &\left.f(\theta)\right|_{\theta=\pi}=0,&\left.f'(\theta)\right|_{\theta=\pi}=-1,&  \left.f'(\theta)\right|_{\theta=\pi}=0\nonumber\\
\end{array}
\]\\
\[
\begin{array}{cccccc}
  \Rightarrow\cos\theta\approx\left\{\matrix{\displaystyle 1-\frac{\theta^2}{2}&\textsf{about}~\theta=0\cr
  \displaystyle-1+\frac{\left(\theta-\pi\right)^2}{2}&\textsf{about}~\theta=\pi}\right.
  ~~&\textsf{and}&~~\sin\theta\approx\left\{\matrix{\displaystyle\theta &\textsf{about}~\theta=0\cr
  -\left(\displaystyle\theta-\pi\right)&\textsf{about}~\theta=\pi}\right.\nonumber\\
  \end{array}
\]
\begin{eqnarray}\label{s65}
\Rightarrow\cos^3\theta\approx\left\{\matrix{\displaystyle \left(1-\frac{\theta^2}{2}\right)^3&\approx&\displaystyle1-\frac{3\theta^2}{2}&~~~&\textsf{about}~\theta=0\cr
  \displaystyle\left[-1+\frac{\left(\theta-\pi\right)^2}{2}\right]^3&\approx&\displaystyle-1+\frac{3\left(\theta-\pi\right)^2
  }{2}&~~~&\textsf{about}~\theta=\pi}\right.
  \end{eqnarray}
  \begin{eqnarray}\label{s66}
\sin\theta\cos^2\theta\approx\left\{\matrix{\displaystyle \theta\left(1-\frac{\theta^2}{2}\right)^2&\approx&\displaystyle\theta&~~~&\textsf{about}~\theta=0\cr
  \displaystyle-\left(\theta-\pi\right)\left[-1+\frac{\left(\theta-\pi\right)^2}{2}\right]^2&\approx&
  \displaystyle-\left(\theta-\pi\right)&~~~&\textsf{about}~\theta=\pi}\right.
  \end{eqnarray}\\
Using Eqs.(\ref{s65}) and (\ref{s66}), we evaluate Eq.(\ref{s64}) for $\lambda>0$, $-1<\lambda<0$ and
  $\lambda<-1$:\\
  \begin{itemize}
  \item $\lambda>0$ (about $\theta_1=0$ and $\theta_2=\pi$)
\begin{eqnarray}
\frac{E_{\displaystyle\perp{\textsf{AB}}}}{E_{\displaystyle\parallel{\textsf{AB}}}}
&\approx& \frac{\displaystyle\theta_1+\left[-\left(\theta_2-\pi\right)\right]}
{\displaystyle\left[1-\frac{3\theta_1^2}{2}\right]+
\left[-1+\frac{3\left(\theta_2-\pi\right)^2}{2}\right]}\nonumber\\
&=&\frac{\displaystyle\theta_1-\left(\theta_2-\pi\right)}
{\displaystyle\frac{3}{2}\left[\theta_1^2-\left(\theta_2-\pi\right)^2\right]}\nonumber\\
&=&\frac{\displaystyle\theta_1-\left(\theta_2-\pi\right)}
{\displaystyle\frac{3}{2}\left[\theta_1-\left(\theta_2-\pi\right)\right]
\left[\theta_1+\left(\theta_2-\pi\right)\right]}\nonumber\\
&=&\fbox{$\displaystyle-\frac{2}{3}\left(\frac{1}{\pi-\theta_1-\theta_2}\right)$}\label{s67}
\end{eqnarray}\\
\item $-1<\lambda<0$ (about $\theta_1=0$ and $\theta_2=0$)
\begin{eqnarray}
\frac{E_{\displaystyle\perp{\textsf{AB}}}}{E_{\displaystyle\parallel{\textsf{AB}}}}
&\approx& \frac{\displaystyle\theta_1-\theta_2}
{\displaystyle\left[1-\frac{3\theta_1^2}{2}\right]-
\left[1-\frac{3\theta_2^2}{2}\right]}\nonumber\\
&=&\frac{\displaystyle-\left(\theta_2-\theta_1\right)}
{\displaystyle\frac{3}{2}\left[\theta_2^2-\theta_1^2\right]}\nonumber\\
&=&\frac{\displaystyle-\left(\theta_2-\theta_1\right)}
{\displaystyle\frac{3}{2}\left(\theta_2-\theta_1\right)\left(\theta_2+\theta_1\right)}
\nonumber\\
&=&\fbox{$\displaystyle-\frac{2}{3}\left(\frac{1}{\theta_1+\theta_2}\right)$}\label{s68}
\end{eqnarray}\\
\item $\lambda<-1$ (about $\theta_1=\pi$ and $\theta_2=\pi$)
\begin{eqnarray}
\frac{E_{\displaystyle\perp{\textsf{AB}}}}{E_{\displaystyle\parallel{\textsf{AB}}}}
&\approx& \frac{\displaystyle\left[-\left(\theta_1-\pi\right)\right]-\left[-\left(\theta_2-\pi\right)\right]}
{\displaystyle\left[-1+\frac{3\left(\theta_1-\pi\right)^2}{2}\right]-
\left[-1+\frac{3\left(\theta_2-\pi\right)^2}{2}\right]}\nonumber\\
&=&\frac{\displaystyle\left(\theta_2-\theta_1\right)}
{\displaystyle\frac{3}{2}\left[\left(\theta_1-\pi\right)^2-\left(\theta_2-\pi\right)^2\right]}\nonumber\\
&=&\frac{\displaystyle-\left(\theta_1-\theta_2\right)}
{\displaystyle\frac{3}{2}\left(\theta_1-\theta_2\right)
\left(\theta_1+\theta_2-2\pi\right)}\nonumber\\
&=&\fbox{$\displaystyle-\frac{2}{3}\left(\frac{1}{\theta_1+\theta_2-2\pi}\right)$}\label{s69}
\end{eqnarray}
  \end{itemize}}
  \vspace{0.3cm}
  \item \textsf{Setting $\theta_1=0$, $\theta_2=\pi$ in Eq.(\ref{s67}), $\theta_1=0$, $\theta_2=0$ in Eq.(\ref{s68}) and $\theta_1=\pi$, $\theta_2=\pi$ in Eq.(\ref{s69}) gives
  $$\frac{E_{\displaystyle\perp{\textsf{AB}}}}{E_{\displaystyle\parallel{\textsf{AB}}}}\rightarrow\infty
  \Rightarrow\tan^{-1}\left( \frac{E_{\displaystyle\perp{\textsf{AB}}}}{E_{\displaystyle\parallel{\textsf{AB}}}}\right)
     =\tan^{-1}(\infty)=\fbox{$\displaystyle\frac{\pi}{2}$}$$
     Therefore, the limiting line of force through N intersects  $\overline{\textsf{AB}}$ at right angles.}
  \end{enumerate}
   \vspace{0.6cm}
  \item
  \begin{enumerate}
  \item\textsf{Here refer to Figures 36a, 36b and 36c. Recall that the coordinates of A, B and N are
  $$\textsf{A}:(a,0,0),~~~\textsf{B}:(\mu a,0,0),~~~\textsf{N}:\left(\left[\frac{\mu\pm\sqrt{|\lambda|}}{1\pm\sqrt{|\lambda|}}\right]a,0,0\right)
  \equiv\left(x_{_\textsf{N}},0,0\right)~~\textsf{for}~~\left\{\matrix{\lambda>0\cr
  \lambda<0}\right.$$
  Let \textsf{S}(x,y,z) be any point on an equipotential surface. Then
  \begin{eqnarray}
  r^2&=&\left(x-x_{_\textsf{N}}\right)^2+y^2+z^2\label{s69'}\\
  r\cos\theta&=&x-x_{_\textsf{N}}\label{s69''}
\end{eqnarray}
The potentials,$\Phi$ at the points S and N are given by
\begin{eqnarray}
\Phi_{_\textsf{S}}=\frac{q}{\textsf{AS}}+\frac{\pm(|\lambda|q)}{\textsf{BS}},&~~~&
\Phi_{_\textsf{N}}=\frac{q}{\textsf{AN}}+\frac{\pm(|\lambda|q)}{\textsf{BN}}\nonumber\\
\Rightarrow \Phi_{_\textsf{S}}-\Phi_{_\textsf{N}}&=&\frac{q}{\textsf{AS}}\pm\frac{|\lambda|q}{\textsf{BS}}
-\left(\frac{q}{\textsf{AN}}\pm\frac{|\lambda|q}{\textsf{BN}}\right)\nonumber\\
\Rightarrow\frac{ \Phi_{_\textsf{S}}-\Phi_{_\textsf{N}}}{q}&=&\left(\frac{1}{\textsf{AS}}-\frac{1}{\textsf{AN}}\right)
\pm|\lambda|\left(\frac{1}{\textsf{BS}}-\frac{1}{\textsf{BN}}\right)\label{s70}
\end{eqnarray}
  We next compute AS and BS. To that end,
  \[
  \begin{array}{ccccc}
  \textsf{AS}^2&=&\textsf{AN}^2+r^2-2r\textsf{AN}\left[\matrix{\cos(\pi-\theta)\cr
  \cos\theta}\right]&\textsf{for}~&\left\{\matrix{\lambda>0,&-1<\lambda<0\cr
  \lambda<-1&}\right.\nonumber\\
  &=&\displaystyle\textsf{AN}^2\left[1+\left(\frac{r}{\textsf{AN}}\right)^2\pm 2\left(\frac{r}{\textsf{AN}}\right)
  \cos\theta\right]&\textsf{for}~&\left\{\matrix{\lambda>0,&~-1<\lambda<0\cr
  \lambda<-1&}\right.\nonumber\\
  \Rightarrow\displaystyle\frac{1}{\textsf{AS}}&=&\displaystyle\frac{1}{\textsf{AN}}\left[1+
  \left(\frac{r}{\textsf{AN}}\right)^2\pm 2\left(\frac{r}{\textsf{AN}}\right)
  \cos\theta\right]^{-\frac{1}{2}}
  &\textsf{for}&\left\{\matrix{\lambda>0,&~-1<\lambda<0\cr
  \lambda<-1&}\right.
  \end{array}
  \]
  We now expand the expression for $\displaystyle\frac{1}{\textsf{AS}}$ in the limit $r\ll\textsf{AN}$ (i.e. S close to N) using $\left(1+\epsilon\right)^n=\displaystyle 1+ n\epsilon+\frac{n(n-1)}{2}\epsilon^2+...$~. Setting, $\epsilon=\displaystyle
  \pm 2\left(\frac{r}{\textsf{AN}}\right)\cos\theta+\left(\frac{r}{\textsf{AN}}\right)^2
  $, we get
  \[
  \begin{array}{ccccc}
  \displaystyle\frac{1}{\textsf{AS}}&=&\displaystyle\frac{1}{\textsf{AN}}\left\{1+
  \left(-\frac{1}{2}\right)\left[\pm2\frac{r}{\textsf{AN}}\cos\theta+\left(\frac{r}{\textsf{AN}}\right)^2\right]
  \right.\nonumber\\
  &&\left.\displaystyle+\frac{\left(-\frac{1}{2}\right)\left(-\frac{3}{2}\right)}{2}
  \left[\pm2\frac{r}{\textsf{AN}}\cos\theta+\left(\frac{r}{\textsf{AN}}\right)^2\right]^2+....\right\}
  &\textsf{for}&\left\{\matrix{\lambda>0,&~-1<\lambda<0\cr
  \lambda<-1&}\right.\nonumber\\
  &=&\displaystyle\frac{1}{\textsf{AN}}\left\{1\mp
\frac{r}{\textsf{AN}}\cos\theta-\frac{1}{2}\left(\frac{r}{\textsf{AN}}\right)^2\right.\nonumber\\
&&\displaystyle\left.+\frac{3}{8}
  \left[4\left(\frac{r}{\textsf{AN}}\right)^2\cos^2\theta
  +\left(\frac{r}{\textsf{AN}}\right)^4\pm4\left(\frac{r}{\textsf{AN}}\right)^3\cos\theta\right]+....\right\}
  &\textsf{for}&\left\{\matrix{\lambda>0,&~-1<\lambda<0\cr
  \lambda<-1&}\right.\nonumber\\
  \end{array}
  \]
  Neglecting terms $\left(r/\textsf{AN}\right)^3$ and higher powers of $\left(r/\textsf{AN}\right)$ , we get
  \begin{eqnarray}\label{s71}
  \displaystyle\frac{1}{\textsf{AS}}\approx\displaystyle\frac{1}{\textsf{AN}}\left[
  1\mp
\frac{r}{\textsf{AN}}\cos\theta+\frac{1}{2}\left(\frac{r}{\textsf{AN}}\right)^2\left(3\cos^2\theta-1\right)
  \right]~\textsf{for}~\left\{\matrix{\lambda>0,~-1<\lambda<0\cr
  \lambda<-1}\right.
  \end{eqnarray}\\
  Similarly,
\[
  \begin{array}{ccccc}
  \textsf{BS}^2&=&\textsf{BN}^2+r^2-2r\textsf{AN}\left[\matrix{\cos\theta\cr\cos(\pi-\theta)}
  \right]&\textsf{for}~&\left\{\matrix{\lambda>0,&~\lambda<-1\cr
  -1<\lambda<0&}\right.\nonumber\\
  &=&\displaystyle\textsf{BN}^2\left[1+\left(\frac{r}{\textsf{AN}}\right)^2\mp 2\left(\frac{r}{\textsf{BN}}\right)
  \cos\theta\right]&\textsf{for}~&
  \left\{\matrix{\lambda>0,&~\lambda<-1\cr
  ~-1<\lambda<0&}\right.\nonumber\\
  \end{array}
  \]
  This expression for $\textsf{BS}^2$ is the same as $\textsf{AS}^2$ given on the previous page, provided we make the following transformation
  $$\textsf{AN}\rightarrow\textsf{BN},~~~~\textsf{AS}\rightarrow\textsf{BS},~~~~
  \mp\cos\theta\rightarrow\pm\cos\theta.$$
 Hence this transformation in Eq.(\ref{s71}) gives
\begin{eqnarray}\label{s72}
  \displaystyle\frac{1}{\textsf{BS}}\approx\displaystyle\frac{1}{\textsf{BN}}\left[
  1\pm
\frac{r}{\textsf{BN}}\cos\theta+\frac{1}{2}\left(\frac{r}{\textsf{BN}}\right)^2\left(3\cos^2\theta-1\right)
  \right]~\textsf{for}~\left\{\matrix{\lambda>0,~\lambda<-1\cr
  -1<\lambda<0}\right.
  \end{eqnarray}\\
  We now evaluate Eq.(\ref{s70}) using Eqs.(\ref{s71}) and (\ref{s72}) for each of cases where $\lambda>0$, $-1<\lambda<0$ and $\lambda<-1$:\\
\[
  \begin{array}{ccccc}
  \displaystyle\frac{\Phi_{_\textsf{S}}-\Phi_{_\textsf{N}}}{q}&\approx&\left\{\matrix{
  \displaystyle\frac{1}{\textsf{AN}}\left[
  1-
\frac{r}{\textsf{AN}}\cos\theta+\frac{1}{2}\left(\frac{r}{\textsf{AN}}\right)^2\left(3\cos^2\theta-1\right)
  \right]-\frac{1}{\textsf{AN}}&\matrix{\cr\cr\left(\lambda>0\right)}&\cr
  \displaystyle+\frac{|\lambda|}{\textsf{BN}}\left[1+
\frac{r}{\textsf{BN}}\cos\theta+\frac{1}{2}\left(\frac{r}{\textsf{BN}}\right)^2\left(3\cos^2\theta-1\right)
  \right]-\frac{|\lambda|}{\textsf{BN}}\cr
  \\ \\
  \displaystyle\frac{1}{\textsf{AN}}\left[
  1-
\frac{r}{\textsf{AN}}\cos\theta+\frac{1}{2}\left(\frac{r}{\textsf{AN}}\right)^2\left(3\cos^2\theta-1\right)
  \right]-\frac{1}{\textsf{AN}}&\matrix{\cr\cr\left(-1<\lambda<0\right)}&\cr
  \displaystyle-\frac{|\lambda|}{\textsf{BN}}\left[1-
\frac{r}{\textsf{BN}}\cos\theta+\frac{1}{2}\left(\frac{r}{\textsf{BN}}\right)^2\left(3\cos^2\theta-1\right)
  \right]+\frac{|\lambda|}{\textsf{BN}}
  \cr
  \\ \\
  \displaystyle\frac{1}{\textsf{AN}}\left[
  1+
\frac{r}{\textsf{AN}}\cos\theta+\frac{1}{2}\left(\frac{r}{\textsf{AN}}\right)^2\left(3\cos^2\theta-1\right)
  \right]-\frac{1}{\textsf{AN}}&\matrix{\cr\cr\left(\lambda<-1\right)}&\cr
  \displaystyle-\frac{|\lambda|}{\textsf{BN}}\left[1+
\frac{r}{\textsf{BN}}\cos\theta+\frac{1}{2}\left(\frac{r}{\textsf{BN}}\right)^2\left(3\cos^2\theta-1\right)
  \right]+\frac{|\lambda|}{\textsf{BN}}
  }\right.\\
  \end{array}
  \]
  \[
  \begin{array}{ccccc}
 \displaystyle\frac{\Phi_{_\textsf{S}}-\Phi_{_\textsf{N}}}{q}&\approx& &\approx&\left\{\matrix{
  \displaystyle\frac{1}{\textsf{AN}}\left[
  -
\frac{r}{\textsf{AN}}\cos\theta+\frac{1}{2}\left(\frac{r}{\textsf{AN}}\right)^2\left(3\cos^2\theta-1\right)
  \right]&\matrix{\cr\cr\left(\lambda>0\right)}&\cr
  \displaystyle+\frac{|\lambda|}{\textsf{BN}}\left[
\frac{r}{\textsf{BN}}\cos\theta+\frac{1}{2}\left(\frac{r}{\textsf{BN}}\right)^2\left(3\cos^2\theta-1\right)
  \right]\cr
  \\ \\
  \displaystyle\frac{1}{\textsf{AN}}\left[
  -
\frac{r}{\textsf{AN}}\cos\theta+\frac{1}{2}\left(\frac{r}{\textsf{AN}}\right)^2\left(3\cos^2\theta-1\right)
  \right]&\matrix{\cr\cr\left(-1<\lambda<0\right)}&\cr
  \displaystyle-\frac{|\lambda|}{\textsf{BN}}\left[-
\frac{r}{\textsf{BN}}\cos\theta+\frac{1}{2}\left(\frac{r}{\textsf{BN}}\right)^2\left(3\cos^2\theta-1\right)
  \right]
  \cr
  \\ \\
  \displaystyle\frac{1}{\textsf{AN}}\left[
  \frac{r}{\textsf{AN}}\cos\theta+\frac{1}{2}\left(\frac{r}{\textsf{AN}}\right)^2\left(3\cos^2\theta-1\right)
  \right]&\matrix{\cr\cr\left(\lambda<-1\right)}&\cr
  \displaystyle-\frac{|\lambda|}{\textsf{BN}}\left[
\frac{r}{\textsf{BN}}\cos\theta+\frac{1}{2}\left(\frac{r}{\textsf{BN}}\right)^2\left(3\cos^2\theta-1\right)
  \right]
  }\right.
  \end{array}
  \]\\
  \\
On using Eq.(\ref{s60}): $\textsf{BN}=\textsf{AN}\sqrt{|\lambda|}$ above, we get\\
\\
 \[
  \begin{array}{ccccc}
  \displaystyle\left(\frac{\Phi_{_\textsf{S}}-\Phi_{_\textsf{N}}}{q}\right)\textsf{AN}&\approx&\left\{\matrix{
  \displaystyle
  -
\frac{r}{\textsf{AN}}\cos\theta+\frac{1}{2}\left(\frac{r}{\textsf{AN}}\right)^2\left(3\cos^2\theta-1\right)
  &\matrix{\cr\cr\left(\lambda>0\right)}&\cr
  \displaystyle+
\frac{r}{\textsf{AN}}\cos\theta+\frac{1}{2}\frac{1}{\sqrt{|\lambda|}}
\left(\frac{r}{\textsf{AN}}\right)^2\left(3\cos^2\theta-1\right)
  \cr
  \\ \\
  \displaystyle
  -
\frac{r}{\textsf{AN}}\cos\theta+\frac{1}{2}\left(\frac{r}{\textsf{AN}}\right)^2\left(3\cos^2\theta-1\right)
  &\matrix{\cr\cr\left(-1<\lambda<0\right)}&\cr
  \displaystyle+
\frac{r}{\textsf{AN}}\cos\theta-\frac{1}{2}\frac{1}{\sqrt{|\lambda|}}
\left(\frac{r}{\textsf{AN}}\right)^2\left(3\cos^2\theta-1\right)
  \cr
  \\ \\
  \displaystyle
\frac{r}{\textsf{AN}}\cos\theta+\frac{1}{2}\left(\frac{r}{\textsf{AN}}\right)^2\left(3\cos^2\theta-1\right)
  &\matrix{\cr\cr\left(\lambda<-1\right)}&\cr
  \displaystyle -
\frac{r}{\textsf{AN}}\cos\theta-\frac{1}{2}\frac{1}{\sqrt{|\lambda|}}
\left(\frac{r}{\textsf{AN}}\right)^2\left(3\cos^2\theta-1\right)
  }\right.
   \end{array}
  \]
  \[
  \begin{array}{ccccc}
 \displaystyle\left(\frac{\Phi_{_\textsf{S}}-\Phi_{_\textsf{N}}}{q}\right)\textsf{AN} &\approx&\left\{\matrix{ \displaystyle
  \frac{1}{2}
\left(\frac{r}{\textsf{AN}}\right)^2\left(3\cos^2\theta-1\right)\left[1+\frac{1}{\sqrt{|\lambda|}}\right]
&\matrix{\left(\lambda>0\right)}&\cr
  \cr
  \\
  \displaystyle
  \frac{1}{2}
\left(\frac{r}{\textsf{AN}}\right)^2\left(3\cos^2\theta-1\right)\left[1-\frac{1}{\sqrt{|\lambda|}}\right]
&\matrix{\left(\lambda<0\right)}&\cr
  }\right.
  \end{array}
  \]
 One can now write this last expression as
 $$ 3r^2\cos^2\theta-r^2=\frac{2~\textsf{AN}^3}{q}\left(\frac{\sqrt{|\lambda|}}{\sqrt{|\lambda|\pm 1}}\right)
 \left(\Phi_{_\textsf{S}}-\Phi_{_\textsf{N}}\right)~~~~~\left\{\matrix{\lambda>0\cr
 \lambda<0}\right.$$\\
 The expression for AN can be calculated from Eq.(\ref{s18'}):
\begin{eqnarray}
     \textsf{AN}&=&\left| x\textsf{-coordinate of N} \left(x_{_\textsf{N}}\right)-x\textsf{-coordinate of A}\right|\nonumber\\
      &=&\left|\frac{\mu\pm\sqrt{|\lambda|}}{1\pm\sqrt{|\lambda|}}a-a\right|
      \nonumber\\
      &=&\left|\frac{\mu-1}{1\pm\sqrt{|\lambda|}}\right|a~~~~~\left\{\matrix{\lambda>0\cr
 \lambda<0}\right.\label{s73}
      \end{eqnarray}\\
On using Eqs.(\ref{s69'}), (\ref{s69''}) and (\ref{s73}),
$$3\left(x-x_{_\textsf{N}}\right)^2-\left[\left(x-x_{_\textsf{N}}\right)^2+y^2+z^2\right]
=\frac{2a^3}{q}\left|\frac{\mu-1}{1\pm\sqrt{|\lambda|}}\right|^3\left(\frac{\sqrt{|\lambda|}}{\sqrt{|\lambda|\pm 1}}\right)
 \left(\Phi_{_\textsf{S}}-\Phi_{_\textsf{N}}\right)~~~~~\left\{\matrix{\lambda>0\cr
 \lambda<0}\right.
$$
or
\begin{eqnarray}
2\left(x-x_{_\textsf{N}}\right)^2-y^2-z^2
&=&\kappa\label{s74}\\
\textsf{where}~~~~\kappa&\equiv&\frac{2a^3}{q}\left|\frac{\mu-1}{1\pm\sqrt{|\lambda|}}\right|^3\left(\frac{\sqrt{|\lambda|}}{\sqrt{|\lambda|\pm 1}}\right)
 \left(\Phi_{_\textsf{S}}-\Phi_{_\textsf{N}}\right)~~~~~\left\{\matrix{\lambda>0\cr
 \lambda<0}\right.\nonumber\\
 \textsf{and}~~~~x_{_\textsf{N}}&\equiv&\left[\frac{\mu\pm\sqrt{|\lambda|}}{1\pm\sqrt{|\lambda|}}\right]a
 ~~~~~\left\{\matrix{\lambda>0\cr
 \lambda<0}\right.\nonumber
\end{eqnarray}\\
\begin{itemize}
\item CASE 1. $\kappa=0$\\
\\
Eq.(\ref{s74}) gives
$$2\left(x-x_{_\textsf{N}}\right)^2-y^2-z^2=0$$
\begin{eqnarray}\label{s75}
\fbox{$\displaystyle2\left(x-x_{_\textsf{N}}\right)^2=y^2+z^2$}
\end{eqnarray}
This is an equation for a \emph{right circular cone} with it vertex at $x=x_{_\textsf{N}}$.\\
\item CASE 2. $\kappa>0$\\
\\
Replacing $\kappa$ by $|\kappa|$ in Eq.(\ref{s74}) we get
$$2\left(x-x_{_\textsf{N}}\right)^2-y^2-z^2=|\kappa|$$
\begin{eqnarray}
\fbox{$\displaystyle\frac{\left(x-x_{_\textsf{N}}\right)^2}{\displaystyle\frac{|\kappa|}{2}}-\frac{y^2}{|\kappa|}
-\frac{z^2}{|\kappa|}=1$}
\end{eqnarray}
This is an equation for a \emph{hyperboloid of revolution of \underline{two} sheets} about the line of charges, $\overline{\textsf{AB}}$.\\
\item CASE 3. $\kappa<0$\\
\\
Replacing $\kappa$ by $-|\kappa|$ in Eq.(\ref{s74}) we get
$$2\left(x-x_{_\textsf{N}}\right)^2-y^2-z^2=-|\kappa|$$
\begin{eqnarray}
\fbox{$\displaystyle\frac{y^2}{|\kappa|}
+\frac{z^2}{|\kappa|}-\frac{\left(x-x_{_\textsf{N}}\right)^2}{\displaystyle\frac{|\kappa|}{2}}=1$}
\end{eqnarray}
This is an equation for a \emph{hyperboloid of revolution of \underline{one} sheet} about the line of charges, $\overline{\textsf{AB}}$.
\end{itemize}
All the three cases are demonstrated in Figure 37.
}
\vspace{0.3cm}
          \item \textsf{In the $xy$-plane ($z=0$), Eq.(\ref{s75}) implies
          $$2\left(x-x_{_\textsf{N}}\right)^2-y^2=0\Rightarrow y=\pm\sqrt{2}x\mp\sqrt{2}x_{_\textsf{N}}$$
          Identifying this last equation with that of straight line:$y=\left(\pm\tan\alpha\right) x+b$ gives
          $$\tan\alpha=\sqrt{2}\Rightarrow \fbox{$\displaystyle\alpha=\tan^{-1}(\sqrt2)$}$$ }
         \textsf{See Figure 37. }
         \item \textsf{In the $xy$-plane ($z=0$), Eq.(\ref{s74}) implies
         $$2\left(x-x_{_\textsf{N}}\right)^2-y^2=|\kappa|$$
         Taking the derivative of this last equation w.r.t. $x$,
          $$4\left(x-x_{_\textsf{N}}\right)-2y\frac{dy}{dx}=0\Rightarrow \frac{dy}{dx}=\frac{2\left(x-x_{_\textsf{N}}\right)}{y}$$
          This is the slope of a tangent line to an equipotential curve \emph{near} N at the point ($x$,$y$). Therefore, the slope of a tangent line to a line of force at the same point
          will be
          $$\frac{dy}{dx}=\frac{-1}{\frac{\displaystyle2\left(x-x_{_\textsf{N}}\right)}{\displaystyle y}}$$
          This is because line of force is perpendicular to its corresponding equipotential curve.
          Thus,
          \begin{eqnarray}
          \int\frac{dy}{y}&=&-\frac{1}{2}\int\frac{dx}{x-x_{_\textsf{N}}}\nonumber\\
          \Rightarrow \ln y&=&-\frac{1}{2}\ln\left(x-x_{_\textsf{N}}\right)+\ln(\textsf{Constant})\nonumber\\
          y&=&\left(x-x_{_\textsf{N}}\right)^{-\frac{1}{2}}\cdot \textsf{Constant}\nonumber
          \end{eqnarray}
          Using the expression for $x_{_\textsf{N}}$, we obtain
          $$\fbox{$\displaystyle y^2\left[x-\left(\frac{\mu\pm\sqrt{|\lambda|}}{1\pm \sqrt{|\lambda|}}\right)a\right]=\textsf{Constant}$}$$}
          \end{enumerate}
  \end{enumerate}
  \vspace{0.5cm}
\begin{center}
\textbf{PROBLEM 2}
\end{center}
\vspace{0.5cm}
  \begin{enumerate}
\item
  \begin{enumerate}
  \item
  \begin{enumerate}
  \item  \textsf{Refer to Figure 38. The equations of motion for the charges ($m_1=m$, $q_1=q$) and  ($m_2=\sigma m$, $q_2=\lambda q$) are
  $$m_2\frac{d^2x_2}{dt^2}=\frac{q_1q_2}{\left(x_2-x_1\right)^2}$$
   $$m_1\frac{d^2x_1}{dt^2}=-\frac{q_1q_2}{\left(x_2-x_1\right)^2}$$
   Multiplying the first equation by $m_1$ and second by $m_2$ and subtracting we get
   $$m_1m_2\frac{d^2}{dt^2}\left(x_2-x_1\right)=\left(m_1+m_2\right)\frac{q_1q_2}{\left(x_2-x_1\right)^2}$$
    $$\textsf{Defining}~~x\equiv x_2-x_1 ~~\textsf{and noting that}~~\frac{m_1+m_2}{m_1m_2}=\frac{1+\sigma}{m\sigma}, ~~q_1q_2=\lambda q^2,$$
    the last equation can be written as
    \begin{eqnarray}\label{s76}
    \frac{d^2x}{dt^2}=\frac{\omega}{x^2},~~~~\omega\equiv\frac{\lambda q^2}{m}\frac{\lambda(\sigma+1)}{\sigma}
    \end{eqnarray}
    Now,
    $$\frac{d^2x}{dt^2}=\frac{d}{dt}\left(\frac{dx}{dt}\right)=\frac{dv}{dt}=\frac{dx}{dt}\frac{dv}{dx}
    =v\frac{dv}{dx}=\frac{d}{dx}\left(\frac{1}{2}v^2\right)$$
    With this expression for $\displaystyle\frac{d^2x}{dt^2}$ in Eq.(\ref{s76}), we have
    $$\int d\left(\frac{1}{2}v^2\right)=\omega\int \frac{dx}{x^2}$$
    $$\Rightarrow \frac{1}{2}v^2=-\frac{\omega}{x}+C_1$$
    The constant of integration $C_1$ is determined from the given initial conditions: $v=0$ at $x=x_0\equiv x_{02}-x_{01}=a(\mu-1)$. This gives $\displaystyle C_1={\omega}/{x_0}$. Therefore,
    $$\frac{1}{2}v^2=\frac{1}{2}\left(\frac{dx}{dt}\right)^2=\left(\frac{1}{x_0}-\frac{1}{x}\right)$$
    $$\Rightarrow\left(\frac{dx}{dt}\right)^2=2\omega\left(\frac{x-x_0}{xx_0}\right)$$
    \begin{itemize}
    \vspace{0.5cm}
    \item \underline{CASE 1: $\lambda<0~(\lambda=-|\lambda|)$}
    $$\omega=-\frac{|\lambda| q^2}{m}\frac{\lambda(\sigma+1)}{\sigma}=-|\omega|$$
    $$\frac{dx}{dt}=\pm \sqrt{2|\omega|\left(\frac{x_0-x}{xx_0}\right)}$$
    Choosing the negative sign, we get
    $$\displaystyle\int \frac{\displaystyle\sqrt{x}}{\displaystyle\sqrt{x_0-x}}~dx=-\displaystyle\sqrt{\frac{2|\omega|}{x_0}}\int dt$$
    The result for the left hand side integral is proven in the appendix (see Eq.()). Performing the integration in the last equation above, we obtain,
    $$x_0~\sin^{-1}\left(\sqrt{\frac{x}{x_0}}\right)+\sqrt{x\left(x_0-x
        \right)}=-\displaystyle\sqrt{\frac{2|\omega|}{x_0}}t+C_2.$$
    The constant of integration $C_2$ can easily be found from the information that $t=0$, $x=x_0$. This gives $C_2=x_0\pi/2$. Thus
    $$\displaystyle\sqrt{\frac{2|\omega|}{x_0}}t=\frac{x_0\pi}{2}-x_0~\sin^{-1}\left(\sqrt{\frac{x}{x_0}}\right)
    -\sqrt{x\left(x_0-x
        \right)}$$
        Inserting expressions for $x_0=a(\mu-1)$ and $|\omega|=\displaystyle\frac{\displaystyle|\lambda| q^2}{m}\frac{\displaystyle\lambda(\sigma+1)}{\sigma}$, we obtain
        \begin{eqnarray}\label{s77}
        \fbox{$\displaystyle t=\sqrt{\frac{ma}{2q^2}\frac{\sigma(\mu-1)}{|\lambda|(\sigma+1)}}\left\{
        \displaystyle\frac{a}{2}(\mu-1)\left[\pi-\displaystyle
        \sin^{-1}\left(\sqrt{\frac{2x}{a(\mu-1)}}~\right)\right]
        -\sqrt{x\left[a(\mu-1)-x\right]}
        \right\}$}\nonumber\\
        \end{eqnarray}
    \item \underline{CASE 2: $\lambda>0~(\lambda=|\lambda|)$}
    $$\omega=\frac{|\lambda| q^2}{m}\frac{\lambda(\sigma+1)}{\sigma}=|\omega|$$
    $$\frac{dx}{dt}=\pm \sqrt{2|\omega|\left(\frac{x-x_0}{xx_0}\right)}$$
    Choosing the positive sign, we get
    $$\displaystyle\int \frac{\displaystyle\sqrt{x}}{\displaystyle\sqrt{x-x_0}}~dx=\displaystyle\sqrt{\frac{2|\omega|}{x_0}}\int dt$$
    The result for the left hand side integral is proven in the appendix (see Eq.()). Performing the integration in the last equation above, we obtain,
    $$\frac{1}{2}x_0\cosh^{-1}\left[\frac{2x}{x_0}-1\right]+\sqrt{x\left[x-x_0\right]}
    =\sqrt{\frac{2|\omega|}{x_0}}t+C_3.$$
    Again the constant of integration, $C_3$ is obtained using the initial condition: at $t=0$, $x=x_0$. This gives $C_3=0$. As before, inserting the expression for $x_0$ and $|\omega|$, we finally obtain,
    $$\fbox{$\displaystyle t=\sqrt{\frac{ma}{2q^2}\frac{\sigma(\mu-1)}{|\lambda|(\sigma+1)}}\left\{
     \displaystyle\frac{1}{2}a(\mu-1)\cosh^{-1}\left[\displaystyle\frac{2x}{a(\mu-1)}-1\right]+\sqrt{x\left[x-a(\mu-1)
        \right]}
    \right\}$}$$
    \end{itemize}}
      \item \textsf{On collision, $x=0$. With this value for $x$, the time for collision is easily obtained from Eq.(\ref{s77}):
          $$\fbox{$\displaystyle t_c=\frac{a(\mu-1)}{2}\pi\sqrt{\displaystyle\frac{ma}{2q^2}\frac{\displaystyle
          \sigma(\mu-1)}{|\lambda|(\sigma+1)}}$}$$ }
\end{enumerate}
\item
  \begin{enumerate}
  \item \textsf{With the information provided in the statement of the problem, one can write the equations of motion for the charges as
      $$\beta_2\frac{dx_2}{dt}=\frac{q_1q_2}{\left(x_2-x_1\right)^2}$$
   $$\beta_1\frac{dx_1}{dt}=-\frac{q_1q_2}{\left(x_2-x_1\right)^2}$$
   Here $\beta_2=\beta$ and $\beta=1$. Multiplying the first equation by $\beta_1$ and second by $\beta_2$ and subtracting we get
   $$\frac{d}{dt}\left(x_2-x_1\right)=\left(\frac{\beta_1+\beta_2}{\beta_1\beta_2}\right)
   \frac{q_1q_2}{\left(x_2-x_1\right)^2}$$
      $$\Rightarrow \frac{dx}{dt}=q^2\frac{\lambda(1+\beta)}{\beta}\frac{1}{x^2}$$
      \begin{itemize}
    \vspace{0.5cm}
    \item \underline{CASE 1: $\lambda<0~(\lambda=-|\lambda|)$}
    $$\int x^2~dx=-q^2\frac{|\lambda|(1+\beta)}{\beta}\int dt$$
    $$\Rightarrow \frac{x^3}{3}=-q^2\frac{|\lambda|(1+\beta)}{\beta}t+C_4.$$
    Using the initial condition: $t=0,~x=x_0$, gives $C_4=x_0^3/3$. With this value for $C_4$, we get
    \begin{eqnarray}\label{s78}
     \fbox{$\displaystyle t=\frac{1}{3q^2}\frac{\beta}{|\lambda|(1+\beta)}
     \left[\displaystyle\frac{a^3}{8}(\mu-1)^3-x^3\right]$}
   \end{eqnarray}
        \vspace{0.5cm}
        \item \underline{CASE 2: $\lambda>0~(\lambda=|\lambda|)$}
    $$\int x^2~dx=q^2\frac{|\lambda|(1+\beta)}{\beta}\int dt$$
    $$\Rightarrow \frac{x^3}{3}=q^2\frac{|\lambda|(1+\beta)}{\beta}t+C_5.$$
       Using the initial conditions gives for $C_5=x_0^3/3$. Thus
      $$ \fbox{$\displaystyle t=\frac{1}{3q^2}\frac{\beta}{|\lambda|(1+\beta)}
     \left[\displaystyle x^3-\frac{a^3}{8}(\mu-1)^3\right]$} $$
      \end{itemize}}
\item \textsf{On collision, $x=0$. With this value for $x$, the time for collision is easily obtained from Eq.(\ref{s78}):
    $$\fbox{$\displaystyle t_c=\frac{a^3}{24q^2}\frac{\beta\left(\mu-1\right)^3}{|\lambda|\left(\beta+1\right)}$}$$}
\end{enumerate}
 \end{enumerate}
 \item
  \begin{enumerate}
  \item  \textsf{For this part of the problem we set: $\lambda=|\lambda|$. The situation is sketched in Figure 39a. The free-body diagram of the charge $|\lambda| q$ is shown in Figure 39b.
      Application of Newton's second law immediately gives
      $$T-\left(F+\sigma mg\cos\theta\right)=m\frac{v^2}{l}.$$
      Here $v$ is the tangential velocity of the charge  at the instant the string makes an angle $\theta$ with the downward vertical. $l$ is the length of the string given by $a(\mu-1)$. $F$ is the coulomb force between the charges and it is given by $\displaystyle\frac{|\lambda|q^2}{l^2}$. At the highest point B$_h$ where $\theta=\pi$, the tension, $T_{B_h}$ and the velocity $v_{B_h}$ are related by
      $$T_{B_h}=\frac{mv_{B_h}^2}{l}-\sigma mg +\frac{|\lambda|q^2}{l^2}.$$
      In order to find the minimum speed of the charge $|\lambda| q$ at the point B$_h$ to make one complete revolution, we set $\left.T_{B_h}\right|_{\textsf{min}}=0$ in the last equation to get
      $$0=\frac{m\left.v_{B_h}^2\right|_{\textsf{min}}}{l}-\sigma mg +\frac{|\lambda|q^2}{l^2}$$
      \begin{eqnarray}\label{s79}
      \Rightarrow\left.v_{B_h}^2\right|_{\textsf{min}}=\sigma g l-\frac{|\lambda|q^2}{ml}
      \end{eqnarray}
      Conservation of mechanical energy between the points B$_0$ and B$_h$ (see Figures 40a and 40b) gives
      $$\frac{1}{2}\sigma m \left.v_{B_0}^2\right|_{\textsf{min}}+\sigma mg l(1-\cos\alpha)=\frac{1}{2}\sigma m \left.v_{B_h}^2\right|_{\textsf{min}}+2\sigma mgl$$
      $$\Rightarrow \left.v_{B_0}\right|_{\textsf{min}}=\sqrt{\left.v_{B_h}^2\right|_{\textsf{min}}+2gl(1+\cos\alpha)}$$
      On using Eq.(\ref{s79}) we get,
      $$\left.v_{B_0}\right|_{\textsf{min}}=\sqrt{\sigma g l-\frac{|\lambda|q^2}{ml}+2gl(1+\cos\alpha)}$$
      Finally using $l=a(\mu-1)$ we get
      $$\fbox{$\displaystyle\left.v_{B_0}\right|_{\textsf{min}}=\sqrt{ga(\mu-1)(\sigma+2+2\cos\alpha)-\frac{|\lambda|q^2}{ma(\mu-1)}}$}.$$
  }
  \item  \textsf{The situation is depicted in Figure 41a. Using the free-body (Figure 41b) for the system one can write down the Newton's second law:
      \begin{eqnarray}
      F^{\textsf{net}}_{\textsf{system}}=qE-|\lambda|qE=(m+\sigma m)\textsf{a}\nonumber\\
      \Rightarrow \textsf{a}=\frac{qE\left(1-|\lambda|\right)}{m\left(1+\sigma\right)}\label{s80}
      \end{eqnarray}
      Here, a represents the common acceleration of the two charges. Next consider the net force on anyone of the charges (say $q$). From Figure 41c we have
      \begin{eqnarray}
      F^{\textsf{net}}_{m}=qE-\frac{|\lambda|q^2}{l^2}=m\textsf{a}\nonumber\\
      \Rightarrow \textsf{a}=\frac{qE}{m}-\frac{|\lambda|q^2}{ml^2}\label{s81}
      \end{eqnarray}
      Eqs.(\ref{s80}) and (\ref{s81}) gives
      $$\frac{qE\left(1-|\lambda|\right)}{m\left(1+\sigma\right)}=\frac{qE}{m}-\frac{|\lambda|q^2}{ml^2}$$
      simplifying,
      $$\frac{|\lambda|q}{l^2}=E\left(\frac{\sigma+|\lambda|}{1+\sigma}\right)$$
      On using $l=a(\mu-1)$, we get
      $$\fbox{$\displaystyle\mu=1+\frac{1}{a}\sqrt{\frac{q}{E_0}\frac{|\lambda|(1+\sigma)}{\sigma+|\lambda|}}$}.$$
      }
 \end{enumerate}
 \item \textsf{See Figure 42. Newton's second law in angular form gives
 $$\sum \tau=I(-\alpha)$$
 Here, the left hand side represents sum of the torques about the pivot O. $I$ is the moment of  inertia of the system about O and perpendicular to the length of the rod and $\alpha$ represents the angular acceleration of the system. The negative sign signifies that the angular displacement is opposite to the angular acceleration. Thus,
 $$|\lambda|qE\left(\frac{l}{2}\sin\theta\right)+qE\left(\frac{l}{2}\sin\theta\right)=-I\frac{d^2\theta}{dt^2}$$ The moment of inertia is calculated to be
 $$I=m\left(\frac{l}{2}\right)^2+\sigma m\left(\frac{l}{2}\right)^2=\frac{ml^2}{4}(1+\sigma)$$
 Thus,
 $$\frac{d^2\theta}{dt^2}+\left[\frac{2qE(1+|\lambda|)}{ml(1+\sigma)}\right]\sin\theta=0$$
 For small $\theta$, the angular frequency, $\omega$ is
 $$\omega^2=\frac{2qE(1+|\lambda|)}{ml(1+\sigma)}$$
 and the period of small oscillations, $T (\displaystyle\equiv\frac{2\pi}{\omega})$ is
$$\fbox{$\displaystyle T=2\pi\sqrt{\frac{ma}{2qE_0}\frac{(\mu-1)(\sigma+1)}{|\lambda+1|}}$}$$
     }
\end{enumerate}

          \newpage
\begin{center}
          \LARGE\bf\textsf{Appendix}
          \end{center}

         \begin{equation}\label{a1}
        \fbox{$ \displaystyle\int\displaystyle\frac{ d\alpha}{\displaystyle\left(\alpha^2+\beta^2\right)
              ^{\frac{3}{2}}}=\frac{1}{\beta^2}\frac{\alpha}{\displaystyle \sqrt{\alpha^2+\beta^2}}$}
              \end{equation}
              \textsf{\underline{Proof}}:
              \begin{eqnarray}
              \int\displaystyle\frac{ d\alpha}{\displaystyle\left(\alpha^2+\beta^2\right)
              ^{\frac{3}{2}}}&=&\int\frac{\beta\sec^2\gamma}{\left(\beta^2\tan^2\gamma+
              \beta^2\right)^{\frac{3}{2}}}\,d\gamma~~~~~~~~~~\left(\alpha\equiv\beta\tan\gamma\Rightarrow
              d\alpha=\beta\sec^2\gamma ~d\gamma\right)\nonumber\\
              &=&\int\frac{\beta\sec^2\gamma}{\beta^3\sec^3\gamma}\,d\gamma\nonumber\\
              &=&\frac{1}{\beta^2}\int\cos\gamma\,d\gamma\nonumber\\
              &=&\frac{1}{\beta^2}\sin\gamma\nonumber\\
              &=&\frac{1}{\beta^2}\frac{\alpha}{\displaystyle \sqrt{\alpha^2+\beta^2}}~~~~~~~~~~\left(\textsf{see Figure 43}\right)\nonumber
              \end{eqnarray}
              \vspace{0.5cm}
              \begin{equation}\label{a2}
                  \fbox{$ \displaystyle\int\displaystyle\frac{\displaystyle\sqrt{\alpha^2-\gamma^2}}{\displaystyle\beta^2
               +\gamma^2}\,d\gamma
              =-\sin^{-1}\left(\displaystyle\frac{\gamma}{\alpha}\right)-\frac{\displaystyle\sqrt{\alpha^2+\beta^2}}
              {\beta}\sin^{-1}
              \left(\frac{\displaystyle\beta}{\alpha}\sqrt{\frac{\displaystyle\alpha^2-\gamma^2}{\displaystyle
              \beta^2+\gamma^2}}\right)$}
              \end{equation}
              \textsf{\underline{Proof}}:
              \begin{eqnarray}
              \displaystyle\int\displaystyle\frac{\displaystyle\sqrt{\alpha^2-\gamma^2}}{\displaystyle\beta^2
               +\gamma^2}\,d\gamma&=&\int\frac{\sqrt{\alpha^2-\gamma^2}}{\displaystyle\beta^2
               +\gamma^2}.\frac{\sqrt{\alpha^2-\gamma^2}}{\sqrt{\alpha^2-\gamma^2}}\,d\gamma\nonumber\\
               &=&\int\frac{1}{\sqrt{\alpha^2-\gamma^2}}\left[\frac{{\alpha^2-\gamma^2}}
               {\beta^2+\gamma^2}\right]\,d\gamma\nonumber\\
               &=&\int\frac{1}{\sqrt{\alpha^2-\gamma^2}}\left[-1+\frac{\alpha^2+\beta^2}{\gamma^2+\beta^2}
               \right]\,d\gamma\nonumber\\
               &=&-\int\frac{1}{\sqrt{\alpha^2-\gamma^2}}\,d\gamma+\left(\alpha^2+\beta^2\right)
               \int\frac{1}{\left(\beta^2+\gamma^2\right)\sqrt{\alpha^2-\gamma^2}}\,d\gamma\nonumber\\
               &=&I_1+I_2\nonumber
              \end{eqnarray}
              \textsf{where,}
              $$I_1=-\int\frac{1}{\sqrt{\alpha^2-\gamma^2}}\,d\gamma
              ~~~\textsf{and}~~~I_2=\left(\alpha^2+\beta^2\right)\int\frac{1}{\left(\beta^2+\gamma^2\right)
              \sqrt{\alpha^2-\gamma^2}}\,d\gamma$$
               \textsf{Therefore,}
              \begin{eqnarray}
              I_1&=&-\int\frac{1}{\sqrt{\alpha^2-\gamma^2}}\,d\gamma\nonumber\\
              &=&-\int\frac{\alpha\cos\theta}{\alpha\cos\theta}\,d\theta~~~~~~~~~~\left(\gamma\equiv\alpha\sin\theta
              \Rightarrow d\gamma=\alpha\cos\theta~d\theta\right)\nonumber\\
              &=&-\theta\nonumber\\
              &=&-\sin^{-1}\left(\frac{\gamma}{\alpha}\right)\nonumber
              \end{eqnarray}
              \textsf{To evaluate $I_2$, we use the substitution
              $$\eta=\sqrt{\frac{\alpha^2-\gamma^2}{\beta^2+\gamma^2}}$$
              \begin{eqnarray}
              \Rightarrow\frac{d\eta}{d\gamma}&=&\displaystyle\frac{\sqrt{\beta^2+\gamma^2}\left(\displaystyle
              \frac{1}{2}\cdot
              \displaystyle\frac{-2\gamma}{\sqrt{\alpha^2-\gamma^2}}\right)-\sqrt{\alpha^2-\gamma^2}
              \displaystyle\left(\frac{1}{2}\cdot
              \displaystyle\frac{2\gamma}{\sqrt{\beta^2+\gamma^2}}\right)
              }{\beta^2+\gamma^2}\nonumber\\
              &=&-\left(\alpha^2+\beta^2\right)\cdot\frac{\gamma}{\sqrt{\beta^2+\gamma^2}}
              \cdot\frac{1}{\left(\beta^2+\gamma^2\right)
              \sqrt{\alpha^2-\gamma^2}}\nonumber\\
              \frac{d\gamma}{\left(\beta^2+\gamma^2\right)
              \sqrt{\alpha^2-\gamma^2}}&=&-\frac{1}{\alpha^2+\beta^2}\left(\frac{\sqrt{\beta^2+\gamma^2}}
              {\gamma}\right)d\eta\nonumber
              \end{eqnarray}
              Further,
              \begin{eqnarray}
              \eta^2&=&{\frac{\alpha^2-\gamma^2}{\beta^2+\gamma^2}}\nonumber\\
              \Rightarrow\gamma^2&=&\frac{\alpha^2-\beta^2\eta^2}{1+\eta^2}\nonumber\\
              \Rightarrow\beta^2+\gamma^2&=&\frac{\alpha^2+\beta^2}{1+\eta^2}\nonumber\\
              \Rightarrow\frac{\beta^2+\gamma^2}{\gamma^2}&=&\frac{\alpha^2+\beta^2}{1+\eta^2}\cdot
              \frac{1+\eta^2}{\alpha^2-\beta^2\eta^2}\nonumber\\
              \Rightarrow\frac{\sqrt{\beta^2+\gamma^2}}{\gamma}&=&\sqrt{\frac{\alpha^2+\beta^2}{\alpha^2-\beta^2
              \eta^2}}\nonumber
              \end{eqnarray}
              Thus,
              \begin{eqnarray}
              \frac{d\gamma}{\left(\beta^2+\gamma^2\right)
              \sqrt{\alpha^2-\gamma^2}}&=&-\frac{1}{\alpha^2+\beta^2}\sqrt{\frac{\alpha^2+\beta^2}{\alpha^2-\beta^2
              \eta^2}}~d\eta\nonumber
              \end{eqnarray}
             \begin{eqnarray}
             ~I_2&=&\left(\alpha^2+\beta^2\right)\int\frac{1}{\left(\beta^2+\gamma^2\right)
              \sqrt{\alpha^2-\gamma^2}}\,d\gamma\nonumber\\
             &=&\left(\alpha^2+\beta^2\right)\int-
             \frac{1}{\alpha^2+\beta^2}\sqrt{\frac{\alpha^2+\beta^2}{\alpha^2-\beta^2
              \eta^2}}~d\eta\nonumber\\
              &=&-\frac{\sqrt{\alpha^2+\beta^2}}{\beta}\int\displaystyle\frac{1}{\sqrt{\displaystyle\left(
              \alpha/\beta
              \right)^2-\eta^2}}\,d\eta\nonumber\\
              &=&-\frac{\sqrt{\alpha^2+\beta^2}}{\beta}\int\displaystyle\frac{
              \alpha/\beta\cos\theta'}
              {\alpha/\beta\cos\theta'}\,d\theta'~~~~~~~~~~\left(\eta=\frac{\alpha}{\beta}\sin\theta'
              \Rightarrow d\eta=\frac{\alpha}{\beta}\cos\theta'd\theta'\right)\nonumber\\
              &=&-\frac{\sqrt{\alpha^2+\beta^2}}{\beta}\theta'\nonumber\\
              &=&-\frac{\sqrt{\alpha^2+\beta^2}}{\beta}\sin^{-1}\left(\frac{\beta}{\alpha}\eta\right)\nonumber\\
              &=&-\frac{\sqrt{\alpha^2+\beta^2}}{\beta}\sin^{-1}\left(\frac{\beta}{\alpha}
              \sqrt{\frac{\alpha^2-\gamma^2}{\beta^2+\gamma^2}}\right)\nonumber
              \end{eqnarray}
              }
              \vspace{0.5cm}
             \begin{equation} \label{a3}
             \fbox{$\displaystyle\int\displaystyle\frac{\displaystyle\sqrt{\alpha}}{\displaystyle\sqrt{\beta
               -\alpha}}\,d\alpha=\beta~\sin^{-1}\left(\sqrt{\frac{\alpha}{\beta}}\right)+\sqrt{\alpha
               \left(\beta-\alpha
        \right)}$}
        \end{equation}
        \textsf{\underline{Proof}}:
        \begin{eqnarray}
        \int\displaystyle\frac{\displaystyle\sqrt{\alpha}}{\displaystyle\sqrt{\beta
               -\alpha}}\,d\alpha&=&\int\frac{\sqrt{\beta}\sin\theta\left(2\beta\sin\theta\cos\theta
               \right)}{\sqrt{\beta}\cos\theta}d\theta~~~~~~~~~~\left(\alpha\equiv\beta\sin^2\theta
               \Rightarrow d\alpha=2\beta\sin\theta\cos\theta~d\theta\right)\nonumber\\
               &=&2\beta\int\sin^2\theta\,d\theta\nonumber\\
               &=&2\beta\int\left(\frac{1-\cos2\theta}{2}\right)d\theta\nonumber\\
        &=&\beta\left(\theta+\frac{\sin2\theta}{2}\right)\nonumber\\
        &=&\beta\left(\theta+\sin\theta\cos\theta\right)\nonumber\\
        &=&\beta\sin^{-1}\left(\sqrt{\frac{\alpha}{\beta}}\right)+\beta\sqrt{\frac{\alpha}{\beta}}
        \cdot\frac{\sqrt{\beta-\alpha}}{\sqrt{\beta}}~~~~~~~~~~\left(\textsf{see Figure} 44\right)\nonumber\\
        &=&\beta~\sin^{-1}\left(\sqrt{\frac{\alpha}{\beta}}\right)+\sqrt{\alpha
               \left(\beta-\alpha
        \right)}\nonumber
        \end{eqnarray}
        \vspace{0.5cm}
       \begin{equation} \label{a4}
       \fbox{$\displaystyle\int\displaystyle\frac{\displaystyle\sqrt{\alpha}}{\displaystyle\sqrt{\alpha
               -\beta}}\,d\alpha=\frac{1}{2}\beta\cosh^{-1}\left(\frac{2\alpha}{\beta}-1\right)+\sqrt{\alpha
               \left(\alpha-\beta
        \right)}$}
        \end{equation}
     \textsf{\underline{Proof}}:
        \begin{eqnarray}
\int\displaystyle\frac{\displaystyle\sqrt{\alpha}}{\displaystyle\sqrt{\alpha
               -\beta}}\,d\alpha&=&\int\displaystyle\frac{\displaystyle\sqrt{\alpha}}{\displaystyle\sqrt{\alpha
               -\beta}}\frac{\sqrt{\alpha}}{\alpha}\,d\alpha\nonumber\\
               &=&\int\frac{\alpha}{\sqrt{\alpha^2-\beta\alpha}}\nonumber
\end{eqnarray}
\textsf{Note that
$$\frac{d}{d\alpha}\left(\alpha^2-\beta\alpha\right)=2\alpha-\beta$$
Need to write:
$$\alpha\equiv\lambda\left(2\alpha-\beta\right)+\mu$$
Comparing coefficients we get: $\lambda=1/2$ and $\mu=1/2\beta$.
Thus,}
\begin{eqnarray}
\int\displaystyle\frac{\displaystyle\sqrt{\alpha}}{\displaystyle\sqrt{\alpha
               -\beta}}\,d\alpha&=&\int\frac{1/2\left(2\alpha-\beta\right)+1/2\beta}{\sqrt{\alpha^2-\beta\alpha}}\nonumber\\
&=&\frac{1}{2}\beta\int\frac{1}{\sqrt{\alpha^2-\beta\alpha}}\,d\alpha+\frac{1}{2}\int\frac{2\alpha-\beta}
{\sqrt{\alpha^2-\beta\alpha}}\,d\alpha\nonumber\\
&=&I_1+I_2\nonumber
\end{eqnarray}
where,
$$I_1=\frac{1}{2}\beta\int\frac{1}{\sqrt{\alpha^2-\beta\alpha}}\,d\alpha~~~\textsf{and}~~~
I_2=\frac{1}{2}\int\frac{2\alpha-\beta}
{\sqrt{\alpha^2-\beta\alpha}}\,d\alpha$$
Therefore,
\begin{eqnarray}
I_1&=&\frac{1}{2}\beta\int\frac{1}{\sqrt{\alpha^2-\beta\alpha}}\,d\alpha\nonumber\\
&=&\int\frac{1}{\sqrt{\displaystyle\left(\alpha-\frac{1}{2}\alpha\right)^2-\left(\frac{1}{2}\beta\right)^2}}
\,d\alpha\nonumber\\
&=&\frac{1}{2}\beta\int\frac{1}{\displaystyle\sqrt{v^2-\left(\frac{1}{2}\beta\right)^2}}\,dv
~~~~~~~~~~\left(v\equiv\alpha-\frac{1}{2}\beta\Rightarrow dv=d\alpha\right)\nonumber\\
&=&\frac{1}{2}\beta\int\frac{1/2\beta\sinh\theta}{1/2\beta\sinh\theta}\,d\theta
~~~~~~~~~~\left(v\equiv\frac{1}{2}\beta\cosh\theta\Rightarrow dv=\frac{1}{2}\beta\sinh\theta~d\theta\right)\nonumber\\
&=&\frac{1}{2}\beta\theta\nonumber\\
&=&\frac{1}{2}\beta\cosh^{-1}\left(\frac{2v}{\beta}\right)\nonumber\\
&=&\frac{1}{2}\beta\cosh^{-1}\left(\frac{2\alpha}{\beta}-1\right)\nonumber
\end{eqnarray}
and
\begin{eqnarray}
I_2&=&\frac{1}{2}\int\frac{2\alpha-\beta}
{\sqrt{\alpha^2-\beta\alpha}}\,d\alpha\nonumber\\
&=&\frac{1}{2}\int\frac{1}{\sqrt{u}}\,du~~~~~~~~~~\left(u\equiv\alpha^2-\beta\alpha\Rightarrow
du=\left(2\alpha-\beta\right)\right)\nonumber\\
&=&\sqrt{u}\nonumber\\
&=&\sqrt{\alpha\left(\alpha-\beta\right)}\nonumber
\end{eqnarray}

\end{document}